\documentclass[twocolumn,tighten,twocolappendix]{aastex701}
\usepackage{wrapfig}
\usepackage{hyperref}
\hypersetup{hypertex=true,
colorlinks=true,
linkcolor=blue,
anchorcolor=blue,
citecolor=blue}
\usepackage{graphics,graphicx}
\usepackage{epstopdf}
\usepackage[utf8]{inputenc}
\usepackage{tikz}
\usetikzlibrary{positioning}
\usetikzlibrary{shapes,arrows}
\usepackage{amssymb, amsmath, amsthm}
\usepackage[normalem]{ulem}
\usepackage{float}
\usepackage{color}
\usepackage{xcolor}
\usepackage{subfigure}
\definecolor{ultramarine}{rgb}{0.01, 0.64, 0.86} 
\def\green#1 {{\textcolor{ultramarine}{#1}}\ }

\usepackage{tikz}
\usetikzlibrary{shapes,arrows}
\definecolor{deepgreen}{RGB}{0,100,0}

\newcommand{\Revise}[1]{{ #1}}

\begin{document}

% \pagestyle{empty}
% Define block styles
\tikzstyle{start} = [rectangle, draw, text width=1.5cm, rounded corners, text badly centered, node distance=3cm, inner sep=2pt, minimum height=3.5em]
\tikzstyle{process1} = [rectangle, draw,  
text width=2cm, text badly centered, node distance=2cm, inner sep=2pt, minimum height=4em]
\tikzstyle{process2} = [rectangle, draw,  
text width=2.5cm, text badly centered, node distance=2cm, inner sep=2pt, minimum height=4em]
\tikzstyle{result1} = [trapezium, trapezium left angle = 60,trapezium right angle = 120, draw, text width=1cm, text badly centered, node distance=1cm, inner sep=4pt, minimum height=1.5em]
\tikzstyle{result2} = [trapezium, trapezium left angle = 60,trapezium right angle = 120, draw, text width=1.5cm, text badly centered, node distance=1cm, inner sep=4pt, minimum height=1.5em]
\tikzstyle{choice} = [diamond,aspect = 1.2, draw, text width = 1.5cm, text badly centered, node distance = 1cm, inner sep=4pt, minimum height=0.8em]
\tikzstyle{line} = [draw, -latex']
\tikzstyle{point} = [coordinate, on grid]

\title{Trace the Self-Gravitating Gas Using CO Isotopologues}

\shorttitle{Trace bound gas using CO}

\correspondingauthor{Jingwen Wu, Sihan Jiao}
\email{jingwen@nao.cas.cn, sihanjiao@nao.cas.cn}

\author[0009-0003-4821-5502]{Linjing Feng}
\affiliation{National Astronomical Observatories, Chinese Academy of Sciences, 20A Datun Road, Chaoyang District, Beijing 100012, China}
\affiliation{University of Chinese Academy of Sciences, Beijing 100049, China}
\email{ljfeng@nao.cas.cn}

\author[0000-0001-7808-3756]{Jingwen Wu}
\affiliation{University of Chinese Academy of Sciences, Beijing 100049, China}
\affiliation{National Astronomical Observatories, Chinese Academy of Sciences, 20A Datun Road, Chaoyang District, Beijing 100012, China}
\email{jingwen@nao.cas.cn}

\author[orcid=0000-0002-9151-1388]{Sihan Jiao}
\affiliation{National Astronomical Observatories, Chinese Academy of Sciences, 20A Datun Road, Chaoyang District, Beijing 100012, China}
\affiliation{Max Planck Institute for Astronomy, Konigstuhl 17, D-69117 Heidelberg, Germany}
\email[]{sihanjiao@nao.cas.cn}

\author[0000-0002-7299-2876]{Zhi-Yu Zhang}
\affiliation{School of Astronomy and Space Science, Nanjing University, Nanjing 210093, China}
\affiliation{Key Laboratory of Modern Astronomy and Astrophysics, Ministry of Education, Nanjing 210093, China}
\email{zzhang@nju.edu.cn}

\author[0000-0001-6106-1171]{Junzhi Wang}
\affiliation{School of Physical Science and Technology, Guangxi University, Nanning 530004, China}
\email{junzhiwang@gxu.edu.cn}

\author[0000-0002-9390-9672]{Chao-Wei Tsai}
\affiliation{National Astronomical Observatories, Chinese Academy of Sciences, 20A Datun Road, Chaoyang District, Beijing 100012, China}
\affiliation{Institute for Frontiers in Astronomy and Astrophysics, Beijing Normal University,  Beijing 102206, China}
\affiliation{University of Chinese Academy of Sciences, Beijing 100049, China}
\email{cwtsai@nao.cas.cn}

\author[0000-0003-3010-7661]{Di Li}
\affiliation{New Cornerstone Science Laboratory, Department of Astronomy, Tsinghua University, Beijing 100084, China}
\affiliation{National Astronomical Observatories, Chinese Academy of Sciences, 20A Datun Road, Chaoyang District, Beijing 100012, China}
\email{dili@mail.tsinghua.edu.cn}

\author[0000-0003-2300-2626]{Hauyu Baobab Liu}
\affiliation{Department of Physics, National Sun Yat-Sen University, No. 70, Lien-Hai Road, Kaohsiung City 80424, Taiwan, R.O.C.}
\affiliation{Center of Astronomy and Gravitation, National Taiwan Normal University, Taipei 116, Taiwan}
\email[]{baobabyoo@gmail.com}

\author[0000-0002-3904-1622]{Yan Sun}
\affiliation{Purple Mountain Observatory, Chinese Academy of Sciences, Nanjing 210008, People’s Republic of China}
\email{yansun@pmo.ac.cn}

\author[0000-0001-5175-1777]{Neal J. Evans II}
\affiliation{Department of Astronomy, The University of Texas at Austin, 2515 Speedway, Stop C1400, Austin, 78712-1205, USA}
\email[]{nje@astro.as.utexas.edu}

\author[0000-0001-9299-5479]{Yuxin Lin}
\affiliation{Max-Planck-Institut f\"ur Extraterrestrische Physik, Giessenbachstr. 1, D-85748 Garching bei M\"unchen, Germany}
\email{ylin@mpe.mpg.de}

\author[orcid=0009-0003-7307-6209
]{Hao Ruan}
\affiliation{University of Chinese Academy of Sciences, Beijing 100049, China}
\affiliation{National Astronomical Observatories, Chinese Academy of Sciences, 20A Datun Road, Chaoyang District, Beijing 100012, China}
\email{ruanhao@bao.ac.cn}

\author[orcid=0009-0003-9223-2297]{Fangyuan Deng}
\affiliation{University of Chinese Academy of Sciences, Beijing 100049, China}
\affiliation{National Astronomical Observatories, Chinese Academy of Sciences, 20A Datun Road, Chaoyang District, Beijing 100012, China}
\email{dengfangyuan21@mails.ucas.ac.cn}

\author[orcid=0009-0000-8457-8720]{Yuanzhen Xiong}
\affiliation{University of Chinese Academy of Sciences, Beijing 100049, China}
\affiliation{National Astronomical Observatories, Chinese Academy of Sciences, 20A Datun Road, Chaoyang District, Beijing 100012, China}
\email{xiongyuanzhen23@mails.ucas.ac.cn}

\author[orcid=0009-0009-7371-1739]{Ruofei Zhang}
\affiliation{University of Chinese Academy of Sciences, Beijing 100049, China}
\affiliation{National Astronomical Observatories, Chinese Academy of Sciences, 20A Datun Road, Chaoyang District, Beijing 100012, China}
\email{zhangruofei24@mails.ucas.ac.cn}

\begin{abstract}

Recent studies have shown that the star formation rate (SFR) correlates tightly and linearly with the mass of gravitationally bound gas, which can be delineated from the power-law tail of the column-density probability distribution function ($N$-PDF) derived from dust emission observations.
This relationship holds across four orders of magnitude within the Milky Way—spanning low-mass to high-mass star-forming regions and encompassing the extreme environment of the Central Molecular Zone.
Building on this framework, we present a new approach for estimating the mass of gravitationally bound gas in molecular clouds using multi-line CO isotopologue observations.
Our sample includes 16 molecular clouds with robust detections in $^{12}$CO, $^{13}$CO, and C$^{18}$O $J$ = 1-0, spanning both massive inner Galaxy clouds and nearby star-forming regions. We find that the $N$-PDFs derived from combined CO isotopologue data recover the characteristic log-normal plus power-law profiles seen in dust-based studies. 
The mass and spatial distribution of the self-gravitating structures estimated from both dust-based and CO-based methods agree well throughout the sample. This indicates that the CO isotopologue combination can robustly trace the self-gravitating component via the $N$-PDF method and provides a reliable, scalable, and velocity-resolved alternative to dust emission for identifying the star-forming gas in molecular clouds.

\end{abstract}

\section{Introduction}
% SF law
% note: SFR important, complex

% Jiao 2025: importantce of Msgb
% note: begin: empirical relation & physical factor - gravity bound. high light: Msgb control SFR, constant SFE.

% deriving Msgb
% note: $N$-PDF, dust emission

% weakness of dust emission, and strength of CO

% weakness of single line, request of wide dynamical range (NPDF)

% In this work, we attempt to ...  pilot project of the large survey

Stars form in molecular clouds where gravity, turbulence, and magnetic fields interact in a complex balance that governs the formation and collapse of dense cores \citep[e.g.,][]{Shu1991ApJ,McKee2007ARA&A,Beuther2025arXiv}.
Motivated by this complexity, many studies have sought empirical relationships between the star formation rate (SFR) and the luminosities of various gas tracers on galactic scales, using such relationships to investigate how different phases of the interstellar medium relate to star formation activity \citep[e.g.,][]{Kennicutt_1998,Bigiel_2008,Gao_2004,Kennicutt_Evans_2012}.
%On galactic scales, empirical relationships between the star formation rate (SFR) and various gas tracers have long served as diagnostics of the star-forming interstellar medium \citep[e.g.,][]{Kennicutt_1998,Bigiel_2008,Gao_2004,Kennicutt_Evans_2012}. 
%In nearby molecular clouds, such relations become tighter when the gas mass is restricted to dense components \citep{Lada_2012,Evans_2014}, suggesting that not all molecular gas contributes equally to star formation.
More refined studies reveal that these relationships become much tighter when focusing only on the dense gas, for both local star-forming regions \citep[e.g.,][]{Lada_2012,Evans_2014} and distant galaxies \citep[e.g.,][]{Gao_2004,Wu_2005}, indicating that star formation is governed primarily by the amount of dense molecular gas rather than by the total gas reservoir \citep[e.g.,][]{Gao_2004,Wu_2005,Lada2010ApJ...724..687L,Lada_2012,Evans_2014}.

While studies have shown that star formation is primarily governed by the amount of dense molecular gas, it appears that dense gas alone does not always correlate directly with star formation. This highlights the need for more refined methods to isolate the star-forming portion of molecular clouds.
A promising avenue for achieving this is the use of the column-density probability distribution function \citep[$N$-PDF,][]{Kainulainen2009A&A,Federrath2013ApJ,Lombardi2015A&A,Schneider2015MNRAS}. 
Observationally, $N$-PDFs typically exhibit a log-normal form at low column densities, consistent with a turbulence-dominated component, while the high-column-density regime often shows excess emission above the log-normal \citep[][]{Kainulainen2013A&A...549A..53K,Schneider2015MNRAS,Chen_2018}. 
\Revise{This excess may appear as one or multiple power-law–like tails \citep[][]{Schneider2015MNRAS}. While such high-column-density tails are commonly associated with gravity-dominated structures \citep[][]{Lin2016,Lin2017ApJ...840...22L,Burkhart_2017}, their slopes may also be affected by stellar feedback \citep[][]{Tremblin_2014A&A...564A.106T} and magnetic fields \citep[][]{Schneider_2022A&A...666A.165S}.}
By identifying these power-law tails as the specific tracers of self-gravitating gas \citep{Ballesteros-Paredes2011MNRAS,Girichidis_2014ApJ...781...91G,Burkhart_2017,Burkhart_2019}, \citet{Jiao2025A&A} established a tight linear correlation between the mass of gravitationally bound gas and the SFR across a broad range of Galactic environments.
Their analysis shows that this bound gas mass vs. SFR relation holds over nearly five orders of magnitude in cloud mass, and presents a consistent star formation efficiency ($\sim0.4\%$ per million years) to convert bound gas into stars. That work further provides a natural explanation for the suppressed star formation in the Galactic Central Molecular Zone (CMZ): only a small fraction of dense gas in CMZ is truly gravitationally bound due to the high turbulence.

Most $N$-PDF studies use dust emission or extinction to trace gas column densities \citep[][]{Lombardi_2008A&A...489..143L,Kainulainen2009A&A,Froebrich_2010MNRAS.406.1350F,Stutz_2015A&A...577L...6S,Lin2016,Lin2017ApJ...840...22L,Schneider_2022A&A...666A.165S}. 
The low optical depth and wide dynamical range of dust thermal emission make it particularly effective for identifying the power-law regime of the $N$-PDF.
However, dust-based analyses are affected by foreground/background contamination along the line of sight (LOS), and are limited in coverage by the high cost of multi-band far-infrared observations.
\Revise{Column-density-screen subtraction can mitigate contamination along the LOS in dust-based column-density maps \citep[][]{Schneider_2015A&A...575A..79S, Ossenkopf_2016A&A...590A.104O}. However, when multiple molecular clouds overlap along the same LOS, subtracting a constant foreground or background column density may not fully recover the intrinsic column-density distribution of the target cloud.
For example, in \cite{Jiao2025A&A}, they have to select sources against overlapping dense clumps via molecular lines, then subtracted a constant column-density-screen estimated from nearby regions with uniform dust emission to generate the final column density map.
}

\Revise{Molecular line tracers provide an alternative means of probing gas column densities and constructing N-PDFs. 
Species such as CO, HCN, HCO$^+$, N$_2$H$^+$, and CS preferentially trace different density regimes owing to their different excitation requirements and critical densities \citep[e.g.,][]{Schneider2016A&A, Wang_2020A&A...641A..53W}. 
Among them, carbon monoxide (CO) is the most widely used tracer of molecular gas, owing to its relatively high abundance and the comparatively broader knowledge of its abundance in the Milky Way \citep[e.g.,][]{Goldsmith2008ApJ...680..428G, Goodman2009ApJ, Lo_2009MNRAS.395.1021L}.
Together with its isotopologues, CO provides a valuable complement to dust-based column density measurements, while its velocity information enables the separation of overlapping structures along the LOS in the Galactic disk.}
% As a widely used tracer of molecular gas, carbon monoxide (CO) and its isotopologues provide a valuable complement to dust  \citep[e.g.,][]{Goldsmith2008ApJ...680..428G, Goodman2009ApJ, Lo_2009MNRAS.395.1021L}, and their velocity information enables the separation of overlapping structures along the LOS in the Galactic disk. 
Individual CO isotopologues only probe column density over a limited dynamical range, owing to optical depth effects, excitation requirements, and abundance variations, but a combination of multiple CO isotopologues may remove this limitation. 
The existing large-scale surveys of CO isotopologues, e.g., FOREST unbiased Galactic plane imaging survey with the Nobeyama 45 m telescope \citep[FUGIN,][]{Umemoto2017PASJ}, Structure, Excitation, and Dynamics of the Inner Galactic InterStellar Medium \citep[SEDIGISM,][]{Schuller2017A&A}, the Milky Way Imaging Scroll Painting \citep[MWISP,][]{Yang_2026ApJS..282...65Y}, have delivered uniform, extensive CO coverage of the Milky Way.
Leveraging their ability to mitigate foreground and background contamination along the LOS, these surveys provide a channel for statistical studies of gravitationally bound gas across large samples of molecular clouds in the Galactic plane by using CO isotopologues.

Nevertheless, previous $N$-PDF studies based on a single CO isotopologue often yield purely log-normal shapes or multiple peaks, with little or no evidence for a power-law tail \citep[e.g.,][]{Schneider2016A&A,Ma_2022,Murase_2023MNRAS.523.1373M_CygnusX_COnpdf}.
This limitation arises naturally from the limited dynamical range of an individual CO line: at low column densities, CO is readily photodissociated or may remain subthermally excited, while at high column densities, the emission saturates due to high optical depth \citep[e.g.,][]{Goodman2009ApJ,Pineda2008ApJ,Galvan_2013}.
CO isotopologues such as $^{13}$CO and C$^{18}$O have lower optical depths, providing better sensitivity to the high-column-density regimes \citep[e.g.,][]{Pineda2008ApJ}. However, their emission is intrinsically weaker and still limited by a restricted dynamical range in the lower-density regimes.

To overcome these limitations, we utilize an optical depth-correction method that combines multiple CO isotopologues.
By jointly using several CO isotopologue transitions, the new approach allows us to trace the gas continuously across a wider range of column densities in molecular clouds than a single CO line, thus leads to a better recovery of the full $N$-PDF shape, including its power-law tail. This will enable a CO-based estimate of the self-gravitationally bound gas mass of molecular clouds. 
In this paper, we apply this method to a sample of 16 molecular clouds in the Milky Way, to investigate whether it can reliably identify the self-gravitating component of molecular clouds. 
%We demonstrate that the combined CO-based $N$-PDFs exhibit well-defined power-law tails consistent with dust-based results, and that the derived bound gas masses correlate tightly with SFR, providing a scalable alternative for tracing star-forming gas in the Milky Way.

% Given that C$^{18}$O is well detected across all clouds in our sample, and that $^{12}$CO is largely optically thick over extended regions, we adopt a combination of $^{13}$CO and C$^{18}$O J=1-0 lines to derive gas column densities. 
% Specifically, we use the $^{13}$CO and C$^{18}$O J=1-0 lines to derive optical depth-corrected $^{13}$CO column densities, using the line ratio as a diagnostic \citep[][]{Galvan_2013}. 
% The $^{12}$CO line is employed to estimate the excitation temperature.
% The method is, in principle, also applicable to sources with only $^{12}$CO and $^{13}$CO detections: in such cases, the $^{12}$CO/$^{13}$CO line ratio can be used to estimate the $^{13}$CO optical depth, allowing a modified implementation without requiring C$^{18}$O data.
The source selection and observational datasets are described in Section~\ref{sec_data}.
In Section~\ref{sec_method}, we present our method for constructing optical-depth-corrected column density maps by combining multiple CO isotopologues.
The resulting $N$-PDFs and the derivation of bound gas structures and masses are given in Section \ref{sec_result}.
The Discussion in Section~\ref{sec_discussion} addresses three aspects: a comparison between our CO-based method and conventional dense-gas tracers, its possible extension to other tracers, and the relation between bound gas and star formation rate.
A summary of our main findings is provided in Section \ref{sec_summary}.

\section{Data}
\label{sec_data}

\subsection{Sample selection}
\label{sub_source}

% To evaluate the feasibility of fitting the $N$-PDF using multiple CO lines, we selected molecular clouds from the sample presented in \citet{Jiao2025A&A}, whose dust-based $N$-PDFs have been analyzed in detail.
% This parent sample includes both nearby low-mass and distant high-mass star-forming regions, spanning a wide range of masses.
% Their star formation rates have been well characterized, providing a well-established reference for star-forming activity.
% An additional advantage of drawing the target clouds from this sample is that these clouds have been shown to be largely free of severe LOS overlap from multiple star-forming clumps \citep[][]{Jiao2025A&A}, making them well suited for our scientific goal.
% This allows us to perform a controlled comparison between the CO-based and dust-based approaches in relatively clean environments, minimizing contamination from unrelated structures along LOS.
% All selected targets have comprehensive coverage in the $^{12}$CO, $^{13}$CO, and C$^{18}$O $J=1$–$0$ transitions, enabling a direct comparison between the dust- and CO-based results.

To evaluate the feasibility of deriving $N$-PDFs using multiple CO lines, we selected molecular clouds from the sample presented in \citet{Jiao2025A&A}, for which dust-based $N$-PDFs have been analyzed in detail.
\Revise{This parent sample includes both nearby low-mass and distant high-mass star-forming regions, providing a broad basis for testing the method.}
All selected targets have comprehensive coverage in the $^{12}$CO, $^{13}$CO, and C$^{18}$O $J$=1-0 transitions, enabling a direct comparison between the dust- and CO-based results.

We began with three well-studied, nearby molecular clouds in the solar neighborhood: Orion A, Orion B, and Aquila as the benchmark clouds for our method.
These nearby, high-Galactic-latitude clouds have well-characterized gas distributions and star formation activity \citep[][]{Konyves2020A&A...635A..34K_HGBS_OrionB,Konyves2015A&A...584A..91K_HGBS_Aquila,Roy2013ApJ...763...55R_HGBS_OrionA,Lada2010ApJ...724..687L}.
They are also among the few nearby regions with complete and homogeneous $^{12}$CO, $^{13}$CO, and C$^{18}$O $J$=1--0 coverage, making them ideal test cases for evaluating the multi-line column density reconstruction.
Their locations help minimize contamination from foreground and background dust emission along the line of sight.
Moreover, the available observations provide high spatial resolution, enabling sufficient sampling of the $N$-PDF. 
These advantages make them ideal laboratories for comparing dust-based and CO-based $N$-PDF analyses.

To broaden our sample, we additionally included 13 molecular clouds in the Galactic disk.
These sources were selected from the disk sample of \citet{Jiao2025A&A} based on the availability of suitable $^{12}$CO, $^{13}$CO, and C$^{18}$O $J$=1--0 data.
For these regions, the gas distribution, gravitationally bound structures, and star formation rates have been well studied \citep[][]{Wu_2010,Jiao2025A&A}, providing a solid basis for testing the method under diverse Galactic conditions.
Our final sample spans more than two orders of magnitude in mass, from $\sim5\times10^{3}\,M_\odot$ to $1\times10^{6}\,M_\odot$, and covers distances from 0.4 to 11 kpc.
\Revise{This diversity enables a comparison of the dust- and CO-based $N$-PDFs across clouds with different masses, environments, and distances.}
The selected target list and a summary of their properties are given in Table~\ref{tab_sources}.

\begin{table*}[ht]
\centering
\caption{CO- and dust-based cloud and bound gas masses.}
    \begin{tabular}{lcccccccccc}
        \hline
        \hline
        Source & Distance & $R_{\rm gc}$ &
        $M_{\rm cloud}^{\rm CO_{comb}}$ & $M_{\rm bound}^{\rm CO_{comb}}$ &
        $M_{\rm cloud}^{\rm dust}$ & $M_{\rm bound}^{\rm dust}$ &
        ${\rm ^{13}CO/C^{18}O}$ & $V_{\rm LSR}$ & CO Data \\
        & [pc] & [kpc] & [$10^3\,{\rm M_\odot}$] & [$10^3\,{\rm M_\odot}$] &
        [$10^3\,{\rm M_\odot}$] & [$10^3\,{\rm M_\odot}$] & & [${\rm km\,s^{-1}}$] & Source \\
        (1) & (2) & (3) & (4) & (5) & (6) & (7) & (8) & (9) & (10) \\
        \hline
        Aquila & $278\pm13$ & 8.03 & 24 & $8.3_{-0.8}^{+0.8}$ & 19 & $6.3_{-0.6}^{+0.6}$ & 11.7 & $0.3\sim13.6$ & MWISP \\
        OrionA & $399\pm19$ & 8.63 & 35 & $17_{-2}^{+2}$ & 13 & $7.1_{-0.7}^{+0.7}$ & 18.7 & $2.8\sim13.8$ & NRO-SF \\
        OrionB & $400\pm20$ & 8.64 & 10 & $4.1_{-0.5}^{+0.4}$ & 3.6 & $1.6_{-0.2}^{+0.2}$ & 20.8 & $6.3\sim13.4$ & IRAM-OB \\
        G10.6$-$0.4 & $5000\pm500$ & 3.48 & 89 & $38_{-8}^{+8}$ & 49 & $31_{-6}^{+6}$ & 6.2 & $-9.7\sim4.3$ & FUGIN \\
        G34.26+0.15 & $3800\pm300$ & 5.56 & 63 & $16_{-3}^{+3}$ & 93 & $49_{-8}^{+8}$ & 5.2 & $48.6\sim65.8$ & FUGIN \\
        G35.20$-$0.74 & $2200\pm200$ & 6.60 & 24 & $9_{-2}^{+2}$ & 30 & $17_{-3}^{+3}$ & 8.1 & $26.7\sim42.6$ & FUGIN \\
        W51M & $5400\pm300$ & 6.29 & 170 & $113_{-13}^{+13}$ & 200 & $133_{-15}^{+15}$ & 8.0 & $44.3\sim74.3$ & FUGIN \\
        W49N & $11100\pm900$ & 7.60 & 360 & $247_{-45}^{+44}$ & 580 & $364_{-59}^{+59}$ & 10.5 & $-4.1\sim25.9$ & FUGIN \\
        W43M & $5300\pm500$ & 4.61 & 430 & $103_{-19}^{+19}$ & 220 & $85_{-16}^{+16}$ & 7.0 & $78.0\sim108.0$ & FUGIN \\
        W33 & $4500\pm400$ & 4.01 & 290 & $55_{-10}^{+10}$ & 430 & $121_{-21}^{+21}$ & 6.5 & $26.1\sim44.1$ & FUGIN \\
        W31 & $3200\pm100$ & 5.14 & 110 & $43_{-3}^{+3}$ & 83 & $35_{-2}^{+2}$ & 5.7 & $1.0\sim22.0$ & FUGIN \\
        DR21S & $1500\pm100$ & 8.19 & 16 & $7.8_{-1.0}^{+1.0}$ & 16 & $10.5_{-1.4}^{+1.4}$ & 12.3 & $-10.0\sim12.0$ & NRO-CX \\
        W75N & $1300\pm100$ & 8.19 & 8.0 & $4.3_{-0.7}^{+0.7}$ & 9.0 & $4.6_{-0.7}^{+0.7}$ & 10.2 & $-10.0\sim18.0$ & NRO-CX \\
        G9.62+0.10 & $5200\pm500$ & 3.27 & 17 & $2.0_{-0.4}^{+0.4}$ & 29 & $7_{-1.4}^{+1.4}$ & 7.4 & $-3.7\sim13.1$ & MWISP \\
        G12.89+0.49 & $2500\pm300$ & 5.86 & 6.8 & $4_{-1.0}^{+1.0}$ & 9.4 & $5.1_{-1.2}^{+1.2}$ & 6 & $27.7\sim38.1$ & FUGIN \\
        G59.78+0.06 & $2200\pm100$ & 7.42 & 14 & $5.4_{-0.6}^{+0.8}$ & 8.3 & $3.6_{-0.3}^{+0.3}$ & 14.3 & $16.8\sim25.4$ & MWISP \\
        \hline
        \hline
    \end{tabular}
    \label{tab_sources}
    \tablecomments{
    Columns:
    (1) Name of the molecular cloud. 
    (2) Distance to the source, with uncertainties, compiled from \citealt{Zucker2020A&A...633A..51Z}.
    (3) Galactocentric radius $R_{\rm gc}$ in kiloparsecs.
    (4) Total cloud mass estimated from CO isotopologue data.
    (5) gravitationally bound gas mass ($M_{\rm bound}^{\rm CO}$), derived from the power-law tail of the $N$-PDF obtained using CO data (see Section~\ref{sec_method}).
    (6) Total cloud mass estimated from dust emission.
    (7) gravitationally bound gas mass ($M_{\rm bound}^{\rm dust}$) derived from the dust-based $N$-PDF analysis.
    (8) $^{13}$CO/C$^{18}$O abundance ratio, estimated using our stacking line method in optically thin regions (Section~\ref{subsub_1318ratio}).
    (9) Velocity range ($V_{\rm LSR}$) over which CO emission is integrated to construct column density maps.
    (10) Source of the CO isotopologue data: MWISP = Milky Way Imaging Scroll Painting survey (Section \ref{sub_pmo}); 
    FUGIN = FOREST Unbiased Galactic Plane Imaging survey (Section \ref{sub_NRO}); 
    NRO-SF = NRO Star Formation Project (Section \ref{sub_NRO}); 
    IRAM-OB = IRAM 30\,m Orion B Survey (Section \ref{sub_iram}); 
    NRO-CX = NRO Cygnus-X Survey (Section \ref{sub_NRO}).
    All mass values are in units of $10^3\,{\rm M_\odot}$. Velocity ranges are in km~s$^{-1}$.
    Uncertainties in $M_{\rm bound}$ reflect propagation from optical depth corrections and abundance ratio estimates.
    }
\end{table*}

\subsection{CO isotopologue spectral lines}
\label{sub_COdata}

\Revise{All CO isotopologue data used in this work correspond to the $J$=1-0 transitions and were retrieved from archival surveys, conducted with the Nobeyama 45 m telescope, the Purple Mountain Observatory (PMO) 13.7 m telescope, and the IRAM 30 m telescope.}

\subsubsection{Nobeyama 45m Telescope}
\label{sub_NRO}

The majority of the $^{12}$CO, $^{13}$CO, and C$^{18}$O ($J=1$--0) data used in this work were obtained with the 45 m telescope at the Nobeyama Radio Observatory\footnote{Nobeyama Radio Observatory is a branch of the National Astronomical Observatory of Japan.} (NRO), with the exception of the Aquila, Orion B, G9.62+0.10, and G59.78+0.06 molecular clouds.

The data of Orion A are obtained from the NRO Star Formation Project \citep[][]{Shimajiri2011PASJ...63..105S_NobeyamaOrionA12CO,Shimajiri2014A&A...564A..68S_NobeyamaOrionA13COC18O}.
This dataset covers the northern portion of the cloud, including OMC 1-4 and L1641 N. 
The angular resolution is $\sim21\arcsec$, with typical RMS noise levels of $\sim0.9$ K for $^{12}$CO, $\sim0.3$ K for $^{13}$CO, and $\sim0.2$ K for C$^{18}$O.

The data of W75 N and DR21 S are from the NRO Cygnus-X Survey project \citep[][]{Yamagishi_2018_NROCygnusX}. 
To balance between angular resolution and sensitivity for adequate sampling of the $N$-PDF, we use the medium-resolution data products. 
These have a uniform angular resolution of $23\arcsec$ for all lines, with typical RMS noise levels of $\sim1.5$ K for $^{12}$CO and $\sim0.7$ K for both $^{13}$CO and C$^{18}$O.

The remaining datasets are drawn from the FOREST Unbiased Galactic plane Imaging survey with the Nobeyama 45-m telescope (FUGIN, \citealt{Umemoto2017PASJ}). 
The observations were carried out using three different receivers: the 4-beam dual-polarization 100~GHz receiver FOREST \citep{Minamidani2016SPIE.9914E..1ZM_FORESTreceiver}, the 25-beam single-polarization BEARS, and the single-beam dual-polarization TZ receiver. 
%Among these, FOREST is particularly well suited for large-scale mapping due to its sensitivity and multiplexing capability.
%The half-power beam width (HPBW) of the 45m telescope is $14\arcsec$ at 115GHz and $15\arcsec$ at 110GHz. 
The final FUGIN data cubes used in this work have beam sizes of $\sim20\arcsec$ for $^{12}$CO and $\sim21\arcsec$ for $^{13}$CO and C$^{18}$O. 
The average RMS noise levels (in $T_\mathrm{mb}$) are approximately $\sim1.4$ K for $^{12}$CO, and $\sim0.7$ K for both $^{13}$CO and C$^{18}$O.

\subsubsection{PMO 13.7m Telescope}
\label{sub_pmo}

Data for the Aquila, G9.62+0.10, and G59.78+0.06 molecular clouds were obtained from the Milky Way Imaging Scroll Painting (MWISP\footnote{\url{http://english.dlh.pmo.cas.cn/ic/in/}}) project \citep{Su2019ApJS}.
MWISP is a large-scale, unbiased CO survey of the northern Galactic plane, carried out with the Purple Mountain Observatory (PMO) 13.7 m millimeter-wavelength telescope. 
The observations simultaneously mapped the $^{12}$CO, $^{13}$CO, and C$^{18}$O ($J=1$--0) lines using the nine-beam Superconducting Spectroscopic Array Receiver (SSAR). 
The data have an angular resolution of $\sim50\arcsec$ for all three lines. 
At a channel width of 0.16 km s$^{-1}$, the typical RMS noise levels are $\sim0.5$ K for $^{12}$CO and $\sim0.3$ K for $^{13}$CO and C$^{18}$O.

\subsubsection{IRAM 30m Telescope}
\label{sub_iram}

Data for the Orion B molecular cloud are from the Orion-B project \citep{Pety2017A&A...599A..98P_IRAMOrionB1}, conducted with the IRAM 30m telescope. 
The observations cover a $\sim0.8^\circ \times 1^\circ$ region encompassing the dense southern part of Orion B, including the H II regions NGC 2023 and NGC 2024, where interactions between molecular gas and H II regions have been identified \citep{Bik2003A&A...404..249B_OrionBHIIregion}.

The final data products were convolved to a uniform angular resolution of $31\arcsec$ for all three lines. 
At a channel width of 0.5 km s$^{-1}$, the typical RMS noise levels are $\sim0.18$ K for $^{12}$CO and $\sim0.07$ K for $^{13}$CO and C$^{18}$O.

\subsection{Dust-based gas column density}
\label{sub_Herschel}

The gas column density maps of the nearby star-forming clouds Orion A, Orion B, and Aquila were derived from archival data of the {\it Herschel} Gould Belt Survey (HGBS\footnote{http://www.herschel.fr/cea/gouldbelt/en/index.php}; \citealt{Andre2010_HGBS}). The HGBS targets the bulk of the nearest ($d \leq 0.5$ kpc) molecular cloud complexes in the Galaxy, most of which are located within the Gould Belt - a giant ($\sim$700 pc $\times$ 1000 pc), flattened structure inclined by $\sim$$20^\circ$ to the Galactic plane. The survey was conducted with the SPIRE (250-500 $\mu$m) and PACS (70-160 $\mu$m) instruments onboard {\it Herschel}, achieving a typical angular resolution of $18\arcsec$-$36\arcsec$ and a 5$\sigma$ column density sensitivity of $N_{\rm H_2} \sim 10^{21}$ cm$^{-2}$ (or $A_V\sim1$). In this study, we utilized the standard-resolution HGBS column density maps, rather than the high-resolution `hires' products, to ensure a more complete and unbiased sampling of the extended gas structures that are critical for the $N$-PDF analysis.

For the more distant molecular clouds located along the Galactic plane, we derived $\mathrm{H_2}$ column density maps by performing pixel-by-pixel spectral energy distribution (SED) fitting using archival far-infrared images from the {\it Herschel} Infrared Galactic Plane Survey (HI-GAL; \citealt{Molinari2010PASP..122..314M_HIGAL}). HI-GAL mapped the inner Galaxy ($|l| \leq 60^\circ$, $|b| \leq 1^\circ$) with PACS and SPIRE across five photometric bands centered at 70, 160, 250, 350, and 500~$\mu$m, with corresponding beam sizes of approximately 8\farcs5, 13\farcs5, 18\farcs2, 24\farcs9, and 36\farcs3, respectively \citep{Marsh2017MNRAS.471.2730M_HIGAL_PPMAP}.
\Revise{We adopted the Level 2.5 processed products for all bands. The archival SPIRE maps have been corrected for absolute zero-point offsets using {\it Planck} data, whereas the PACS maps do not include such corrections. We therefore calibrated the absolute background levels of the PACS maps by applying a linear transformation derived from {\it IRAS} and {\it Planck} data, following \citet{Bernard_2010A&A...518L..88B} and \citet{Marsh2017MNRAS.471.2730M_HIGAL_PPMAP}.}
% We adopted the Level 2.5 processed extended emission products, which have been corrected for absolute zero-point offsets using {\it Planck} data. To ensure consistent spatial sampling across all wavelengths, all images were first regridded and convolved to a common resolution. 
Assuming optically thin dust emission and a single temperature component, we performed modified blackbody SED fitting at each pixel to derive gas column density $N_{\rm H_2}^{\rm dust}$ and dust temperature $T_{\rm d}$, resulting in final maps at a common angular resolution of 36\farcs3 (see Appendix~\ref{appendix_dustSED} for details on SED fitting).

\Revise{After SED fitting, we applied an LOS contamination correction following \citet{Schneider_2015A&A...575A..79S}. Dust continuum emission can be contaminated by foreground/background diffuse emission along the LOS, which can be corrected by subtracting a constant column density screen. The impact of this correction on the derived $N$-PDFs is discussed in detail by \citet{Ossenkopf_2016A&A...590A.104O}. 
It should be noted, however, some distant Galactic-plane clouds in our sample contain multiple overlapping clouds along the LOS (e.g. CO observations reveal multiple velocity components). This contamination cannot be fully removed by subtracting a constant foreground/background level.}

% \begin{table*}[tp]
%     \begin{center}
%     \caption{Properties of the Observation in Each Line}
%         \begin{tabular}{c c c c c c c}
%             \hline
%             \hline
%             Line & Rest Frequency & Velocity Resolution & Spatial Resolution & $\mathrm{n_{crit}}$ & $\mu$ & $\mathrm{B_0}$\\ 
%              & (GHz) & ($\mathrm{km\; s^{-1}}$) & (arcsec) & $\mathrm{cm^{-3}}$ & Debye & MHz \\
%             \hline
%             \rule{0pt}{12pt}
%             $^{12}$CO & 115.271 & 0.11 & 21.7 & $1\times10^{3}$ & 0.11011 & 57635.968\\
%             $^{13}$CO & 110.201 & 0.11 & 22.1 & $1\times10^{3}$ & 0.11046 & 55101.011\\
%             C$^{18}$O & 109.782 & 0.11 & 22.2 & $1\times10^{3}$ & 0.11079 & 54891.420\\
%             \hline
%         \end{tabular}
%     \end{center}
% \label{tab1}
% \end{table*}

\section{Method}
\label{sec_method}

\begin{figure}[!ht]
\centering
\begin{tikzpicture}[node distance = 1.5cm, auto]
    \node [start,fill=white] (13CO) {$^{13}$CO};
    % \node [process1,fill=white,below of = 13CO, node distance = 2.5cm] (13CO_thin) {Optically thin assumption} ;
    \node [point, below of = 13CO, node distance = 4.5 cm] (Nthin_point1) {};
    \node [point, below of = 13CO, node distance = 4.6 cm] (Nthin_point2) {};
    \node [result2,fill=orange!25,right of = Nthin_point1, node distance = 3cm] (Nthin) {$N_{\rm ^{13}CO,thin}$};
    \draw [line] (Nthin_point1) -- (Nthin);
    \draw [-] (13CO) -- (Nthin_point1) node[midway, above, rotate=90] {Optically thin assumption};

    \node [start,fill=white,right of = 13CO,node distance = 6cm] (C18O) {C$^{18}$O} ;
    \node [process1,fill=blue!20,right of = 13CO,node distance = 3cm] (stacking) {Stack spectral lines in optically thin regions};
    \path [line] (C18O) -- (stacking);
    \path [line] (13CO) -- (stacking);
    \node [result2,fill=blue!20,below of = stacking,node distance = 1.8cm] (iso_ratio){$^{13}$CO/C$^{18}$O};
    \node [result1,fill=blue!20,below of = iso_ratio,node distance = 1.3cm] (tau_13CO) {$\tau_{\rm ^{13}CO}$};
    \draw [line] (stacking) -- (iso_ratio);
    \draw [line] (iso_ratio) -- (tau_13CO);

    \node [point, below of = 13CO, node distance = 0.1cm] (13CO2tau13_0) {};
    \node [point, right of = 13CO2tau13_0, node distance = 0.825cm] (13CO2tau13_1) {};
    \node [point, right of = 13CO2tau13_1, node distance = 0.7cm] (13CO2tau13_2) {};
    \draw [-] (13CO2tau13_1) -- (13CO2tau13_2) ;
    \draw [line] (13CO2tau13_2) |- (tau_13CO);
    \draw [line] (C18O) |- (tau_13CO);

    \node [choice,fill=white,below of = Nthin,node distance = 1.8cm] (ifC18O) {C$^{18}$O detection};
    \node [point, right of = ifC18O, node distance = 3.1cm] (rightofifC18O) {} ;
    \node [result2,fill=blue!20,below of = rightofifC18O,node distance = 0.8cm] (Nthick) {$N_{\rm ^{13}CO,thick}$};
    \node [point, below of = tau_13CO, node distance = 0.75cm] (belowoftau_13CO) {} ;
    \draw [-] (tau_13CO) -- (belowoftau_13CO) ;
    \draw [line] (belowoftau_13CO) -| (Nthick) ;
    % \node [point, below of = Nthin, node distance = 0.3cm] (Nthin2Nthick_0) ;
    % \node [point, right of = Nthin2Nthick_0, node distance = 0.25cm] (Nthin2Nthick_1) ;
    \draw [line] (Nthin) -- (ifC18O) ;
    \draw [-] (ifC18O) -- node[above] {Yes} (rightofifC18O) ;

    \node [process2,fill=green!20,below of = Nthick,node distance = 1.5cm] (12C_13C) {A Galactic gradient model of $^{12}$C/$^{13}$C};
    \draw [line] (Nthick) -- (12C_13C) ;
    \draw [line] (ifC18O) |- (12C_13C);
    \node [point, left of = 12C_13C,node distance = 2.25cm] (leftof12C) {};
    \node [above of = leftof12C,node distance = 0.2cm] {No};
    \node [result2,fill=green!20,below of = 12C_13C,node distance = 1.5cm] (N12CO) {$N_{\rm ^{12}CO}$};
    \draw [line] (12C_13C) -- (N12CO) ;
    \node [process2,fill=green!20,below of = N12CO,node distance = 1.5cm] (CO_H2) {A metallicity dependent CO abundance};
    \draw [line] (N12CO) -- (CO_H2) ;
    \node [start,fill=green!20,left of = CO_H2,node distance = 3cm] (NH2) {$N_{\rm H_2}^{\rm COcomb}$};
    \draw [line] (CO_H2) -- (NH2) ;

    \node [result1,fill=orange!25, below of = Nthin_point2, node distance = 1.5cm] (Tex) {$T_{\rm ex}$};
    % \node [process1,fill=white,below of = Tex, node distance = 1.5cm] (12CO_thick) {Optically thick assumption};
    \node [start,fill=white,below of = Tex,node distance = 5.5cm] (12CO) {$^{12}$CO};
    \draw [line] (12CO) -- (Tex) node[midway, above, rotate=90] {Optically thick assumption};
    \draw [-] (Tex) -- (Nthin_point2);
    \node [point, right of = Nthin_point2, node distance = 1.87 cm] (Nthin_point3) {};
    \draw [line] (Nthin_point2) -- (Nthin_point3);
\end{tikzpicture}
\caption{The flowchart of our method for combining data from three CO isotopologues to obtain the gas column density.
The orange boxes denote the $^{13}$CO optically thin assumption, where the integrated $^{13}$CO intensity is directly converted to $N_{\rm ^{13}CO,thin}$ (Section \ref{subsub_N13CO_thin}).
The purple boxes correspond to the derivation of the $^{13}$CO optical depth from the $^{13}$CO/C$^{18}$O line ratio and the subsequent correction of the $^{13}$CO column density (Section \ref{subsub_N13CO_thick} and \ref{subsub_1318ratio}).
The green boxes illustrate how we employ an isotopic abundance gradient model together with a metallicity-dependent CO abundance prescription to convert $^{13}$CO column densities into $N_{\rm H_2}$ (Section \ref{subsub_NH2_COcomb}).
%\jnote{hard to follow. Maybe could try to link the boxed here with subsections in Section 3.1.}
%\fnote{done.}
}
\label{fg:flowchart}
\end{figure}
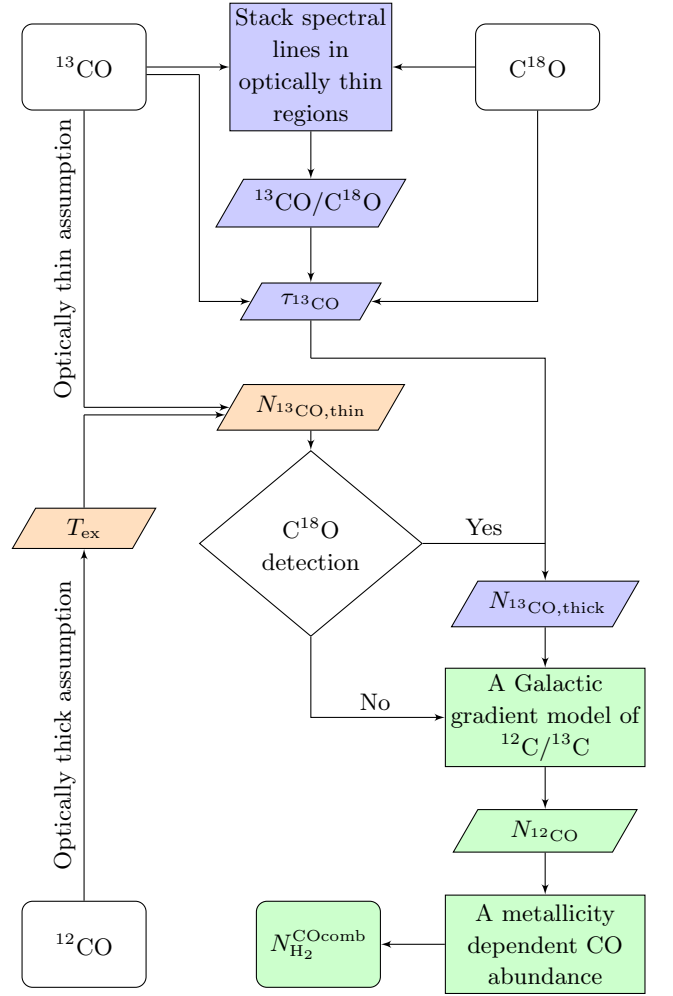

\subsection{CO-based gas column density}
\label{sub_NH2_CO}

\begin{figure*}[htp!]
        \begin{tabular}{ c c }
            \hspace{-0.5cm}\includegraphics[height = 7.6cm]{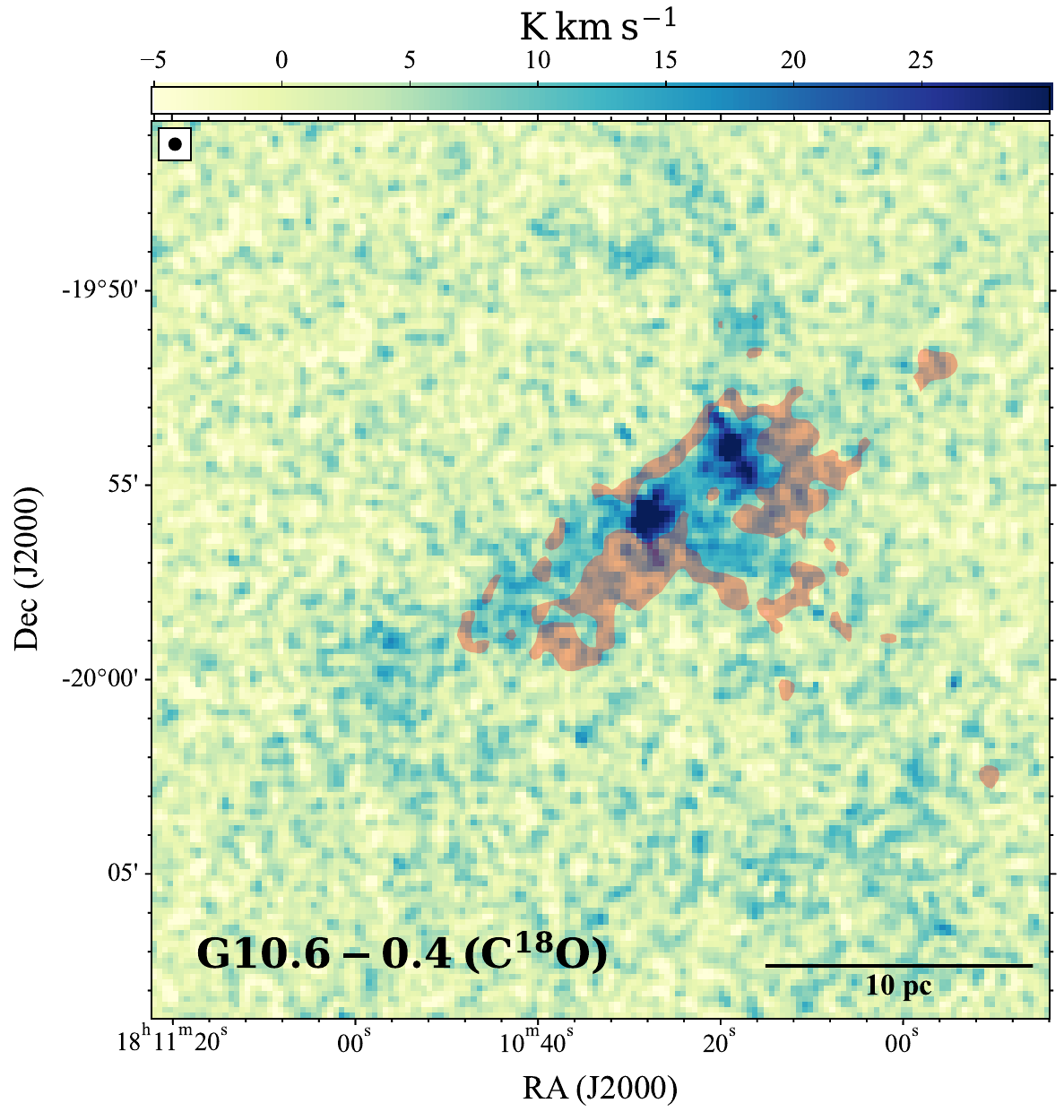} & 
            \includegraphics[height = 7.25cm]{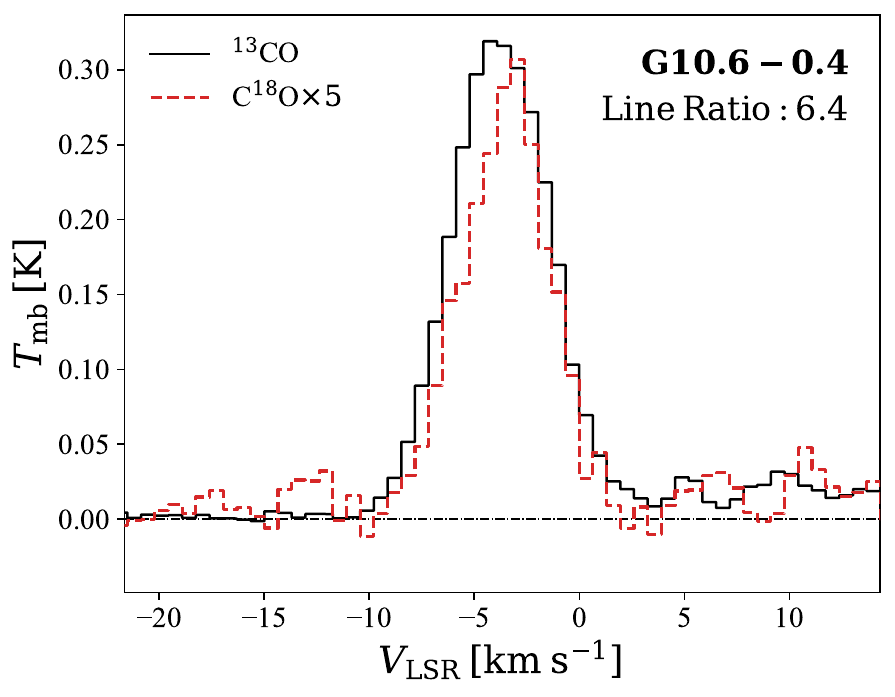} \\
        \end{tabular}
    \caption{Example of the method used to estimate the abundance ratio, applied to the G10.6-0.4 region. 
    {\it Left:} Moment 0 map of C$^{18}$O, with the shaded area indicating the region selected for stacking the $^{13}$CO and C$^{18}$O spectrum.
    {\it Right:} Stacked spectra of $^{13}$CO (black solid step line) and C$^{18}$O (red dashed step line, scaled by a factor of 5 for comparison).
    The line ratio shown in the upper right corner corresponds to the derived $^{13}$CO/C$^{18}$O abundance ratio.
    }
\label{fg_R1318_G10}
\end{figure*}

We estimate molecular gas column densities by combining multiple CO isotopologue $J$=1-0 transitions using a unified, optical-depth–aware framework.
In general, the intensity ratio between two isotopologue lines with different optical depths can be used to estimate the optical depth of the more opaque tracer, allowing their complementary information to be combined to recover the underlying gas column density.
In principle, this procedure can be extended iteratively to incorporate more than two transitions and thereby increase the dynamical range of the derived column densities.
For our scientific objective of tracing gravitationally bound gas, combining a single isotopologue pair is sufficient, and we therefore focus on the two-line case throughout this paper.
The overall workflow of the method is summarized in Figure~\ref{fg:flowchart}.

For the molecular clouds studied here, C$^{18}$O is detected in the main dense regions of the targets, whereas $^{12}$CO emission is largely optically thick over extended regions. 
\Revise{We therefore adopt the $^{13}$CO+C$^{18}$O $J$=1-0 line pair as our primary tracer pair of the gas column density. 
In principle, the same framework can be applied to other CO isotopologue line pairs, including higher-$J$ transitions of $^{13}$CO and C$^{18}$O, or pairs involving different isotopologues, such as $^{12}$CO+$^{13}$CO. 
However, different transitions may preferentially trace gas under different excitation conditions, and the optical-depth ranges of the selected isotopologues must remain suitable for constraining the column density. }
For regions where C$^{18}$O emission is weak or undetected, we demonstrate the application of this method to the $^{12}$CO+$^{13}$CO $J$=1-0 pair using two molecular clouds in Appendix~\ref{appendix_test12+13}.

The velocity range for column-density integration is determined via multi-Gaussian fitting of the $^{13}$CO spectra for each source.

\subsubsection{$N_{\rm ^{13}CO}$ Under the Optically Thin Approximation}
\label{subsub_N13CO_thin}

% \begin{figure*}[htp!]
%         \begin{tabular}{ p{0.47\linewidth}p{0.47\linewidth} }
%         \hspace{-0.8cm}\includegraphics[height = 7.6cm]{Aquila_NH2COcomb_showmap.pdf} & \hspace{-0.4cm}\includegraphics[height = 7.6cm]{Aquila_NH2dust_showmap.pdf} \\
%         \end{tabular}
%     \caption{
%     {\it Left:} The gas column density derived by combining CO isotopologues (Section \ref{sub_NH2_CO}).
%     {\it Right:} The gas column density derived by fitting the gray body spectrum of the dust far-infrared emission (Section \ref{sub_Herschel} and Appendix \ref{appendix_dustSED}).
%     Both panels use the same color bar and dynamic range, enabling a direct visual comparison of the gas structures traced by the two methods.
%     The gray dashed contours indicate the last closed contour adopted in the $N$-PDF fitting analysis.
%     }
% \label{fg_showmap}
% \end{figure*}

\begin{figure}[htbp!]
    \centering
    \includegraphics[width = 8.5cm]{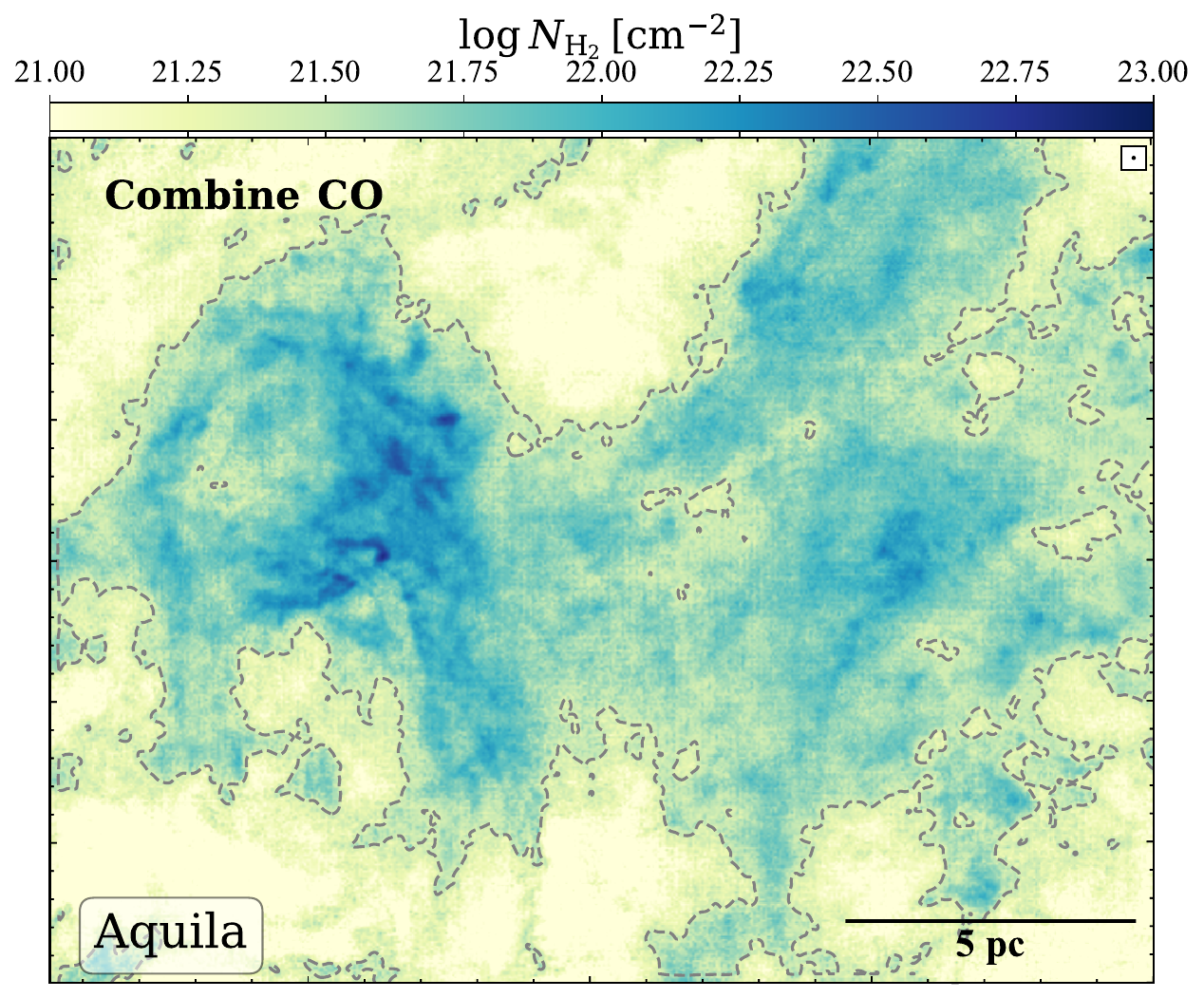}
    \includegraphics[width = 8cm]{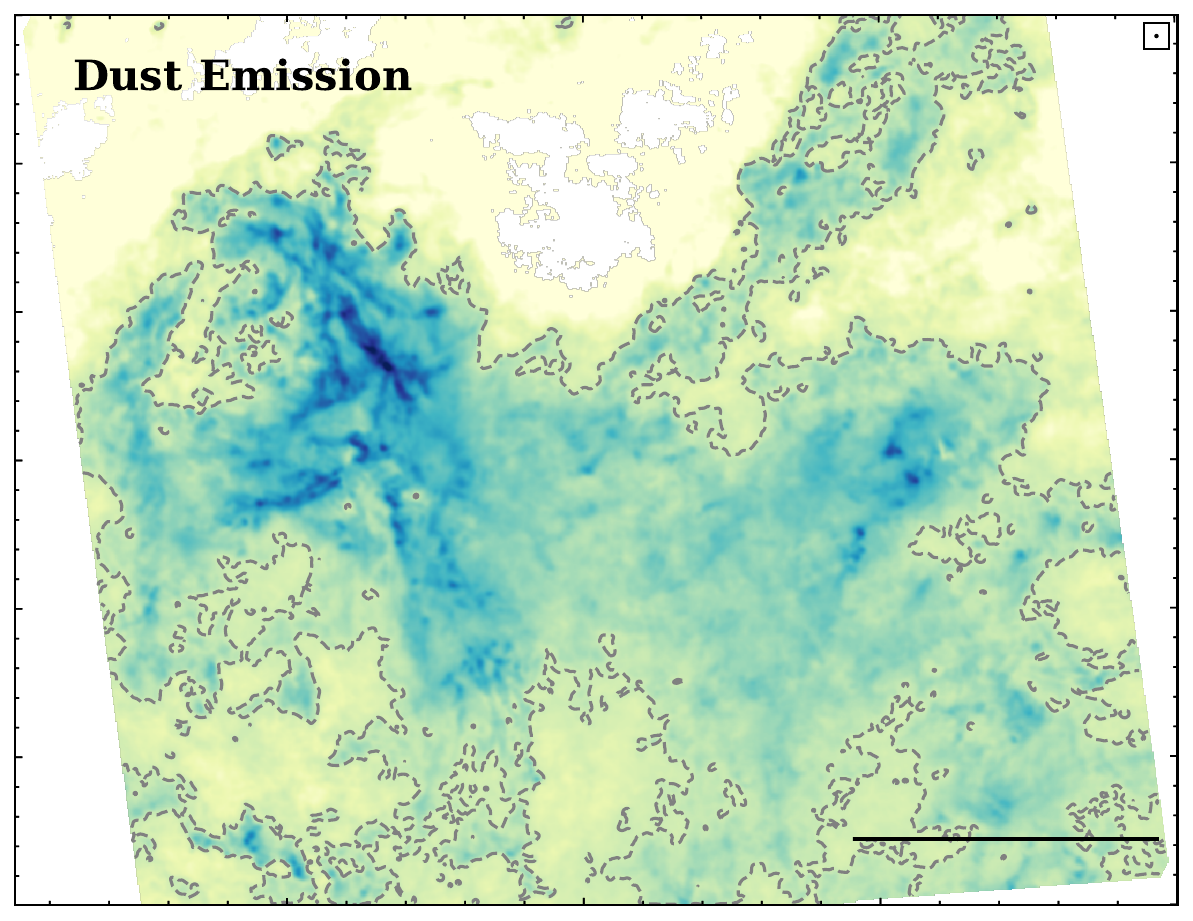}
    \includegraphics[width = 8cm]{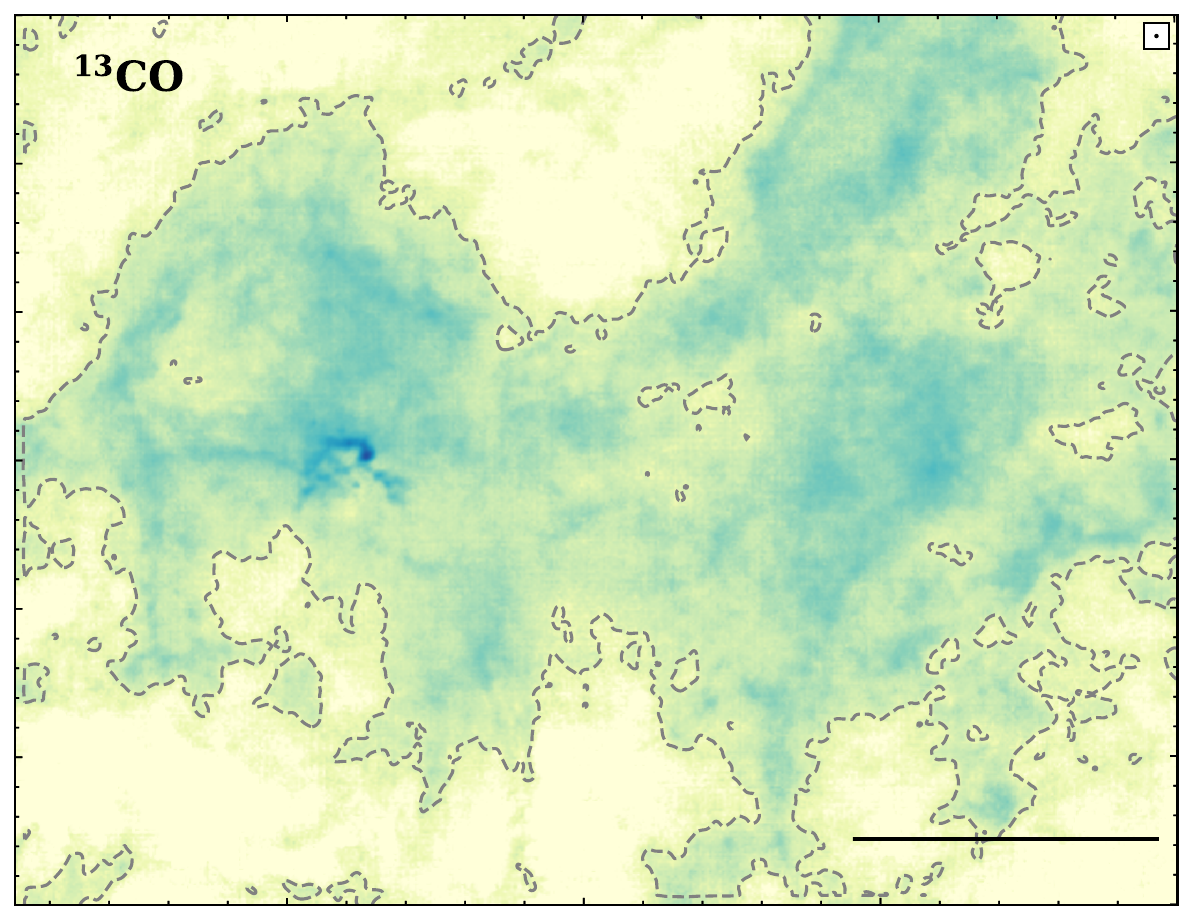}
    \caption{
    {\it Top:} Gas column density derived by combining CO isotopologues (Section~\ref{sub_NH2_CO}).
    {\it Middle:} Gas column density from gray-body fitting of the dust far-infrared emission (Section~\ref{sub_Herschel}; Appendix~\ref{appendix_dustSED}).
    {\it Bottom:} Gas column density derived from $^{13}$CO alone, assuming LTE and optically thin emission (Section~\ref{subsub_N13CO_thin}).
    These panels share the same color bar and dynamic range.
    The gray dashed contours mark the cutoff contours adopted for the $N$-PDF fitting.
    }
    \label{fg_showmap}
\end{figure}

% Start from the radiative transfer:
% \begin{equation}
%     \begin{aligned}
%         T_{\rm R} = [J_\nu(T_{\rm ex}) - J_\nu(T_{\rm bg})](1-e^{-\tau}),
%         \label{eq1}
%     \end{aligned}
% \end{equation}
% where
% \begin{equation}
%     \begin{aligned}
%         J_\nu(T) = \frac{\frac{{\rm  h}\nu}{\rm k}}{{\rm exp}(\frac{{\rm  h}\nu}{{\rm k}T}) - 1}
%         \label{eq2}
%     \end{aligned}
% \end{equation}
% is the Rayleigh-Jeans Equivalent Temperature, which is the equivalent temperature of a black body at temperature T.
% The h and k are the Planck and Boltzmann constants, $\nu$ is the frequency, $\tau$ is the optical depth, $\mathit{T_{\rm R}}$, $\mathit{T_{\rm ex}}$, and $\mathit{T_{\rm bg}}$ represent the radiation temperature, the excitation temperature, and the background temperature, respectively.

For the target clouds, $^{13}$CO is generally abundant enough to be detected with a good signal-to-noise ratio (SNR).
In voxels (position-position-velocity cells) where C$^{18}$O is not detected, $^{13}$CO can be assumed to be optically thin, allowing its column density to be modeled using the $^{12}$CO observations based on radiative transfer theory.

For a linear rigid rotor molecule, the column density of the particles in the upper energy level, $N_{\rm u}$, is related to the optical depth $\tau_\nu$ by the formula \citep[][]{Mangum2017}:
\begin{equation}
    \begin{aligned}
    {\rm \mathit{N_{\rm u}} = \frac{3h(2\mathit{J_{\rm u}}+1)}{8\pi^3\mu^2\mathit{J_{\rm u}}}[e^{h\nu/(k\mathit{T})}-1]^{-1}\int\mathit{\tau_{\nu}d\nu}},
    \label{eq_Nu}
    \end{aligned}
\end{equation}
where $h$ and $k$ are the Planck and Boltzmann constants, $J_{\rm u}$ is the rotational quantum number of the upper energy level, $\mu$ is the dipole moment, and $\nu$ is the rest frequency of the transition.
Assuming detailed balance at a constant temperature defined by a single excitation temperature $T_{\rm ex}$ for all transitions, the total column density $N_{\rm tot}$ is related to $N_u$ by:
\begin{equation}
    \begin{aligned}
        \frac{N_{tot}}{N_u} = \frac{Q_{tot}}{g_u}\mathrm{exp}({\frac{E_u}{\mathrm{k}T_{\rm ex}}}),
        \label{eq_NuNtot}
    \end{aligned}
\end{equation}
where $Q_{\rm rot}$ is the rotational partition function, and $g_u$ and $E_u$ are the degeneracy and energy of the upper level $u$.
For linear molecules, we adopt the high-temperature approximation for $Q_{\rm rot}$ from \cite{Gordy1984}:
\begin{equation}
    \begin{aligned}
        Q_{\rm rot} &= \sum_{J=0}^{\infty}(2J+1)e^{-\frac{E_J}{\mathrm{k}T_{\rm ex}}}\\
        &\approx \frac{\mathrm{k}T_{\rm ex}}{\mathrm{h}B_0} + \frac{1}{3},
        \label{eq_Qrot}
    \end{aligned}
\end{equation}
where $B_0$ is the rigid rotor rotation constant. 
By combining Equations \ref{eq_Nu}, \ref{eq_NuNtot}, and \ref{eq_Qrot}, the total column density can be expressed as:
\begin{equation}
    \begin{aligned}
        N_{tot} =& \frac{\rm 3h}{8\pi^3\mu^2J_u}(\frac{{\rm k}T_{\rm ex}}{{\rm hB_0}} + \frac{1}{3}){\rm exp}(\frac{E_u}{{\rm k}T_{\rm ex}}) \\
        &\times [{\rm exp}(\frac{{\rm h}\nu}{{\rm k}T_{\rm ex}})-1]^{-1}\int \tau_\nu dv.
        \label{eq_Ntot}
    \end{aligned}
\end{equation}

For a given transition, the column density of the molecule can be obtained by substituting the dipole moment, the rotational constant, the energy of the upper energy level, the rest frequency, and the integrated optical depth into Equation \ref{eq_Ntot}. 
Under the optically thin approximation, the radiative transfer equation implies that the optical depth is proportional to the radiation temperature. 
Substituting this relation, together with the molecular constants, into Equation \ref{eq_Ntot} yields the expression for the $^{13}$CO column density (cm$^{-2}$) in each voxel:
\begin{equation}
    \begin{aligned}
        N_{\rm ^{13}CO,thin} =\:&  2.93\times10^{14}(T_{\rm ex} + 0.88)
        \\&\times[\frac{7.09}{{\rm exp}(5.29/T_{\rm ex})} - 1]^{-1} \times T_{\rm ^{13}CO} \delta v,
        \label{eq_N13COthin}
    \end{aligned}
\end{equation}
where $T_{\rm ^{13}CO}$ is the $^{13}$CO brightness temperature and $\delta v$ is the channel width in km s$^{-1}$. 
This expression is applied to all voxels where C$^{18}$O is not detected.

To evaluate Equation \ref{eq_N13COthin}, the excitation temperature $T_{\rm ex}$ must first be determined. Assuming that $^{12}$CO is optically thick at its peak and adopting the Cosmic Microwave Background (CMB) as the background radiation, a local $T_{\rm ex}$ is estimated for each spatial pixel from the peak brightness temperature of $^{12}$CO \citep[][]{Nagahama_1998}:
\begin{equation}
    \begin{aligned}
        \mathit{T_{\rm ex}} = \frac{5.53}{\ln\left(1 + \frac{5.53}{\mathit{T_{\rm peak,^{12}CO}} + 0.819}\right)}.
    \label{eq_Tex}
    \end{aligned}
\end{equation}
In practice, we set a lower limit of $\sim$15 K for the derived $T_{\rm ex}$, since very low peak temperatures may indicate that the optically thick assumption for $^{12}$CO is no longer valid. Given that our sources are relatively warm molecular clouds, adopting 15 K as the lower bound of $T_{\rm ex}$ is a reasonable choice.
With $T_{\rm ex}$ thus determined \Revise{along each LOS}, the radiation temperature $T_{\rm R}$ of a transition can be related to its optical depth $\tau$ through the equation of radiative transfer.

\subsubsection{$^{13}$CO/C$^{18}$O Abundance Estimation}
\label{subsub_1318ratio}

\begin{figure*}[htp!]
    \begin{tabular}{ c c c }
        \hspace{-1.1cm}\includegraphics[height = 7.5 cm]{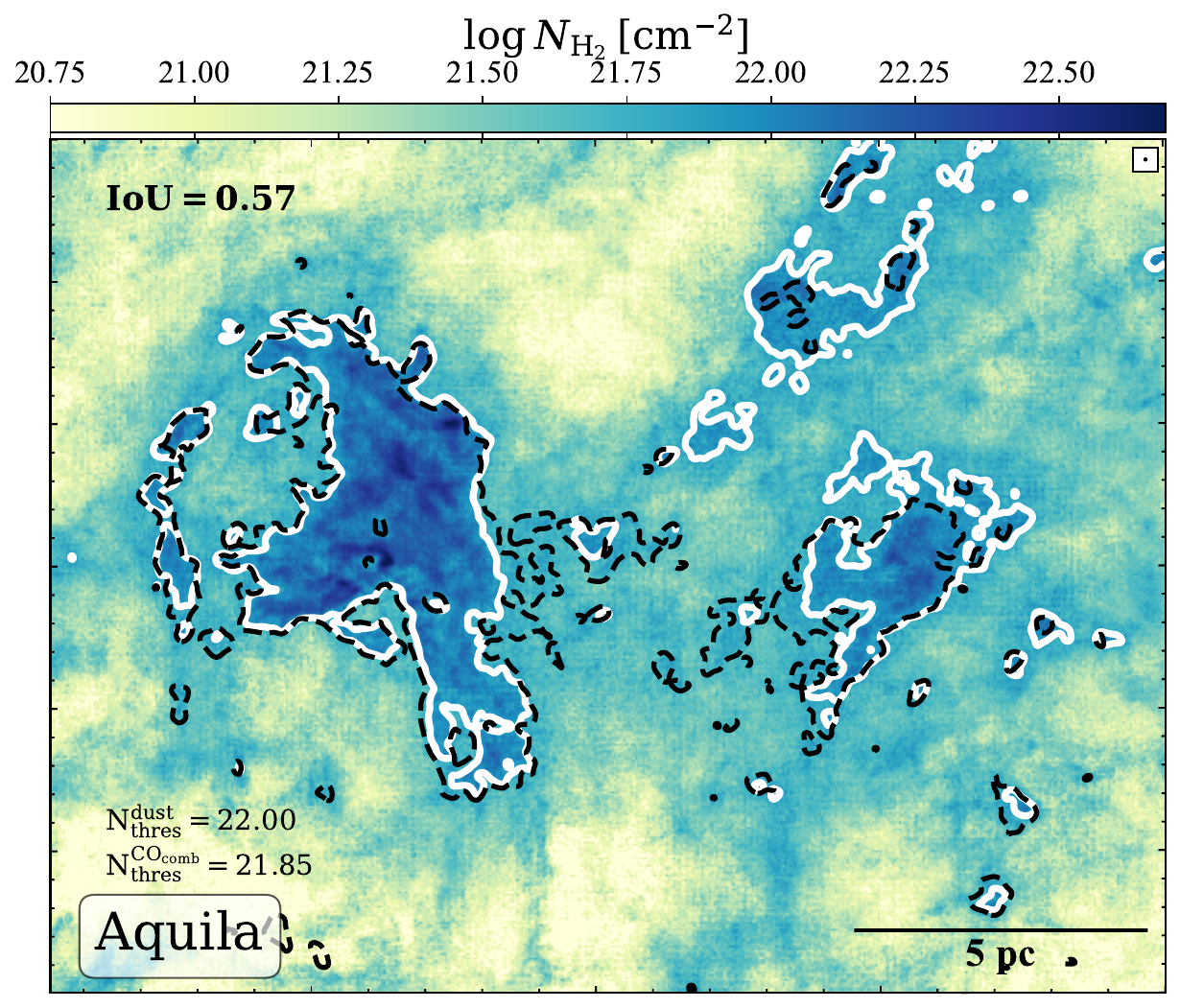} & 
        \hspace{-0.3cm}\includegraphics[height = 7.4 cm]{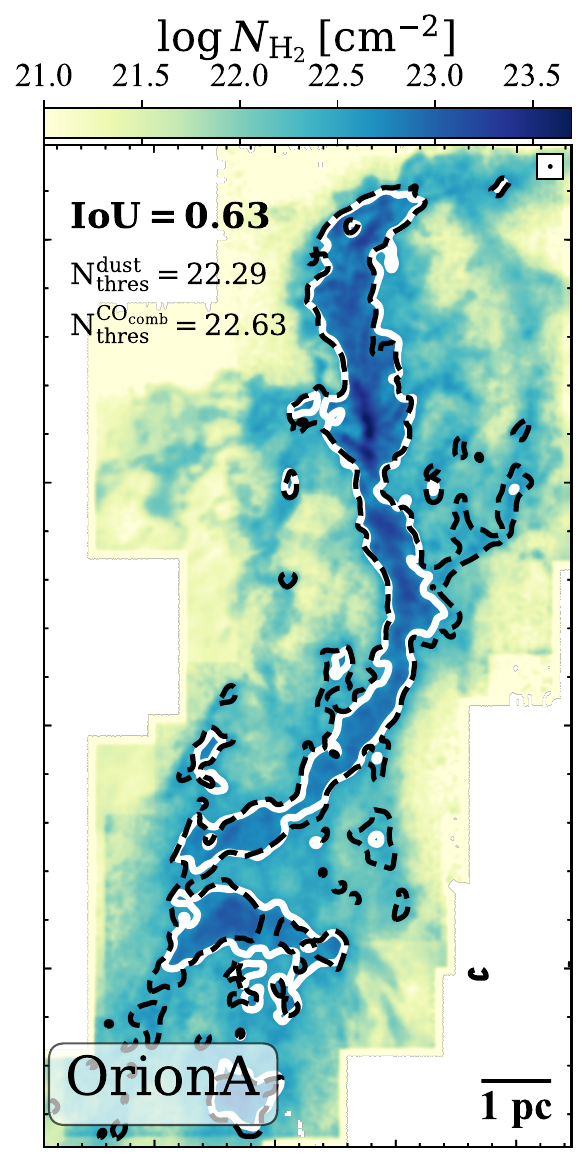} & 
        \hspace{-0.3cm}\includegraphics[height = 7.4 cm]{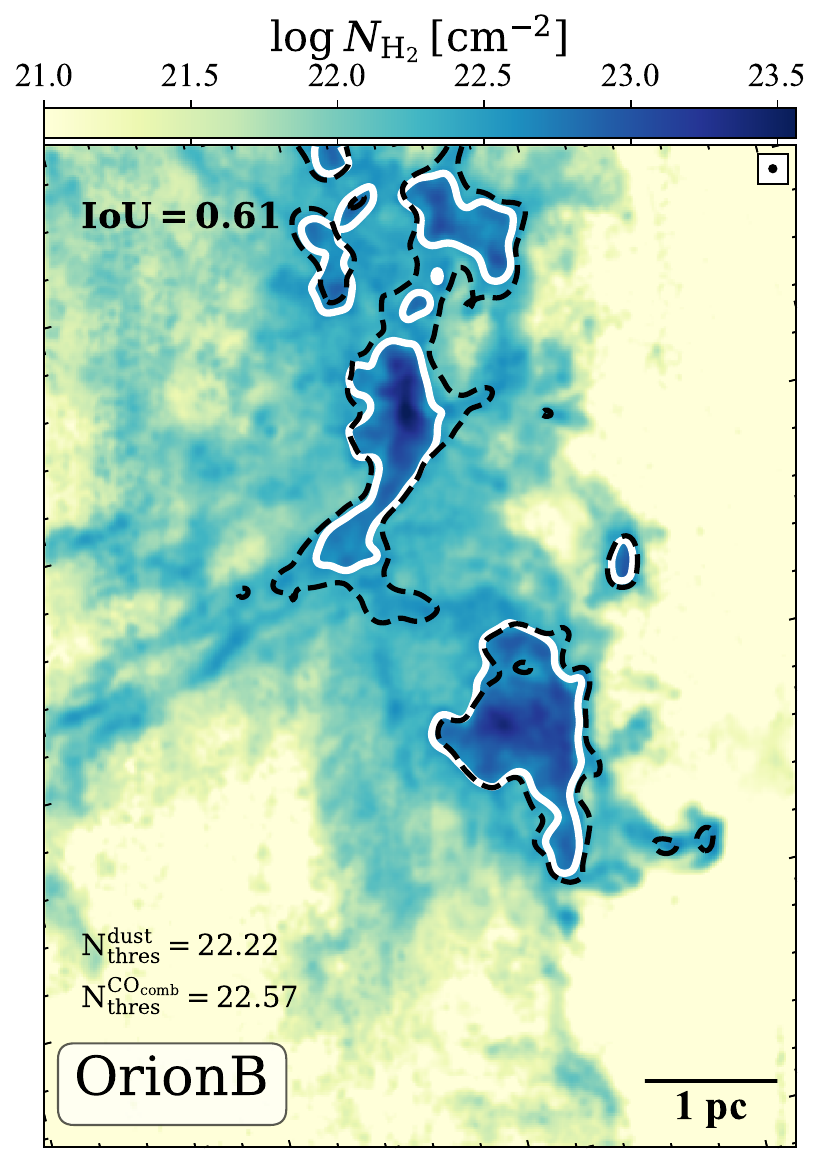}\\
    \end{tabular}
    \caption{
    Maps of gas column density derived from the combined CO isotopologue lines for Aquila, OrionA, and OrionB. The black and white contours represent the self-gravitationally-bound structures identified using the dust-based and CO-based methods, respectively. The labeled IoU values quantify the degree of spatial overlap between the two contour sets in each region, indicating their morphological agreement in tracing gravitationally bound gas.}
\label{fg_NH2map_nearby}
\end{figure*}

\Revise{
To derive the optical depth of $^{13}$CO by combining the $^{13}$CO and C$^{18}$O line intensities, their relative abundance ratio is required.
In optically thin regions, the intensity of spectral lines is proportional to the number of particles in the upper energy level. For $^{13}$CO and C$^{18}$O, we can reasonably assume that they share similar excitation conditions, allowing us to estimate their abundance ratio directly from the line intensity ratio in optically thin gas (Equation \ref{eq_N13COthin}).

For each source, we stacked the spectra of $^{13}$CO and C$^{18}$O in regions where both species are expected to be optically thin. We first estimated $T_{\rm ex}$ from the peak brightness temperature of $^{12}$CO (Equation~\ref{eq_Tex}). Using the radiative transfer equation, we then estimated the preliminary optical depths of $^{13}$CO and C$^{18}$O, and selected regions where both optical depths are below a given threshold. 
In addition, to avoid areas where C$^{18}$O has essentially no detectable emission (SNR $\sim$1), we excluded regions with $^{13}$CO SNRs lower than 10. The rms levels of $^{13}$CO and C$^{18}$O are assumed to be similar, which holds for all sources in our sample, as both were observed simultaneously under the same conditions.
We tested optical-depth thresholds from 0.35 to 0.6, adopted the mean abundance ratio derived over this threshold range as the measured $^{13}$CO/C$^{18}$O ratio, and used the standard deviation as its uncertainty. These thresholds provide a balance between ensuring that $^{13}$CO remains optically thin and maintaining sufficient C$^{18}$O detection. If the threshold is set too low (e.g., $\tau\lesssim 0.35$), the stacked C$^{18}$O spectra become strongly affected by noise, while a much higher threshold would be inconsistent with the goal of stacking spectra in optically thin regions. 

Figure \ref{fg_R1318_G10} illustrates the method applied to the G10.6-0.4 cloud. The stacked C$^{18}$O spectrum closely matches the $^{13}$CO spectrum, with a difference in intensity of approximately a factor of 5. From this, we calculated the abundance ratio of $^{13}$CO to C$^{18}$O in G10.6-0.4 to be 6.4. The abundance ratios for the other sources are listed in Table \ref{tab_sources}.
}

\subsubsection{Derive $\tau_{\rm ^{13}CO}$ from $^{13}CO/C^{18}O$ Ratio}
\label{subsub_N13CO_thick}

In regions where C$^{18}$O is detected with a good SNR, the optically thin assumption for $^{13}$CO may break down due to the high gas column densities.
To account for this, we use C$^{18}$O observations to estimate the optical depth of $^{13}$CO and thereby obtain its optical--depth--corrected column density \citep[][]{Galvan_2013}.

When both $^{13}$CO and C$^{18}$O are optically thin and share similar excitation conditions, their line intensity ratio should approach their abundance ratio. 
Assuming the same excitation temperature for both isotopologues along the LOS, deviations in the observed intensity ratio primarily reflect variations in optical depth. 
Using the radiative transfer and Equation \ref{eq_Ntot}, the optical depth of $^{13}$CO can be related to the line ratio between $^{13}$CO and C$^{18}$O as:
\begin{equation}
    \begin{aligned}
        \frac{T_{\rm ^{13}CO}}{T_{\rm C^{18}O}} = \frac{1-{\rm exp}(-\tau_{\rm ^{13}CO})}{1-{\rm exp}(-\tau_{\rm ^{13}CO}/\chi)},
        \label{eq_lineratio}
    \end{aligned}
\end{equation}
where $\chi$ is the relative abundance of $^{13}$CO and C$^{18}$O.
For voxels where both $^{13}$CO and C$^{18}$O are detected, we then calculate the optical-depth-corrected column density of $^{13}$CO by combining Equation~\ref{eq_N13COthin} with Equation~\ref{eq_lineratio}:
\begin{equation}
    \begin{aligned}
        N_{\rm ^{13}CO,thick}=N_{\rm ^{13}CO,thin}\times \frac{\tau_{\rm ^{13}CO}}{1-{\rm exp}(-\tau_{\rm ^{13}CO})}.
        \label{eq_N13COthick}
    \end{aligned}
\end{equation}
Finally, for all voxels with $^{13}$CO detection, we integrate the derived column densities ($N_{\rm ^{13}CO,thin}$ or $N_{\rm ^{13}CO,thick}$) along the velocity axis to create the final $^{13}$CO gas column density map.

\subsubsection{Molecular Hydrogen Column Density}
\label{subsub_NH2_COcomb}

To convert the $^{13}$CO column density into molecular hydrogen column density, both the isotopic abundance ratio $^{12}$C/$^{13}$C and the CO-to-H$_2$ abundance ratio are required.

The $^{12}$CO $J=1$--0 line is typically saturated, making it difficult to measure the $^{12}$CO/$^{13}$CO ratio directly. 
Instead, we adopt a Galactic gradient model,
\begin{equation}
^{12}{\rm C}/^{13}{\rm C} = 5.87 \times R_{\rm gc} + 13.25,
\end{equation}
to estimate $^{12}$C/$^{13}$C for each source \citep[][]{Jacob_2020}, where $R_{\rm gc}$ is the Galactocentric distance.

% \begin{figure*}[htp!]
%         \includegraphics[width = 18cm]{NPDF_nearby.pdf}
%     \caption{The $N$-PDFs of the three nearby molecular clouds: Aquila, Orion A, and Orion B. The $N$-PDFs derived from combined CO isotopologue lines are shown as orange step lines, and those derived from Herschel dust emission are shown in blue. Vertical dashed lines indicate the best-fit normalized transition column densities $\eta_{\rm t}$ for each tracer. The lower panels display the residuals between the logarithmic probabilities of the two $N$-PDFs, highlighting systematic differences between the CO- and dust-based tracers.}
% \label{fg_npdf_nearby}
% \end{figure*}

The CO abundance in nearby molecular clouds is relatively well-constrained, with reported values of $^{12}$CO/H$_2$ ranging from $\sim0.8$ to $2.5 \times 10^{-4}$ \citep{Lacy2017ApJ...838...66L_COabundance, Kulesa_2002, Frerking_1982, Dickman_1978}.
For the solar neighborhood, we adopt a representative value of $1.67 \times 10^{-4}$, based on recent measurements of CO and H$_2$ absorption lines in local molecular clouds \citep[][]{Lacy2017ApJ...838...66L_COabundance}.

For more distant molecular clouds located in the Galactic disk, it is necessary to account for variations in CO abundance with metallicity, which itself decreases with Galactocentric distance. 
A metallicity-dependent CO-to-H$_2$ conversion factor is commonly used to estimate molecular cloud masses \citep{Evans_2022}.
Following recent observational and theoretical work, we assume the CO abundance scales linearly with gas-phase metallicity \citep[e.g.,][]{Lin2025NatAs...9..406L, Gong_2020}. 
%We apply the same linear relation in our gas column density calculation by using a metallicity-dependent CO abundance.

We adopt the gas phase metallicity gradient derived from oxygen abundances in H II regions \citep[][]{Mendez-delgado_2022}:
\begin{equation}
    \begin{aligned}
    {\rm Z = 10^{-0.044(R_{gc}-R_{gc,\odot})},}
    \label{eq_Z_Rgc}
    \end{aligned}
\end{equation}
where the ${\rm R_{gc}}$ and ${\rm R_{gc,\odot}}$ represent the Galactocentric distance of each molecular cloud and the Sun in kiloparsecs (kpc). 
%Equation \ref{eq_Z_Rgc} allows us to translate the dependence of CO abundance on metallicity into a relationship with the Galactocentric distance. 
%By adding a reference point, we can then apply this relationship to all sources.
%The galactocentric distance of the Sun is adopted as 8.28 kpc \citep[][]{GRAVITY2021A&A...647A..59G}. Following this, we obtain a Galactocentric distance-dependent relationship for CO abundance and convert all the ${\rm N(CO)^{com}}$ maps to ${\rm N(H_2)}$ maps:
Assuming a solar Galactocentric distance of $R_{\mathrm{gc},\odot} = 8.28$ kpc \citep{GRAVITY2021A&A...647A..59G}, we scale the local CO abundance by the metallicity to obtain a Galactocentric distance-dependent relationship:
\begin{equation}
    \begin{aligned}
    {\rm CO/H_2 = 1.67\times10^{-3.64-0.044(R_{gc}/kpc)}}.
    \label{eq_COabundance_Rgc}
    \end{aligned}
\end{equation}
Using this framework, we convert derived molecular hydrogen column density maps using the three CO isotopologues, denoted as $N_{\rm H_2}^{\rm CO_{comb}}$. 
\Revise{The relatively high abundance of CO, together with the comparatively well-studied CO isotopologue and CO/H$_2$ abundance ratios in the Milky Way, makes CO isotopologues advantageous for column density measurements. In comparison, molecular tracers with lower or less well-constrained abundances may introduce larger uncertainties in derived ${\rm H_2}$ column densities \citep[][]{Schneider2016A&A}.}

%Although all three isotopologues are involved, it is worth noting that $^{12}$CO is used solely to estimate the excitation temperature, while the line intensities of $^{13}$CO and C$^{18}$O are directly related to the column density. 

As an example, Figure~\ref{fg_showmap} presents the CO-based column density map of the nearby Aquila cloud alongside a dust-based column density map for comparison. 
\Revise{The CO-combination and dust-based maps show broadly similar gas structures in regions of moderate to high column density.
Compared with the $^{13}$CO-only map, the CO-combination method recovers a larger fraction of the dense structures.}

Noticeable differences between $N_{\rm H_2}^{\rm CO_{comb}}$ and $N_{\rm H_2}^{\rm dust}$ mainly appear in more diffuse regions.
Several factors may contribute to this behavior: 
(i) CO self-shielding becomes inefficient at low column densities, leading to a decline in CO abundance \citep[][]{vanDishoeck1988ApJ...334..771V_COPDR}.
(ii) \Revise{in diffuse environments, dust emission may no longer exclusively trace molecular gas \citep[][]{Bohlin1978ApJ...224..132B,Liszt2014ApJ...780...10L}, as the increasing contribution from atomic gas can produce an additional H{\sc I}-related component in dust-based N-PDFs \citep[][]{Schneider_2022A&A...666A.165S}.}
(iii) calibration uncertainties in far-infrared continuum emission tend to be larger in extended, low-surface-brightness regions.
However, these differences have a negligible impact on our scientific goals, which focus on fitting the power-law tail of the $N$-PDF and estimating the mass of gravitationally bound gas.
Our $N$-PDF analysis is performed within the last closed contour, which naturally excludes a substantial fraction of diffuse gas.
Moreover, the power-law component is dominated by more compact structures well within this boundary, further minimizing the influence of low-density regions.

The procedure for deriving the dust-based column density is described in Appendix~\ref{appendix_dustSED}.

\subsection{Mass of the gravitationally bound Gas}
\label{sec4.1}

\begin{figure*}[htp!]
    \begin{tabular}{ c }
        \hspace{-0.9cm}\includegraphics[width = 18.2 cm]{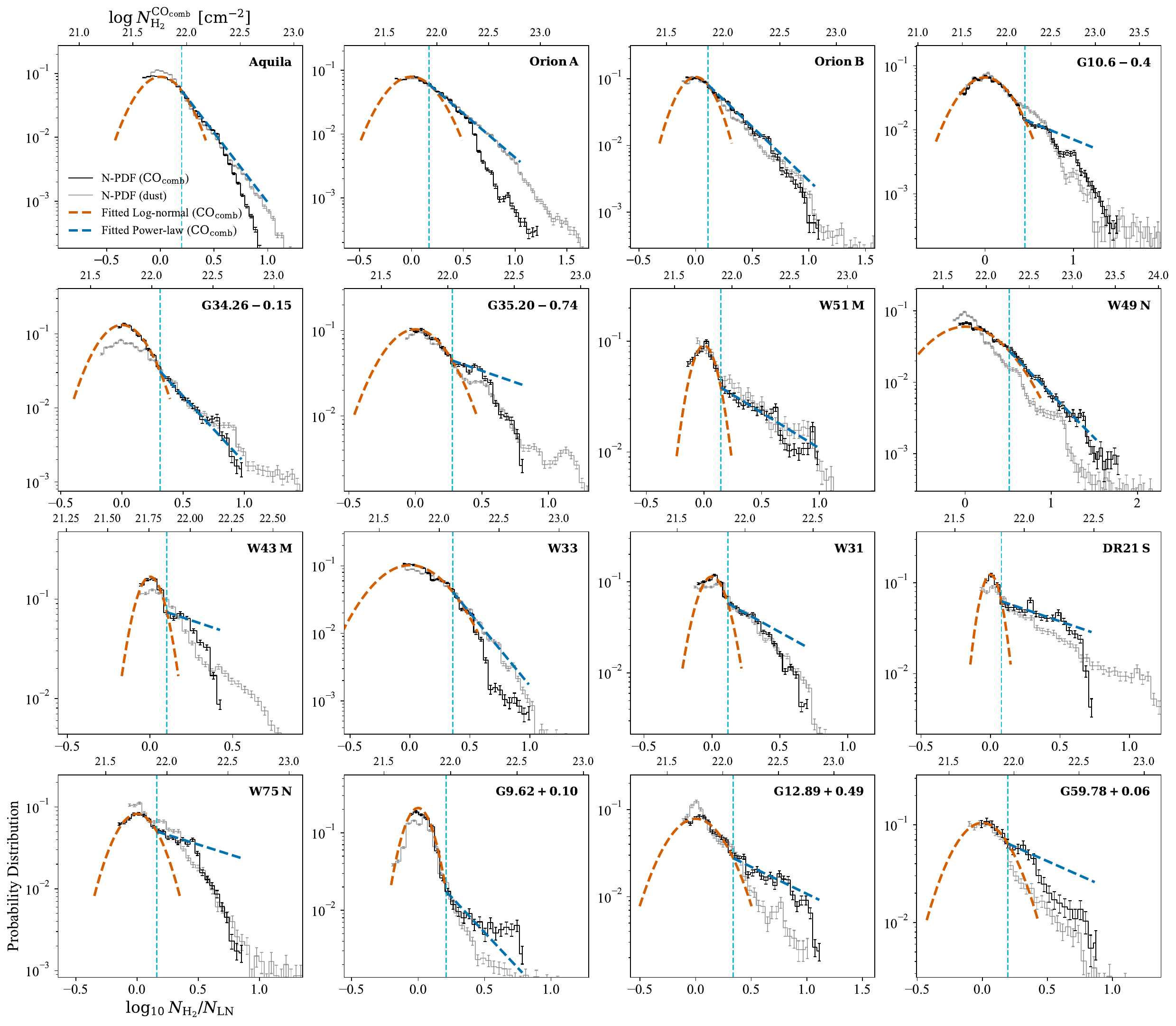}
    \end{tabular}
    \caption{
    \Revise{
    $N$-PDFs for all 16 molecular clouds in this work, derived from dust-based column density maps (gray step lines) and from our combined CO isotopologue method (black step lines). 
    The first three panels show the solar-neighborhood clouds, while the remaining panels show the distant Galactic-plane clouds. 
    All $N$-PDFs are normalized by the central column density of the fitted log-normal component, $N_{\rm LN}$, and the normalized column density is shown on the lower x-axis. 
    For the CO-based $N$-PDFs, the log-normal and power-law fits are shown as red and blue dashed lines, respectively, and the vertical cyan dashed lines mark the corresponding transition column densities.
    The upper x-axis shows the absolute column density of the CO-based $N$-PDF.}
    }
    \label{fg_npdf_all}
\end{figure*}

To quantify the mass of self-gravitationally bound gas, we analyzed the $N$-PDFs of our 16 molecular clouds following the same method described in \cite{Jiao2025A&A}. Prior to fitting, all $N_{\rm H_2}$ maps were convolved to the resolution of the tracer with the larger beam to ensure a reliable comparison, and all maps and analyses presented in the following sections are based on these convolved data.
We fit the $N$-PDFs with a piecewise log-normal-plus-power-law function of the form \citep[e.g.,][]{Schneider2015MNRAS,Burkhart_2017,Chen_2018}:
\begin{equation}
    p(\eta) = 
    \begin{cases}
        \frac{1}{\sqrt{2\pi}\sigma}{\rm exp}[{-\frac{(\eta-\eta_0)^2}{2\sigma^2}}] \quad\quad\quad\quad\quad\quad\:\: \eta<\eta_t \\
        \frac{1}{\sqrt{2\pi}\sigma}{\rm exp}[{-\frac{(\eta_{t}-\eta_0)^2}{2\sigma^2}}+\alpha(\eta-\eta_t)] \quad \eta\geq \eta_t \\
    \end{cases},
    \label{eq_npdf}
\end{equation}
where $\eta = {\rm ln}(N_{\rm H_2}/\langle N_{\rm H_2}\rangle)$ is the logarithm of the normalized gas column density and $p(\eta)$ is the normalized probability density. 
The parameters $\eta_0$ and $\sigma$ describe the log-normal component, while $\eta_t$ is the transition point between the log-normal and power-law regimes, and $\alpha$ is the slope of the power-law tail.

We implemented the fitting in a Bayesian framework, constructing a log-likelihood function based on Equation~\ref{eq_npdf}:
\begin{equation}
    \mathcal{L}(\theta) = \sum_i \log\left[ p_\eta(\eta_i | \theta) \right],
    \label{eq_likelihood}
\end{equation}
where $\theta$ represents the full set of model parameters, $p_\eta$ is the probability density given by the $N$-PDF model (Equation~\ref{eq_npdf}), and $\eta_i$ is the value of $\eta$ at each pixel in the map. 
For the prior distributions, we adopted uniform priors for all parameters except $\alpha$, for which we imposed a uniform prior on $\arctan(\alpha)$ to ensure even sampling in slope angle.

We performed parameter inference using the Markov Chain Monte Carlo (MCMC) method, implemented with the affine-invariant ensemble sampler in the Python package {\tt emcee} \citep{Foreman-Mackey2013PASP_EMCEE}. 
Posterior sampling was carried out in a four-dimensional parameter space corresponding to the $N$-PDF model parameters: $\eta_0$, $\sigma$, $\eta_{\rm t}$, and $\alpha$ (as defined in Equation \ref{eq_npdf}). 
The sampling was initialized with 16 walkers, each starting from a small random perturbation ($\sim10^{-3}$) around a fiducial set of initial values.
To ensure efficient exploration of the posterior distribution, we used a weighted combination of differential-evolution moves in the sampling process. Specifically, each walker randomly selects between {\tt DEMove} (80\%) and {\tt DESnookerMove} (20\%), as recommended in the {\tt emcee} documentation for handling lightly multimodal or correlated parameter spaces. 
This mixture significantly reduces the autocorrelation time and enhances the convergence speed compared to using a single move type alone. 
Additionally, the ensemble sampler's affine-invariant design minimizes the need for manual tuning while facilitating parallel computation. 
Our $N$-PDF fitting code, which implements this sampler configuration, is publicly available \footnote{\url{https://github.com/Linjing2021/NPDF_MLE_MCMC}}.
For each parameter, we adopted the median value of the posterior distribution as the best estimation, and the 16th and 84th percentiles as the lower and upper uncertainties, respectively.
The best-fit parameters are listed in Appendix~\ref{appendix_npdf_para}.

This maximum likelihood approach allows us to avoid pre-binning the column density data, which can introduce additional biases and uncertainties \citep{Yogesh_2012arXiv1208.3524V_prebin_NPDF, Veltchev_2019MNRAS.489..788V}. 
Instead, we directly model the probability distribution of the unbinned data, which is especially beneficial for sources with limited dynamical range or sparse high-density tails.

By identifying the column density threshold that separates the log-normal and power-law components of the $N$-PDF, we estimate the mass of gravitationally bound gas in each cloud by integrating the mass above this threshold, i.e., 
\begin{equation}
    M_{\mbox{\scriptsize bound}} = \int^{+\infty}_{N_{{\mbox{\scriptsize threshold}}}}M(N)dN.
\end{equation}
The contribution from helium and heavier elements is included by adopting a mean molecular weight of $\mu = 2.8$ \citep{Draine2011piim.book.....D}.
This corresponds to the self-gravitating component associated with the power-law tail. 
In the following, we denote this quantity as $M_{\rm bound}^{\rm dust}$ and $M_{\rm bound}^{\rm CO_{comb}}$ when derived from dust emission and from combined CO isotopologue data, respectively.

\section{Result}
\label{sec_result}

\begin{figure}[b!]
    \centering
        \includegraphics[width = 8.5 cm]{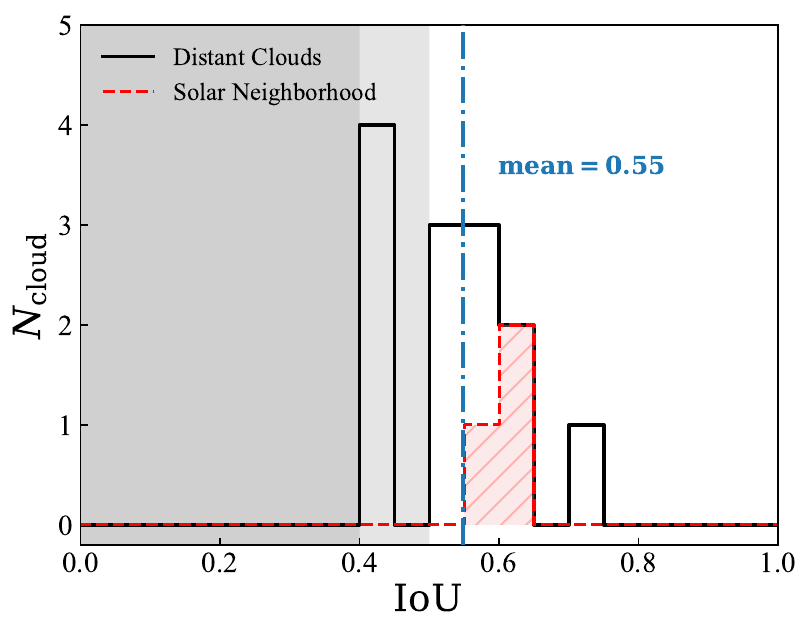}
    \caption{
    Distribution of the IoU values between the bound structures identified by the two methods. The black histogram shows the results for distant clouds, while the red histogram corresponds to solar neighborhood sources. The vertical dot-dashed blue line indicates the mean IoU value of 0.52, suggesting a generally good spatial correspondence between the two methods. The shaded gray areas mark typical IoU thresholds ($\geq$0.4 and $\geq$0.5) that are often used to judge whether two structures are considered a good match.
    }
    \label{fg_result_IOU}
\end{figure}

Figure~\ref{fg_NH2map_nearby} presents the $N_{\rm H_2}^{\rm CO_{comb}}$ maps of the three solar-neighborhood clouds, Orion A, Orion B, and Aquila, together with the gravitationally bound structures identified from the CO- and dust-based $N$-PDFs. 
Owing to their proximity and relatively high Galactic latitudes, these clouds are expected to be less affected by the overlap of multiple unrelated clouds along the LOS, and thus provide benchmark cases for evaluating the consistency between the two tracers. On the other hand, the distant Galactic-plane clouds are affected by more substantial LOS complexity. For the distant clouds in our sample, the main velocity component selected in $^{13}$CO contributes only $\sim$40\%–90\% of the total $^{13}$CO integrated flux, whereas this fraction exceeds $95\%$ for the nearby clouds.

The dust- and CO-based $N$-PDFs for all 16 molecular clouds are presented in Figure~\ref{fg_npdf_all}.
To facilitate comparison among clouds with different characteristic column densities, we normalize the column density of each cloud by the central column density of its fitted log-normal component, $N_{\rm LN}$.
The absolute column density of the CO-based $N$-PDF is shown on the upper x-axis.
The un-normalized N-PDFs are presented in Appendix~\ref{appendix_npdf_NoNor}.

\begin{figure*}[htbp!]
        \hspace{-0.2 cm}\includegraphics[width = 18 cm]{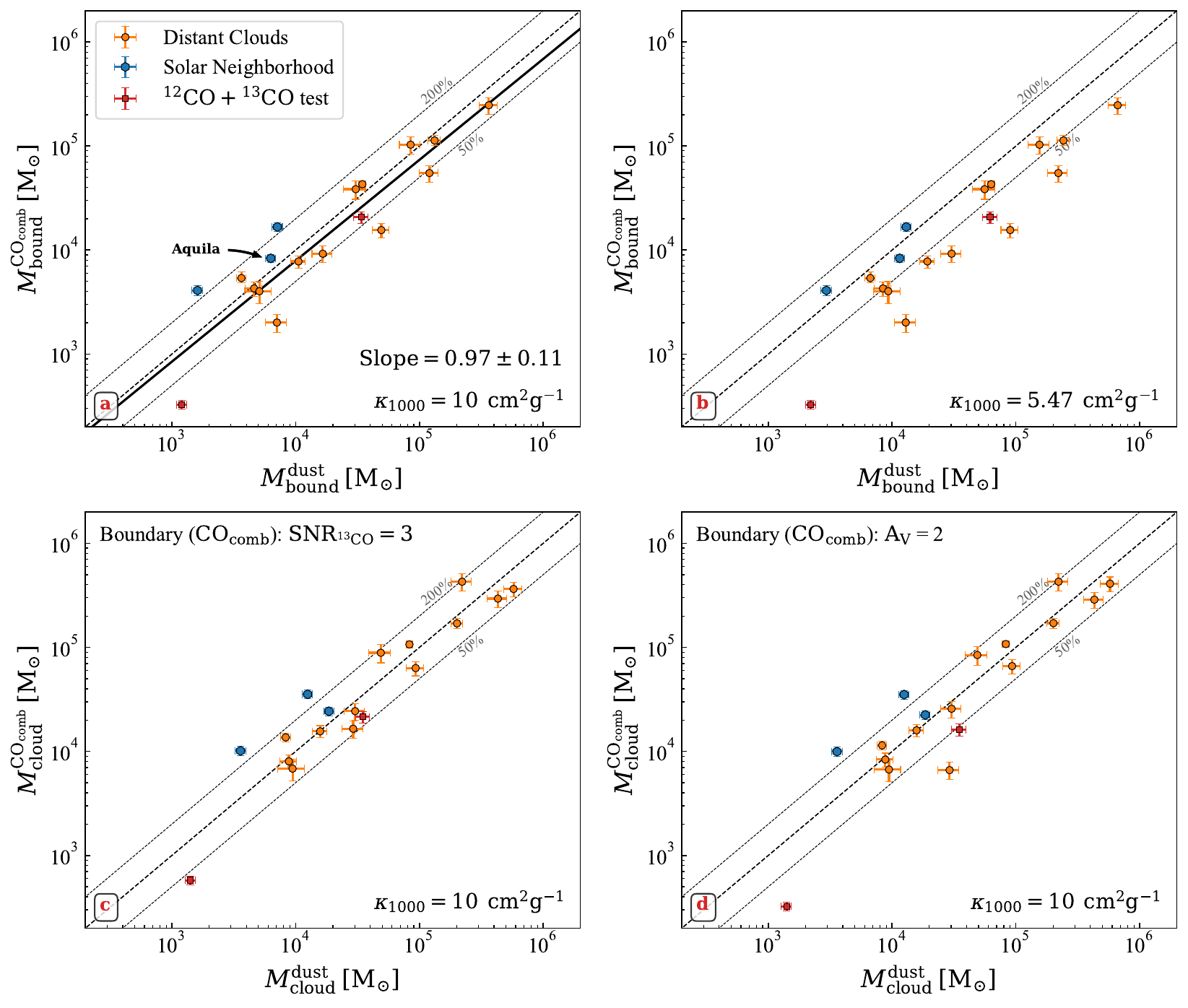}
    \caption{
    Comparison of gas masses derived from our combined CO isotopologue method and from dust continuum emission.
    {\it (a)}: Comparison of the gravitationally bound gas masses, $M_{\rm bound}^{\rm CO_{comb}}$ and $M_{\rm bound}^{\rm dust}$, adopting $\kappa_{1000}=10$~cm$^2$~g$^{-1}$ for the dust-based measurements.
    {\it (b)}: Same comparison as panel {\it (a)}, but adopting the lower dust opacity of $\kappa_{1000}=5.47$~cm$^2$~g$^{-1}$ \citep[][]{Pontoppidan2024RNAAS...8...68P}.
    {\it (c)}: Comparison of the total cloud masses, $M_{\rm cloud}^{\rm CO_{comb}}$ and $M_{\rm cloud}^{\rm dust}$, adopting $\kappa_{1000}=10$~cm$^2$g$^{-1}$. The cloud boundary in the $N_{\rm H_2}^{\rm CO_{\rm comb}}$ maps is defined by the 3$\sigma$ contour of $^{13}$CO.
    {\it (d)}: Same as panel \textit{(c)}, but adopting an $A_V=2$ cloud boundary for the $M_{\rm cloud}^{\rm CO_{comb}}$ calculation.
    Orange circles represent distant Galactic-plane clouds, blue hexagons denote solar-neighborhood clouds, and red squares indicate test cases adopting the $^{12}$CO+$^{13}$CO tracer pair (Appendix~\ref{appendix_test12+13}). The dashed line marks the one-to-one relation, while the dotted lines indicate deviations of 50\% and 200\%. 
    In panel {\it (a)}, the black solid line shows the fit to the data, with its slope indicated in the lower-right corner.
    }
    \label{fg_result_distant}
\end{figure*}

For Orion A, Orion B, and Aquila, the CO-based $N$-PDFs agree well with their dust-based counterparts over both the log-normal component and the first power-law tail. 
The normalized transition column densities $N_{\rm H_2}/N_{\rm LN}$, inferred from the two tracers are also broadly consistent, indicating that the combined CO isotopologue method recovers the onset of the self-gravitating component traced by dust. 
This consistency is further confirmed by the column density maps, in which the power-law components identified by the two tracers correspond to highly similar gravitationally bound structures (Figure~\ref{fg_NH2map_nearby}).
At the highest column densities, however, the CO-based $N$-PDFs of Orion A and Aquila show a decrease relative to the dust-based distributions. 
This difference may arise from limitations in recovering the highest-column-density gas with the adopted molecular tracers, for example because of optical-depth effects, depletion, or excitation-related uncertainties in the densest regions.
Nevertheless, this discrepancy does not significantly affect the identification of gravitationally bound structures.

For the distant clouds, most CO-based $N$-PDFs still exhibit the characteristic combination of a log-normal component and a power-law tail. 
However, their agreement with the dust-based $N$-PDFs is generally poorer than for the nearby clouds, and the level of agreement varies substantially from source to source. 
This behaviour is expected because the distant clouds are affected by overlapping cloud components along the LOS. Such structured contamination cannot be fully removed from the dust-derived maps by subtracting a constant foreground/background.
The dust-based maps integrate emission from all structures along the LOS, whereas the CO-based maps use velocity information to isolate the principal molecular cloud component. 
Consequently, discrepancies between the two tracers in the distant clouds do not necessarily indicate a failure of the CO-based method, but can instead reflect their different sensitivities to structured LOS contamination.

To quantify the spatial consistency of the bound structures, we computed the intersection-over-union (IoU) between the dust- and CO-based contours for all clouds in the sample. 
The IoU is defined as the ratio of the intersection area to the union area of the two structures, with values above 0.5 commonly regarded as indicating substantial overlap \citep{Burke2019MNRAS.490.3952B_IoU,Grishin2023A&A...677A.101G_IoU}. 
For the nearby clouds, all three sources yield IoU values above 0.5, with Orion A reaching 0.63, confirming the strong spatial correspondence visible in Figure~\ref{fg_NH2map_nearby}. 
The IoU distribution for the full sample is shown in Figure~\ref{fg_result_IOU} (See Appendix~\ref{appendix_map_distant}). 
Among the distant clouds, 69.2\% have IoU values above 0.5, and this fraction increases to 75\% when the nearby clouds are included. 
Unlike the well-defined, discrete objects usually analyzed in computer vision tasks, molecular clouds exhibit intrinsically irregular morphologies and poorly defined boundaries, further complicated by observational uncertainties.
It is therefore reasonable to adopt a slightly lower IoU threshold when comparing cloud-scale structures.
Similar relaxations have also been explored in some image-recognition studies \citep[][]{Yanrs11030286}.
If a threshold of 0.4 is taken as the criterion for good agreement, the matching fractions increase to 100\% for the full sample, indicating an overall robust spatial correspondence between the two identification methods.
These results indicate that the bound structures identified from CO and dust show substantial spatial correspondence for the majority of the sample.

Figure~\ref{fg_result_distant} shows the comparison of the gas masses derived from the two methods. The top left panel compares the self-gravitating gas masses.
The two $M_{\rm bound}$ estimates agree well: the fitted relation has a slope close to unity, and most sources lie within an approximately factor-of-two scatter around this relation. 
These results indicate good overall agreement between the two $M_{\rm bound}$ measurements, with no systematic variation in their agreement across the sampled cloud-mass range. 
In particular, there is no evidence that our combined CO isotopologue method systematically overestimates or underestimates $M_{\rm bound}$ toward either the low- or high-mass end of the sample.
We note that the nearby clouds show a mild tendency toward larger $M_{\rm bound}^{\rm CO_{comb}}$ than $M_{\rm bound}^{\rm dust}$.
Given the good agreement between the CO- and dust-based $N$-PDFs (Figure~\ref{fg_NH2map_nearby}) and the spatial distributions (Figure~\ref{fg_result_IOU}) of the bound structures in these regions, this small systematic offset is likely related to the adopted Galactocentric-radius-dependent abundance prescriptions, which may not fully capture the local abundance variations in the solar neighborhood, as also discussed in Section~\ref{sub_1318ratio_Rgc}.

The upper-right panel of Figure \ref{fg_result_distant} illustrates the effect of adopting a different dust opacity. The value of $\kappa_{\rm 1000}=10$~cm$^2$~g$^{-1}$ has been commonly adopted in previous studies based on {\it Herschel} observations \citep[e.g.,][]{Andre2010_HGBS,Palmeirim_2013A&A...550A..38P,Konyves2015A&A...584A..91K_HGBS_Aquila,Schneider_2015A&A...575A..79S,Lin2016,Schneider_2022A&A...666A.165S}. 
However, a dust-opacity model developed to reproduce the NIR--MIR extinction curve suggests a lower opacity, with $\kappa_{\rm 1000}=5.47$~cm$^2$~g$^{-1}$ \citep[][]{Chapman_2009ApJ...690..496C, Pontoppidan2024RNAAS...8...68P}. 
This model has also been found to reproduce recent {\it JWST} spectra of dense molecular cloud regions well \citep[][]{Tyagi_2025ApJ...983..110T}.
Adopting this lower opacity increases the dust-based mass estimates and therefore introduces a systematic offset relative to the CO-based results. 
Throughout this work, our column density maps and mass estimates are based on $\kappa_{\rm 1000}=10$~cm$^2$~g$^{-1}$. 
This alternative case is shown to illustrate the systematic uncertainties associated with the assumed physical parameters, such as the dust opacity, CO abundance, isotopologue abundance ratios, and gas-to-dust ratio.
A detailed calibration of these quantities is beyond the scope of this work. 
We therefore focus primarily on the scatter in the mass comparison and on whether the agreement varies systematically with cloud mass, rather than on a direct comparison of their absolute mass values.
We also compares the total cloud masses derived from the two tracers. $M_{\rm cloud}^{\rm dust}$ is calculated over regions with $A_V>2$.
For $M_{\rm cloud}^{\rm CO_{comb}}$, we consider two cloud-boundary definitions: all pixels with detected $^{13}$CO emission (3$\sigma$ level), and the same $A_V>2$ boundary used for $M_{\rm cloud}^{\rm dust}$. 
The total cloud masses derived from the two methods also show good agreement across the sample.

\begin{figure}[t!]
    \begin{tabular}{ c }
        \hspace{-1cm}\includegraphics[width = 9 cm]{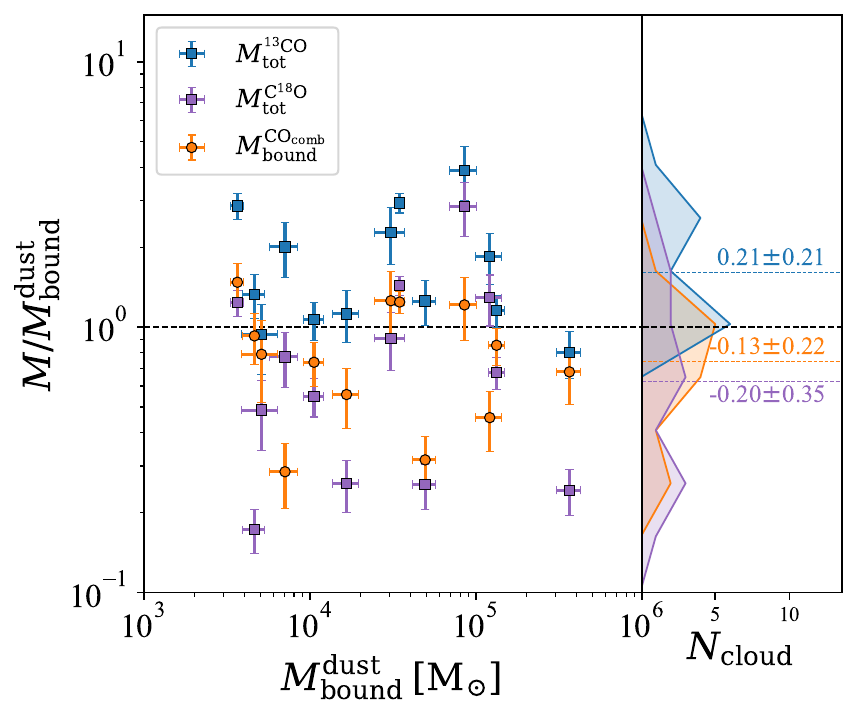}
    \end{tabular}
    \caption{
    Ratio of CO-based mass estimates to the dust-based bound gas mass, $M / M_{\rm bound}^{\rm dust}$, as a function of $M_{\rm bound}^{\rm dust}$.
    Blue and purple squares show the ratios $M_{\rm tot}^{^{13}{\rm CO}} / M_{\rm bound}^{\rm dust}$ and $M_{\rm tot}^{{\rm C}^{18}{\rm O}} / M_{\rm bound}^{\rm dust}$, respectively.
    Orange points indicate $M_{\rm bound}^{\rm CO_{comb}} / M_{\rm bound}^{\rm dust}$
    Black Dashed horizontal line indicate a ratio of unity.
    The right panel shows the corresponding distributions of $\log(M/M_{\rm bound}^{\rm dust})$ for the three mass estimates.
    The annotated values indicate the mean and standard deviation of each distribution.
    }
    \label{fg_MMcorr_13CO}
\end{figure}

Together with the comparisons of the $N$-PDFs and the spatial distributions of the bound structures, the agreement in $M_{\rm bound}$ demonstrates that combining multiple CO isotopologues provides a viable column-density tracer comparable to dust emission, particularly for identifying the power-law component of the $N$-PDF and tracing the self-gravitating structures in molecular clouds.

\section{discussion}
\label{sec_discussion}

\subsection{Comparison with $^{13}$CO- and C$^{18}$O-Derived Total Gas Masses}
\label{sub_13CO}

Low-$J$ transitions of $^{12}$CO are commonly used as tracers of the total molecular gas mass in molecular clouds, while $^{13}$CO and C$^{18}$O have been used in some studies as proxies for relatively higher–column-density gas \citep{Yuan_2022,Torii2019PASJ...71S...2T,Cormier2018MNRAS.475.3909C_nearbyGalaxies_1213CO}.
To assess how our CO-based modeling compares with these conventional methods, we compare $M_{\rm bound}^{\rm CO_{comb}}$ with total gas masses derived directly from $^{13}$CO and C$^{18}$O integrated intensities, denoted as $M_{\rm tot}^{^{13}{\rm CO}}$ and $M_{\rm tot}^{{\rm C}^{18}{\rm O}}$, respectively.
In practice, we simply integrate the line emission over the selected velocity range and convert the resulting integrated intensities to total gas masses assuming optically thin emission, using $T_{\rm ex}$ derived from Equation~\ref{eq_Tex}.
% \Revise{To assess how our CO-based modeling compares with these conventional methods, we compare what fraction of the cloud mass is accounted for by each gas component or tracer.
% Specifically, for all CO-related mass fractions, including $f_{^{13}{\rm CO}}$, $f_{{\rm C}^{18}{\rm O}}$, and $f_{\rm bound}^{\rm CO_{comb}}$, we adopt $M_{\rm cloud}^{\rm CO_{comb}}$ as the denominator.
% For the dust-based bound-gas fraction, $f_{\rm bound}^{\rm dust}$, we instead use the dust-derived cloud mass $M_{\rm cloud}^{\rm dust}$.
% The advantage of using mass fractions, rather than absolute mass values, is that they reduce the impact of systematic uncertainties in the absolute mass scale, which have been discussed in Section~\ref{sec_result}.
% The masses traced by $^{13}$CO and C$^{18}$O are directly derived from the corresponding integrated intensities using Equations~\ref{eq_Nu}–\ref{eq_Tex}, assuming LTE and optically thin emission.}

\Revise{
Figure~\ref{fg_MMcorr_13CO} compares the dust-based bound gas mass with several CO-based mass estimates.
The CO-combination bound gas mass agrees well with the dust-based estimate, as discussed in Section~\ref{sec_result}.
The total gas mass inferred from $^{13}$CO is systematically higher than both bound-gas mass estimates.
The poor spatial overlap between $^{13}$CO emission and the bound structures, with a mean IoU of $\sim 0.18$ and 93.8\% of the sources below 0.4, further indicates that $^{13}$CO traces a gas component that is significantly more extended than the gravitationally bound structures.
On the other hand, C$^{18}$O behaves differently.
The total gas mass derived from C$^{18}$O is typically lower than the bound gas mass, and the ratio $M_{\rm tot}^{\rm C^{18}O}/M_{\rm bound}$ shows a larger cloud-to-cloud dispersion.
The spatial agreement is also heterogeneous: although a few clouds show good overlap with the bound structures (e.g., ${\rm IoU}_{\rm W51M}=0.69$), most sources still have weak overlap, with a mean IoU of $\sim 0.36$.

Therefore, integrating a single CO isotopologue line does not provide a uniformly reliable tracer of the self-gravitating gas.
The $^{13}$CO-based total gas masses systematically exceed the bound-gas masses, likely because of the extended $^{13}$CO emission, whereas the C$^{18}$O-based total gas masses are generally lower and show stronger cloud-to-cloud variations.
The CO-combination method, which uses the N-PDF rather than a single-line integrated intensity, provides a more stable way to identify and quantify the bound gas component.}

\subsection{$M_{\rm bound}$-SFR Relation}
\label{sub_sfr}

\begin{figure}[b!]
    \begin{tabular}{ c }
        \hspace{-1cm}\includegraphics[width = 8.5 cm]{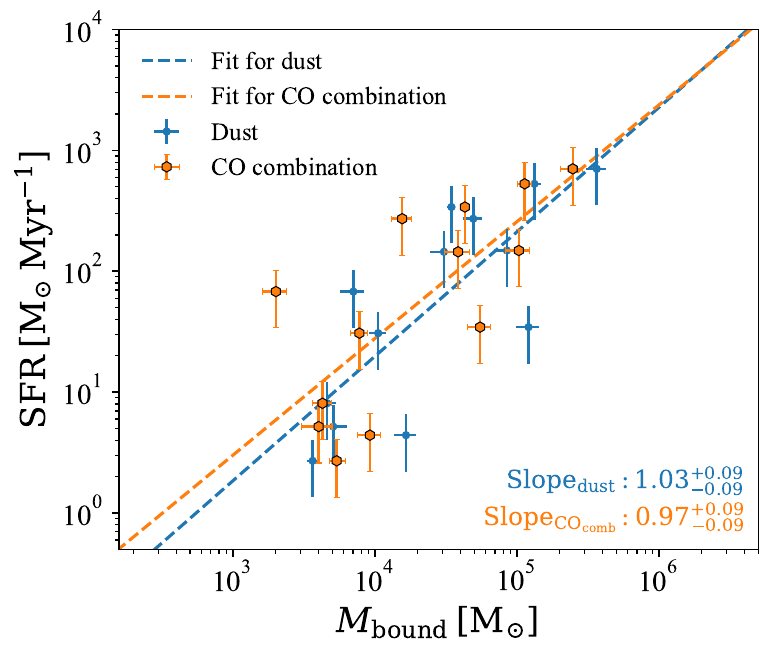}
    \end{tabular}
    \caption{
    Star formation rate versus bound gas mass. Orange points show results from the CO combination method, and blue points from dust. The orange dashed line is the best fit to the CO-based measurements, with slope $0.98\pm0.08$, compared to the dust-based relation from \cite[][blue line]{Jiao2025A&A}.
    }
    \label{fg_Mbound_SFR}
\end{figure}

A key motivation for quantifying the mass of gravitationally bound gas is its tight linear correlation with SFR, as recently established by \citet{Jiao2025A&A}. 
Given the linear relation we find between the CO-based and dust-based estimates of the bound gas mass, it naturally follows that $M_{\rm bound}^{\rm CO_{comb}}$ should also correlate linearly with the SFR.
To verify this consistency, we compared the bound gas masses derived from our multi-line CO method with the SFRs of the same clouds.

Figure~\ref{fg_Mbound_SFR} shows the $M_{\rm bound}$-SFR relation.
SFRs for the target clouds are calculated based on the infrared luminosity measurements using IRAS data, following the same procedure described in \cite{Wu_2010,Jiao2025A&A}.
%for most sources are taken from the infrared luminosity measurements of \citet{Jiao2025A&A}; for additional clouds not included in their sample, we derived SFRs from the IRAS data following the same procedure. 
The CO-based bound gas masses reproduce a linear correlation with SFRs, with a best-fit slope of $1.08_{-0.10}^{+0.11}$, in excellent agreement with the dust-based results \citet{Jiao2025A&A}, which further supports multi-line CO modeling as a robust tracer of gravitationally bound gas.

As a general caveat, the two axes of this comparison (e.g., Figure~\ref{fg_Mbound_SFR}) are not always strictly symmetric.
In rare cases where multiple star-forming clumps overlap along the same LOS but are separated in velocity, the inferred $M_{\rm bound}$–SFR relation may be subject to systematic effects.
In such situations, the star formation rate, typically derived from infrared luminosity, represents a LOS–integrated quantity, while dust-based $M_{\rm bound}$ estimates are influenced by the superposition of $N$-PDFs from clouds at different distances, potentially affecting both the reliability of the $N$-PDF and the derived bound gas mass.
On the other hand, the CO-based $N$-PDF analysis isolates individual velocity components and therefore traces the bound structures of only one cloud along LOS.
Under these circumstances, comparisons between CO-based self-gravitating gas masses and velocity-resolved star formation tracers (e.g., H$\alpha$ emission) may provide a more appropriate framework.
Such cases are largely avoided in our sample owing to the source selection criteria of \citet{Jiao2025A&A}, but this consideration should be kept in mind when extending the $M_{\rm bound}$–SFR analysis to more complex regions.

The consistency between CO- and dust-based analyses indicates that multi-line CO isotopologue data provide a reliable, velocity-resolved alternative for tracing bound gas.

\begin{figure}[t!]
    \centering
        \hspace{-0.6cm}\includegraphics[width = 8.9cm]{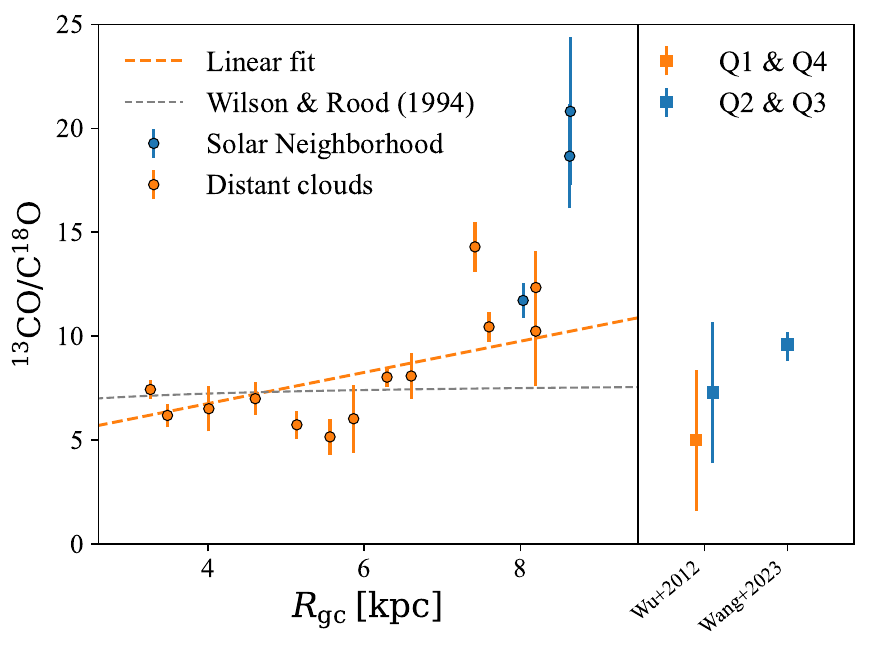}
    \caption{
    The $^{13}$CO/C$^{18}$O abundance ratio as a function of Galactocentric radius $R_{\rm gc}$.
    Blue and orange points denote solar-neighborhood and distant molecular clouds, respectively.
    The dashed orange line shows a linear fit to the distant-cloud sample only, while the gray dashed line indicates result from \citet{Wilson1994}.
    The right panel compares our results with those from \citet{Wu_2012ApJ...756...76W} and \citet{Wang2023AJ....166..121W}.
    Blue squares indicate samples from the second and third Galactic quadrants, which roughly trace the outer Galactic disk, while orange squares indicate samples from the first and forth quadrants, which roughly trace the inner Galactic disk.
    }
\label{fg_R1318_Rgc}
\end{figure}

\subsection{The Galatic gradient of $^{13}CO/C^{18}O$}
\label{sub_1318ratio_Rgc}

\Revise{
We present the derived $^{13}$CO/C$^{18}$O abundance ratios (Section~\ref{subsub_1318ratio}) as a function of Galactocentric radius in Figure~\ref{fg_R1318_Rgc}. 
The resulting values are broadly consistent with previous measurements based on molecular clouds.
The $^{13}$CO/C$^{18}$O ratios in the solar neighborhood vary strongly relative to the large-scale Galactic trend.
Together with the relatively extreme values reported in Solar Neighborhood \citep[][]{Roueff_2021A&A...645A..26R,Shimajiri2014A&A...564A..68S_NobeyamaOrionA13COC18O}, this suggests that nearby clouds may not be fully representative of typical clouds at $R_{\rm gc}\sim8$~kpc.
We therefore fit only the distant clouds in our sample:
\begin{equation}
^{13}\mathrm{CO}/\mathrm{C}^{18}\mathrm{O} = (0.75\pm0.14)\,\frac{R_{\rm gc}}{\rm kpc} + (3.76\pm0.72),
\end{equation}
where $R_{\rm gc}$ is the Galactocentric radius in kpc.
It should be noted that our results are constrained by the limited sampling in Galactocentric radius, and that additional constraints will be required when analyzing sources in the inner ($R_{\rm gc} < 3$\,kpc) and outer ($R_{\rm gc} > 9$\,kpc) Galactic disk.
}

\section{Conclusion}
\label{sec_summary}

We have presented a unified, optical-depth–corrected framework for deriving molecular cloud column densities and self-gravitating gas masses by combining multiple CO isotopologue lines.
By jointly exploiting the complementary optical-depth and detectability regimes of $^{13}$CO and C$^{18}$O $J$=1-0, the method achieves an extended column-density dynamic range and a self-consistent identification of gravitationally bound structures via $N$-PDF analysis.

Applying this approach to nearby and distant molecular clouds, we find that the CO-based $N$-PDFs agree well with the dust-based results specifically in the power-law tail, and that both the derived bound gas masses and the spatial distributions of bound structures are in close agreement between the two methods.

A key advantage of the CO-based method lies in its intrinsic velocity resolution.
In this work, we deliberately apply the framework to a sample that has been shown to suffer minimal LOS confusion, allowing a controlled and fair test of whether the CO-based approach can reproduce dust-based $N$-PDFs and bound gas properties.
Having demonstrated such consistency, our results establish the CO-based method as a powerful tool for future studies of regions with complex velocity structures, where LOS blending can strongly bias dust-based $N$-PDF and bound-mass measurements.
This capability is particularly important for large-sample analyses in the Galactic plane, where LOS confusion is widespread.

\begin{acknowledgements}

This work is supported by the National Natural Science Foundation of China (NSFC) grant numbers 12588202 and 12041302, the New Cornerstone Science Foundation, and additional funding sources acknowledged by individual authors below.
H.B.L. is supported by the National Science and Technology Council (NSTC) of Taiwan (Grant Nos. 111-2112-M-110-022-MY3, 113-2112-M-110-022-MY3).
Z.Y.Z. acknowledges the support of the National Natural Science Foundation of China (NSFC) under grants No. 12041305, 12173016,
1257030642, 12533003.

\end{acknowledgements}

\facility{Nobeyama, \textit{Hersechel}, IRAM-30m, PMO, FCRAO}
\software{Python}

\begin{appendix}

\section{Testing the Optical-depth-corrected Combination Method Based on $^{12}$CO and $^{13}$CO}
\label{appendix_test12+13}

\begin{figure}[htb!]
    \begin{tabular}{ c }
        \hspace{-0.8cm}\includegraphics[width = 8.5 cm]{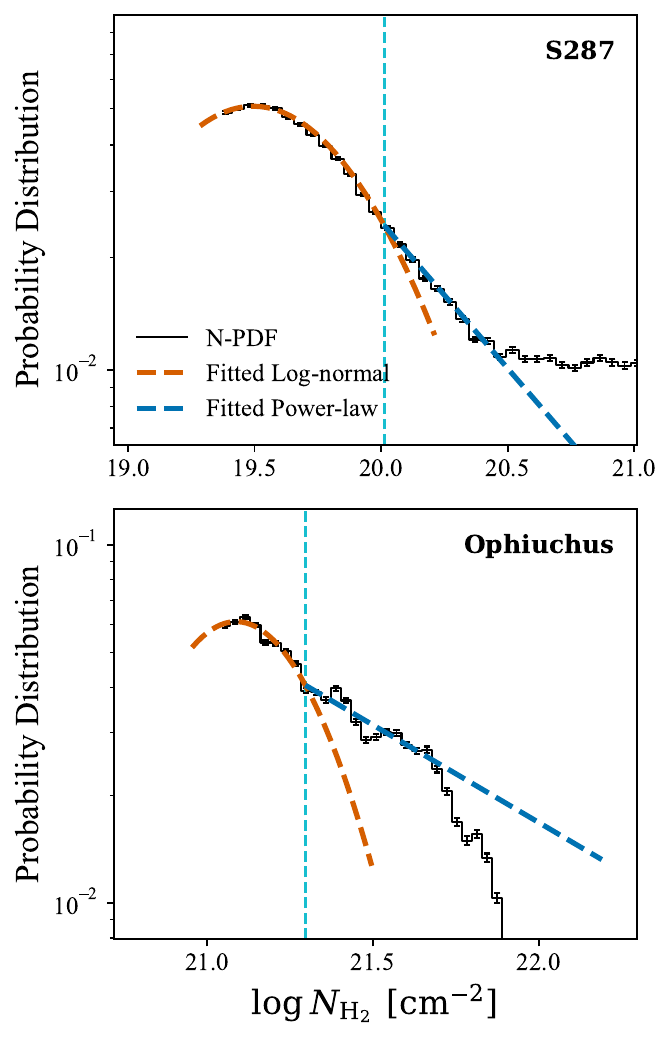}
    \end{tabular}
    \caption{
    $N$-PDFs of Ophiuchus (bottom) and S287 (top), based on column density maps derived from the $^{12}$CO+$^{13}$CO combination with optical-depth correction. The black step lines show the measured distributions, while the orange and blue dashed curves represent the fitted lognormal and power-law components, respectively. The vertical cyan dashed line marks the transition between the two regimes.
    }
    \label{fg_npdf_12COtest}
\end{figure}

CO isotopologue surveys reveal that the Milky Way hosts a large population of low- to intermediate-mass molecular clouds, where $^{12}$CO and $^{13}$CO are typically the dominant tracers, while C$^{18}$O detections are sparse \citep[][]{Wang2023AJ....166..121W,Wang2019ApJS..243...25W,Yuan_2022}. 
% \jnote{Whether do the sensitivities of these surveys influence this statement?}
%In such clouds, $^{12}$CO and $^{13}$CO are typically the dominant tracers, while C$^{18}$O detections are relatively sparse \citep[][]{Wang2023AJ....166..121W,Wang2019ApJS..243...25W,Yuan_2022}. 
To explore whether our optical-depth-corrected combination method can be extended to the $^{12}$CO + $^{13}$CO case, we focused on Perseus and Ophiuchus, the two nearby star-forming regions for which wide-field $^{12}$CO and $^{13}$CO data are available from the Coordinated Molecular Probe Line Extinction and Thermal Emission (COMPLETE; \citealt{Ridge_2006AJ....131.2921R_FCRAO_CO}) survey.
% \jnote{Why do we select these two clouds? This survey only cover these two regions?} \fnote{Yes. I've revised the corresponding content.}
Because the Perseus FCRAO field does not fully enclose the dust-derived bound structure, its $N$-PDFs may be unreliable due to sensitivity to the last closed contour \citep[][]{Alves_2017}. 
We therefore focus on Ophiuchus as the primary example to demonstrate the $^{12}$CO + $^{13}$CO combination method.
To further assess the applicability of this approach to large-scale Galactic plane surveys, we also included S287, a molecular cloud located in the anti-center direction at a Galactocentric radius of $\sim$10.2 kpc, where C$^{18}$O detections are nearly absent.

Using the same method described in Section \ref{sec_method}, we derived optical depths for $^{12}$CO from the $^{12}$CO/$^{13}$CO line ratio.
This allowed us to construct an optical-depth-corrected $^{12}$CO column density map. 
Following Equation \ref{eq_COabundance_Rgc}, we obtained $N_{\rm H_2}^{\rm CO_{comb}}$ maps. 
For Ophiuchus, we adopted the $^{12}$C/$^{13}$C ratio measured from CN absorption \citep[][]{Ritchey2011ApJ...728...36R_1213_nearby}, while for S287 we applied the Galactocentric gradient model from Section~\ref{subsub_NH2_COcomb}.

\begin{figure*}[!t]
    \begin{tabular}{ c c }
        \hspace{-0.9cm}\includegraphics[height = 6.1 cm]{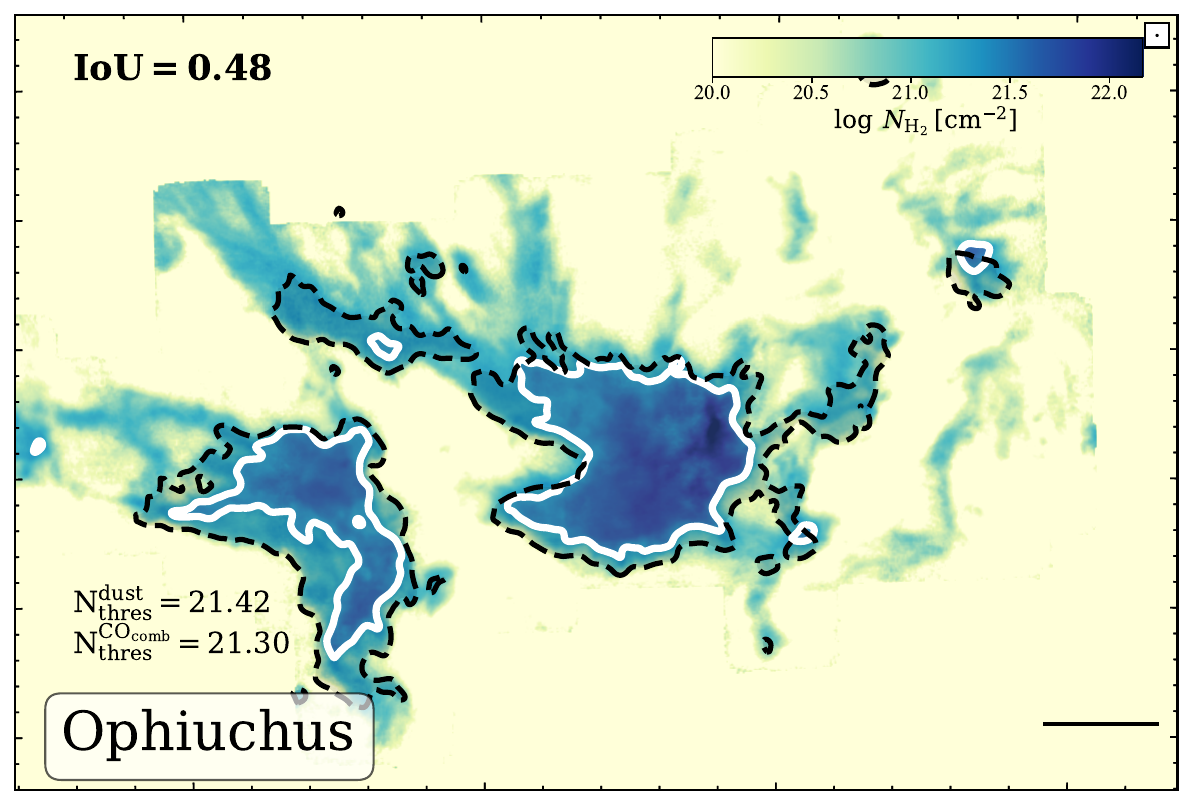} & \hspace{-0.2cm}\includegraphics[height = 6.1 cm]{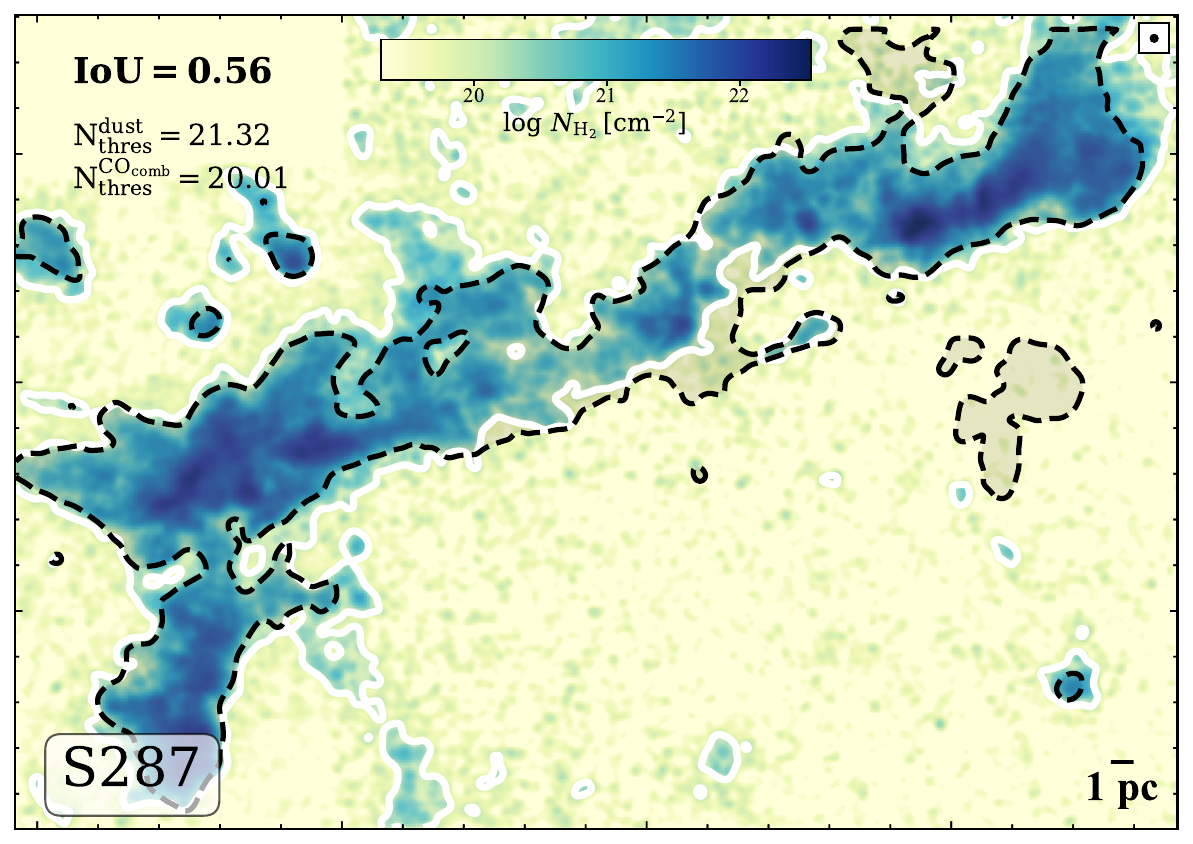} \\
    \end{tabular}
    \caption{
    Same as Figure \ref{fg_NH2map_nearby}, but for the $^{12}$CO+$^{13}$CO combination in Ophiuchus and S287.
    }
\label{fg_NH2map_12COtest}
\end{figure*}

The derived $N$-PDFs for Ophiuchus and S287 (Figure~\ref{fg_npdf_12COtest}) show log-normal components at low column densities and power-law tails at high column densities. 
The overlap between the bound gas structures traced by the CO combination and those derived from dust emission reaches an IoU of $\sim$0.5 in both cases, indicating good spatial consistency (Figure~\ref{fg_NH2map_12COtest}). Together with the agreement in the mass–mass comparison (Figure~\ref{fg_result_distant}), this indicates that the two methods consistently recover both the spatial distribution and the total amount of self-gravitating gas.

These tests demonstrate that combining $^{12}$CO and $^{13}$CO within the optical–depth–corrected framework is feasible, confirming that the method is applicable beyond the $^{13}$CO+C$^{18}$O case.
More broadly, with appropriate treatment of excitation and abundance, the framework can be extended to other tracers, such as HCN or HCO$^+$, to probe higher column densities. 
A caveat is that when $^{12}$CO is used to trace diffuse gas, effects such as abundance reduction in photodissociation region \citep[][]{Liszt2007A&A...476..291L_CO_PDR,vanDishoeck1988ApJ...334..771V_COPDR} and possible departures from LTE \citep[][]{Goldsmith2008ApJ...680..428G} should be considered, although a detailed discussion lies beyond the scope of this work.

\section{Maps and IoU Measurements for the Distant Cloud Sample}
\label{appendix_map_distant}

\begin{figure*}[htp!]
    \begin{tabular}{ c }
        \hspace{-0.9cm}\includegraphics[width = 18.2 cm]{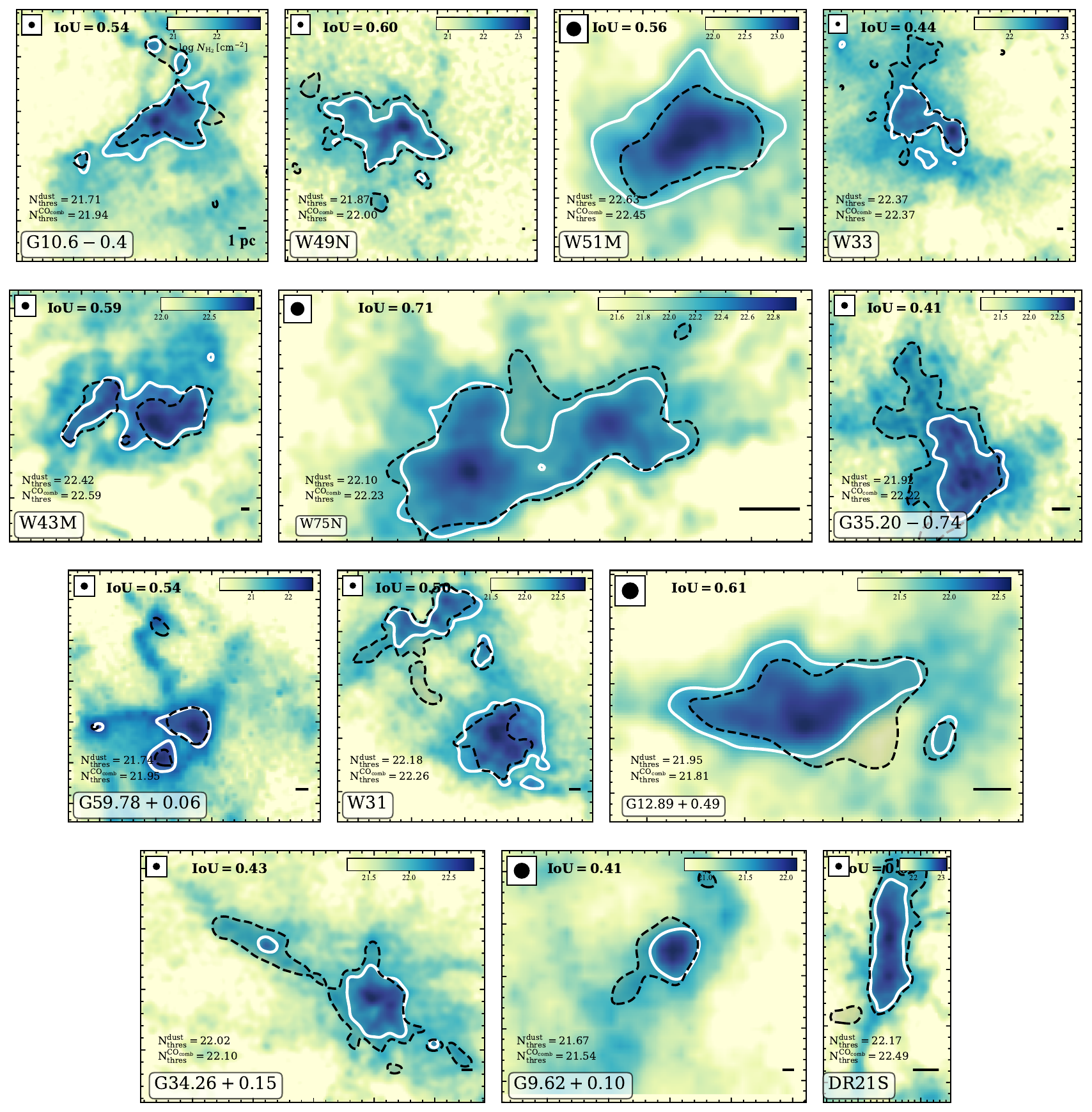}
    \end{tabular}
    \caption{
    Same as Figure~\ref{fg_NH2map_nearby}, but for the distant clouds.
    }
    \label{fg_NH2map_distant}
\end{figure*}

In this appendix, we present the column-density maps of all distant molecular clouds in Figure~\ref{fg_NH2map_distant}.
For each source, we show the contours of the self-gravitating structures traced by two methods, together with the corresponding IoU value.

\section{Dust column density}
\label{appendix_dustSED}

We derived column density and temperature maps for all sources by fitting the spectral energy distributions (SEDs) with a modified graybody model. \citep{Hildebrand1983QJRAS..24..267H},
\begin{equation}
  S_{\nu} = \Omega B_{\nu}(T_{\rm d})(1-e^{-\tau}),
\label{eq1}
\end{equation}
where $S_{\nu}$ is the flux density at frequency $\nu$.
\begin{equation}
  B(T_{\rm d}) = \frac{{\rm 2h}\nu^3}{\rm c^2}\frac{1}{e^{{\rm h}\nu/{\rm k}{T_{\rm d}}}-1}
\label{eq2}
\end{equation}
is the Planck function for a given dust temperature $T_{\rm d}$, and $\Omega$ is the solid angle. 
The optical depth of the dust emission $\tau$ can be expressed by:
\begin{equation}
    \tau_\nu = \Sigma_{\rm dust}\kappa_\nu,
\label{eq3}
\end{equation}
where $\Sigma_{\rm dust}$ is the surface density of dust grains and $\kappa_\nu$ is the dust opacity at frequency $\nu$. The hydrogen column density $N_{\rm H_2}$ can be related to dust surface density as:
\begin{equation}
    N_{\rm H_2} = \frac{\Sigma_{\rm dust} g}{\rm \mu  m_H},
\label{eq4}
\end{equation}
where, ${\rm \mu}=2.8$ is the mean molecule weight in the interstellar medium \citep{Draine2011piim.book.....D}, and $m_H$ is the mass of a hydrogen atom. 
The absorption coefficient at a fixed wavelength, $\kappa_\nu$, can be related with the dust emissivity index $\beta$ as:
\begin{equation}
    \kappa_\nu = \kappa_{\rm 1000}(\frac{\nu}{\rm 1000GHz})^\beta,
\label{eq5}
\end{equation}
where $\kappa_{\rm 1000}$ is the dust absorption coefficient at 1000 GHz. 
We adopted $\kappa_{\rm 1000}=$10 cm$^2$g$^{-1}$ and $\beta\:=1.8$ \citep{Hildebrand1983QJRAS..24..267H}.
The gas-to-dust ratio $g$ was determined using a Galactocentric-radius-dependent model \citep{Giannetti_2017A&A...606L..12G}:
\begin{equation}
    \log(g) = 0.062 (R_{\rm gc}/{\rm kpc}) + 1.65.
\label{eq6}
\end{equation}

\Revise{
For each pixel, the SED fit was first performed using the $160$--$500\mu$m bands. 
The $70\mu$m band was not used as a standard fitting point for all pixels, but was included only as an upper-limit constraint when the $160$--$500\mu$m fit extrapolated to an intensity higher than the observed $70\mu$m value \citep[][]{Feng_2026ApJ...998..224F}. 
This treatment prevents possible excess $70\mu$m emission from warm dust \citep[e.g.][]{Barnes_2017MNRAS.469.2263B} or very small grains \citep[][]{Compiegne_2010ApJ...724L..44C, Desert_1990A&A...237..215D} from biasing the cold-dust column-density estimate, while still constraining cases in which the long-wavelength fit would otherwise overpredict the short-wavelength emission.
}
% \Revise{In the SED fitting, the 70~$\mu$m data were treated as upper limits rather than being directly included as regular fitting points.
% This choice helps reduce the possible bias from excess 70~$\mu$m emission relative to the far-infrared gray-body component, which may arise from warmer dust components  or a contribution from very small grains. }
The fittings were carried out using the {\tt curve\_fit} function of the Python package \textbf{SciPy} \citep{Virtanen2020NatMe..17..261V_SciPy}, which is based on a non-linear least-squares algorithm and accounts for the noise level of each band. 
Our fittings incorporated the {\it Herschel} 70, 160, 250, 350, and 500 $\mu$m images (Section \ref{sub_Herschel}).
Prior to the SED fitting, all Herschel maps were convolved to a common angular resolution corresponding to the largest beam size among the bands, i.e., 36\farcs3 at 500~$\mu$m.

\section{Unnormalized $N$-PDFs}
\label{appendix_npdf_NoNor}

\begin{figure*}[!b]
    \begin{tabular}{ c }
        \hspace{-0.8cm}\includegraphics[width = 18 cm]{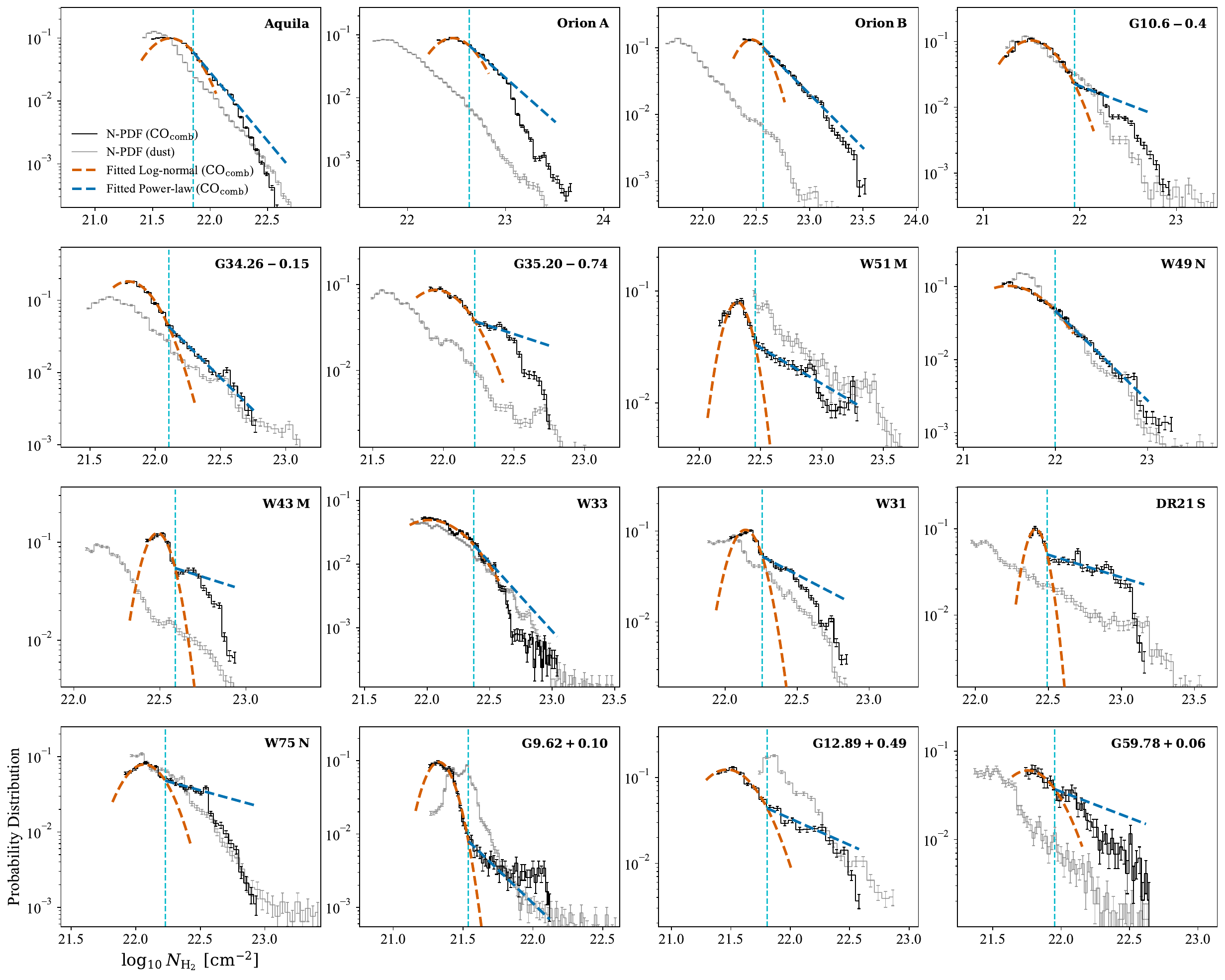}
    \end{tabular}
    \caption{
    Same as Figure~\ref{fg_npdf_all}, but with the x-axis shown in $N_{\rm H_2}$. The dust- and CO-based $N$-PDFs are expressed using the $N_{\rm H_2}$ values independently derived from the corresponding tracers.
    }
    \label{fg_npdf_all_NoNor}
\end{figure*}

This appendix presents a direct comparison of the dust- and CO-based $N$-PDFs for all clouds in the sample without normalization by $N_{\rm LN}$ (Figure~\ref{fg_npdf_all_NoNor}), thereby preserving their absolute column density scales.

\section{Best-fit $N$-PDF Parameters}
\label{appendix_npdf_para}

Table~\ref{tb_npdf_para} lists the best-fit $N$-PDF parameters derived from the dust- and CO${\rm comb}$-based column-density maps for all clouds analyzed in this work. For each tracer, we report the four fitted $N$-PDF parameters, $\mu$, $\sigma\eta$, $\eta_t$, and $\alpha$, together with the corresponding absolute transition column density $N_{\rm thres}$, the cutoff column density $N_{\rm cutoff}$ adopted for the $N$-PDF fitting, and the mean column density $N_{\rm mean}$. All column densities are given in units of $10^{21}{\rm cm}^{-2}$. 

\begin{sidewaystable*}
\caption{N-PDF fitting results and associated column-density parameters.}
\label{tb_npdf_para}
\scriptsize
\setlength{\tabcolsep}{2.2pt}
\hspace{-4.3cm}
\begin{tabular}{lcccccccc|ccccccc}
\hline
\hline
Source &
\multicolumn{8}{c|}{Dust} &
\multicolumn{7}{c}{CO$_{\rm comb}$} \\
\cline{2-9}
\cline{10-16}
&
$\mu$ & $\sigma_\eta$ & $\eta_t$ & $\alpha$ &
$N_{\rm thres}$ & $N_{\rm cutoff}$ & $N_{\rm mean}$ & $N_{\rm subs}$ &
$\mu$ & $\sigma_\eta$ & $\eta_t$ & $\alpha$ &
$N_{\rm thres}$ & $N_{\rm cutoff}$ & $N_{\rm mean}$ \\
&
& & & &
\multicolumn{1}{c}{$(10^{21}\,{\rm cm}^{-2})$} &
\multicolumn{1}{c}{$(10^{21}\,{\rm cm}^{-2})$} &
\multicolumn{1}{c}{$(10^{21}\,{\rm cm}^{-2})$} &
\multicolumn{1}{c|}{$(10^{21}\,{\rm cm}^{-2})$} &
& & & &
\multicolumn{1}{c}{$(10^{21}\,{\rm cm}^{-2})$} &
\multicolumn{1}{c}{$(10^{21}\,{\rm cm}^{-2})$} &
\multicolumn{1}{c}{$(10^{21}\,{\rm cm}^{-2})$} \\
\hline
Aquila & $-0.29\pm0.01$ & $0.38\pm0.01$ & $0.29_{-0.01}^{+0.01}$ & $-2.16\pm0.01$ & $6.1_{-0.1}^{+0.1}$ & 2.5 & 4.5 & 6.5 & $-0.31\pm0.01$ & $0.46\pm0.01$ & $0.15_{-0.01}^{+0.01}$ & $-2.18\pm0.01$ & $7.1_{-0.1}^{+0.1}$ & 3.0 & 6.1 \\
Orion A & $-0.80\pm0.01$ & $0.66\pm0.02$ & $-0.21_{-0.03}^{+0.03}$ & $-1.33\pm0.02$ & $10.6_{-0.3}^{+0.4}$ & 4.4 & 13.0 & 0.8 & $-0.46\pm0.01$ & $0.53\pm0.02$ & $-0.08_{-0.02}^{+0.02}$ & $-1.38\pm0.02$ & $42.5_{-0.8}^{+0.9}$ & 19.5 & 45.8 \\
Orion B & $-0.54\pm0.01$ & $0.36\pm0.01$ & $-0.11_{-0.02}^{+0.03}$ & $-1.80\pm0.05$ & $9.0_{-0.2}^{+0.3}$ & 4.4 & 10.1 & 1.1 & $-0.54\pm0.04$ & $0.34\pm0.06$ & $-0.28_{-0.05}^{+0.05}$ & $-1.61\pm0.04$ & $36.9_{-1.6}^{+2.1}$ & 22.9 & 49.0 \\
G10.6-0.4 & $-0.69\pm0.01$ & $0.52\pm0.01$ & $-0.05_{-0.03}^{+0.04}$ & $-0.95\pm0.12$ & $5.1_{-0.2}^{+0.2}$ & 1.6 & 5.4 & 8.2 & $-0.62\pm0.02$ & $0.59\pm0.02$ & $0.42_{-0.02}^{+0.02}$ & $-0.56\pm0.06$ & $8.8_{-0.2}^{+0.2}$ & 1.6 & 5.8 \\
G34.26-0.15 & $-0.81\pm0.01$ & $0.57\pm0.01$ & $0.05_{-0.02}^{+0.01}$ & $-1.54\pm0.04$ & $10.4_{-0.2}^{+0.1}$ & 2.9 & 9.9 & 9.8 & $-0.53\pm0.02$ & $0.42\pm0.02$ & $0.18_{-0.03}^{+0.03}$ & $-1.79\pm0.10$ & $12.7_{-0.4}^{+0.4}$ & 5.6 & 10.6 \\
G35.20-0.74 & $-0.85\pm0.01$ & $0.52\pm0.01$ & $-0.04_{-0.01}^{+0.01}$ & $-0.37\pm0.07$ & $8.2_{-0.1}^{+0.1}$ & 3.0 & 8.6 & 5.5 & $-0.60\pm0.02$ & $0.49\pm0.02$ & $0.04_{-0.02}^{+0.02}$ & $-0.54\pm0.04$ & $16.7_{-0.3}^{+0.3}$ & 7.8 & 16.1 \\
W51 M & $-1.16\pm0.03$ & $0.28\pm0.05$ & $-0.87_{-0.04}^{+0.04}$ & $-0.82\pm0.16$ & $42.8_{-1.8}^{+1.7}$ & 26.8 & 102.1 & 5.5 & $-0.76\pm0.01$ & $0.25\pm0.02$ & $-0.42_{-0.02}^{+0.02}$ & $-0.64\pm0.09$ & $28.5_{-0.6}^{+0.7}$ & 14.1 & 43.4 \\
W49 N & $-0.85\pm0.01$ & $0.38\pm0.01$ & $-0.35_{-0.02}^{+0.02}$ & $-1.22\pm0.05$ & $7.5_{-0.1}^{+0.1}$ & 3.2 & 10.6 & 10.9 & $-1.19\pm0.02$ & $0.94\pm0.03$ & $-0.01_{-0.14}^{+0.17}$ & $-1.23\pm0.05$ & $9.9_{-1.3}^{+1.8}$ & 2.5 & 10.0 \\
W43 M & $-0.53\pm0.02$ & $0.37\pm0.01$ & $0.12_{-0.02}^{+0.02}$ & $-1.05\pm0.11$ & $26.6_{-0.4}^{+0.5}$ & 11.4 & 23.6 & 4.4 & $-0.26\pm0.01$ & $0.18\pm0.01$ & $-0.02_{-0.01}^{+0.01}$ & $-0.56\pm0.06$ & $38.8_{-0.3}^{+0.3}$ & 25.7 & 39.8 \\
W33 & $-0.63\pm0.02$ & $0.67\pm0.01$ & $0.34_{-0.01}^{+0.01}$ & $-2.16\pm0.03$ & $23.5_{-0.2}^{+0.1}$ & 7.4 & 16.7 & 5.5 & $-0.48\pm0.01$ & $0.61\pm0.01$ & $0.34_{-0.02}^{+0.01}$ & $-2.16\pm0.05$ & $23.7_{-0.4}^{+0.2}$ & 8.9 & 16.8 \\
W31 & $-0.45\pm0.01$ & $0.34\pm0.01$ & $-0.06_{-0.01}^{+0.01}$ & $-0.87\pm0.14$ & $15.2_{-0.2}^{+0.2}$ & 7.4 & 16.1 & 8.2 & $-0.40\pm0.01$ & $0.23\pm0.01$ & $-0.13_{-0.01}^{+0.01}$ & $-0.82\pm0.09$ & $18.1_{-0.2}^{+0.2}$ & 10.5 & 20.6 \\
DR21 S & $-1.14\pm0.02$ & $0.31\pm0.04$ & $-0.83_{-0.03}^{+0.03}$ & $-0.93\pm0.09$ & $15.0_{-0.5}^{+0.5}$ & 9.1 & 34.5 & 5.5 & $-0.66\pm0.01$ & $0.16\pm0.02$ & $-0.48_{-0.02}^{+0.02}$ & $-0.52\pm0.02$ & $31.0_{-0.5}^{+0.5}$ & 22.9 & 50.2 \\
W75 N & $-0.61\pm0.01$ & $0.18\pm0.02$ & $-0.44_{-0.02}^{+0.02}$ & $-0.64\pm0.11$ & $12.6_{-0.2}^{+0.3}$ & 8.9 & 19.7 & 3.8 & $-0.49\pm0.01$ & $0.37\pm0.02$ & $-0.12_{-0.02}^{+0.03}$ & $-0.47\pm0.08$ & $16.9_{-0.4}^{+0.5}$ & 7.9 & 19.1 \\
G9.62+0.10 & $-0.20\pm0.01$ & $0.23\pm0.01$ & $0.27_{-0.01}^{+0.01}$ & $-3.64\pm0.18$ & $4.7_{-0.1}^{+0.1}$ & 1.8 & 3.6 & 3.8 & $-0.26\pm0.01$ & $0.22\pm0.01$ & $0.22_{-0.02}^{+0.02}$ & $-1.83\pm0.43$ & $3.5_{-0.1}^{+0.1}$ & 1.8 & 2.8 \\
G12.89+0.49 & $-0.50\pm0.01$ & $0.20\pm0.01$ & $-0.26_{-0.04}^{+0.02}$ & $-1.18\pm0.24$ & $8.8_{-0.3}^{+0.2}$ & 5.1 & 11.4 & 5.5 & $-0.84\pm0.02$ & $0.54\pm0.03$ & $-0.07_{-0.04}^{+0.05}$ & $-0.62\pm0.07$ & $6.4_{-0.2}^{+0.3}$ & 2.3 & 6.8 \\
G59.78+0.06 & $-0.52\pm0.01$ & $0.38\pm0.01$ & $0.09_{-0.02}^{+0.02}$ & $-1.42\pm0.06$ & $5.5_{-0.1}^{+0.1}$ & 2.3 & 5.1 & 5.5 & $-0.69\pm0.07$ & $0.46\pm0.11$ & $-0.23_{-0.13}^{+0.07}$ & $-0.59\pm0.26$ & $9.0_{-1.1}^{+0.6}$ & 5.4 & 11.3 \\
\hline
\end{tabular}
\tablecomments{
For each tracer, the first four columns give the best-fit $N$-PDF parameters:
$\mu$, $\sigma_\eta$, $\eta_t$, and $\alpha$.
The additional columns give the absolute transition column density $N_{\rm thres}$, the cutoff column density $N_{\rm cutoff}$ adopted for the $N$-PDF fitting, and the mean column density $N_{\rm mean}$.
For the dust-derived maps, $N_{\rm subs}$ denotes the foreground/background column density subtracted during the LOS correction.
All column densities are given in units of $10^{21}\,{\rm cm}^{-2}$.
}
\end{sidewaystable*}

\end{appendix}

\bibliography{ref.bib}

@ARTICLE{Andre2010_HGBS,
       author = {{Andr{\'e}}, Ph. and {Men'shchikov}, A. and {Bontemps}, S. and {K{\"o}nyves}, V. and {Motte}, F. and {Schneider}, N. and {Didelon}, P. and {Minier}, V. and {Saraceno}, P. and {Ward-Thompson}, D. and {di Francesco}, J. and {White}, G. and {Molinari}, S. and {Testi}, L. and {Abergel}, A. and {Griffin}, M. and {Henning}, Th. and {Royer}, P. and {Mer{\'\i}n}, B. and {Vavrek}, R. and {Attard}, M. and {Arzoumanian}, D. and {Wilson}, C.~D. and {Ade}, P. and {Aussel}, H. and {Baluteau}, J. -P. and {Benedettini}, M. and {Bernard}, J. -Ph. and {Blommaert}, J.~A.~D.~L. and {Cambr{\'e}sy}, L. and {Cox}, P. and {di Giorgio}, A. and {Hargrave}, P. and {Hennemann}, M. and {Huang}, M. and {Kirk}, J. and {Krause}, O. and {Launhardt}, R. and {Leeks}, S. and {Le Pennec}, J. and {Li}, J.~Z. and {Martin}, P.~G. and {Maury}, A. and {Olofsson}, G. and {Omont}, A. and {Peretto}, N. and {Pezzuto}, S. and {Prusti}, T. and {Roussel}, H. and {Russeil}, D. and {Sauvage}, M. and {Sibthorpe}, B. and {Sicilia-Aguilar}, A. and {Spinoglio}, L. and {Waelkens}, C. and {Woodcraft}, A. and {Zavagno}, A.},
        title = "{From filamentary clouds to prestellar cores to the stellar IMF: Initial highlights from the Herschel Gould Belt Survey}",
      journal = {\aap},
         year = 2010,
        month = jul,
       volume = {518},
          eid = {L102},
        pages = {L102},
          doi = {10.1051/0004-6361/201014666},
archivePrefix = {arXiv},
       eprint = {1005.2618},
 primaryClass = {astro-ph.GA},
       adsurl = {https://ui.adsabs.harvard.edu/abs/2010A&A...518L.102A}
}

@ARTICLE{Lin2016,
   author = {{Lin}, Y. and {Liu}, H.~B. and {Li}, D. and {Zhang}, Z.-Y. and 
	{Ginsburg}, A. and {Pineda}, J.~E. and {Qian}, L. and {Galv{\'a}n-Madrid}, R. and 
	{McLeod}, A.~F. and {Rosolowsky}, E. and {Dale}, J.~E. and {Immer}, K. and 
	{Koch}, E. and {Longmore}, S. and {Walker}, D. and {Testi}, L.
	},
    title = "{Cloud Structure of Galactic OB Cluster-forming Regions from Combining Ground- and Space-based Bolometric Observations}",
  journal = {\apj},
archivePrefix = "arXiv",
   eprint = {1606.07645},
     year = 2016,
    month = sep,
   volume = 828,
      eid = {32},
    pages = {32},
      doi = {10.3847/0004-637X/828/1/32},
   adsurl = {http://adsabs.harvard.edu/abs/2016ApJ...828...32L}
}

@ARTICLE{Lin2017ApJ...840...22L,
       author = {{Lin}, Yuxin and {Liu}, Hauyu Baobab and {Dale}, James E. and {Li}, Di and {Busquet}, Gemma and {Zhang}, Zhi-Yu and {Ginsburg}, Adam and {Galv{\'a}n-Madrid}, Roberto and {Kov{\'a}cs}, Attila and {Koch}, Eric and {Qian}, Lei and {Wang}, Ke and {Longmore}, Steve and {Chen}, Huei-Ru and {Walker}, Daniel},
        title = "{Cloud Structure of Three Galactic Infrared Dark Star-forming Regions from Combining Ground- and Space-based Bolometric Observations}",
      journal = {\apj},
         year = 2017,
        month = may,
       volume = {840},
       number = {1},
          eid = {22},
        pages = {22},
          doi = {10.3847/1538-4357/aa6c67},
archivePrefix = {arXiv},
       eprint = {1704.06448},
 primaryClass = {astro-ph.GA},
       adsurl = {https://ui.adsabs.harvard.edu/abs/2017ApJ...840...22L}
}

@ARTICLE{Wilson1994,
       author = {{Wilson}, T.~L. and {Rood}, R.},
        title = "{Abundances in the Interstellar Medium}",
      journal = {\araa},
         year = 1994,
        month = jan,
       volume = {32},
        pages = {191-226},
          doi = {10.1146/annurev.aa.32.090194.001203},
       adsurl = {https://ui.adsabs.harvard.edu/abs/1994ARA&A..32..191W}
}

@ARTICLE{Mangum2017,
       author = {{Mangum}, Jeffrey G. and {Shirley}, Yancy L.},
        title = "{How to Calculate Molecular Column Density}",
      journal = {\pasp},
         year = 2015,
        month = mar,
       volume = {127},
       number = {949},
        pages = {266},
          doi = {10.1086/680323},
archivePrefix = {arXiv},
       eprint = {1501.01703},
 primaryClass = {astro-ph.IM},
       adsurl = {https://ui.adsabs.harvard.edu/abs/2015PASP..127..266M}
}

@ARTICLE{Gordy1984,
       author = {{Aliev}, M.~P. and {Krupnov}, A.~F.},
        title = "{Microwave Spectra of Molecules, By Walter Gordy and Robert L. Cook. 3rd edition (Wiley Interscience, New York, 1984)}",
      journal = {Optics and Spectroscopy},
         year = 1985,
        month = nov,
       volume = {59},
       number = {5},
        pages = {699},
       adsurl = {https://ui.adsabs.harvard.edu/abs/1985OptSp..59..699A}
}

@ARTICLE{Kennicutt_Evans_2012,
       author = {{Kennicutt}, Robert C. and {Evans}, Neal J.},
        title = "{Star Formation in the Milky Way and Nearby Galaxies}",
      journal = {\araa},
         year = 2012,
        month = sep,
       volume = {50},
        pages = {531-608},
          doi = {10.1146/annurev-astro-081811-125610},
archivePrefix = {arXiv},
       eprint = {1204.3552},
 primaryClass = {astro-ph.GA},
       adsurl = {https://ui.adsabs.harvard.edu/abs/2012ARA&A..50..531K}
}

@ARTICLE{Frerking_1982,
       author = {{Frerking}, M.~A. and {Langer}, W.~D. and {Wilson}, R.~W.},
        title = "{The relationship between carbon monoxide abundance and visual extinction in interstellar clouds.}",
      journal = {\apj},
         year = 1982,
        month = nov,
       volume = {262},
        pages = {590-605},
          doi = {10.1086/160451},
       adsurl = {https://ui.adsabs.harvard.edu/abs/1982ApJ...262..590F}
}

@ARTICLE{Galvan_2013,
       author = {{Galv{\'a}n-Madrid}, R. and {Liu}, H.~B. and {Zhang}, Z. -Y. and {Pineda}, J.~E. and {Peng}, T. -C. and {Zhang}, Q. and {Keto}, E.~R. and {Ho}, P.~T.~P. and {Rodr{\'\i}guez}, L.~F. and {Zapata}, L. and {Peters}, T. and {De Pree}, C.~G.},
        title = "{MUSCLE W49: A Multi-Scale Continuum and Line Exploration of the Most Luminous Star Formation Region in the Milky Way. I. Data and the Mass Structure of the Giant Molecular Cloud}",
      journal = {\apj},
         year = 2013,
        month = dec,
       volume = {779},
       number = {2},
          eid = {121},
        pages = {121},
          doi = {10.1088/0004-637X/779/2/121},
archivePrefix = {arXiv},
       eprint = {1309.4129},
}

@ARTICLE{Jacob_2020,
       author = {{Jacob}, Arshia M. and {Menten}, Karl M. and {Wiesemeyer}, Helmut and {G{\"u}sten}, Rolf and {Wyrowski}, Friedrich and {Klein}, Bernd},
        title = "{First detection of $^{13}$CH in the interstellar medium}",
      journal = {\aap},
         year = 2020,
        month = aug,
       volume = {640},
          eid = {A125},
        pages = {A125},
          doi = {10.1051/0004-6361/201937385},
archivePrefix = {arXiv},
       eprint = {2007.01190},
 primaryClass = {astro-ph.GA},
       adsurl = {https://ui.adsabs.harvard.edu/abs/2020A&A...640A.125J}
}

@ARTICLE{Dickman_1978,
       author = {{Dickman}, R.~L.},
        title = "{The ratio of carbon monoxide to molecular hydrogen in interstellar dark clouds.}",
      journal = {\apjs},
         year = 1978,
        month = aug,
       volume = {37},
        pages = {407-427},
          doi = {10.1086/190535},
       adsurl = {https://ui.adsabs.harvard.edu/abs/1978ApJS...37..407D}
}

@PHDTHESIS{Kulesa_2002,
       author = {{Kulesa}, Craig Alan},
        title = "{Molecular hydrogen and its ions in dark interstellar clouds and star forming regions}",
       school = {University of Arizona},
         year = 2002,
        month = nov,
       adsurl = {https://ui.adsabs.harvard.edu/abs/2002PhDT........28K}
}

@ARTICLE{Mendez-delgado_2022,
       author = {{M{\'e}ndez-Delgado}, J.~E. and {Amayo}, A. and {Arellano-C{\'o}rdova}, K.~Z. and {Esteban}, C. and {Garc{\'\i}a-Rojas}, J. and {Carigi}, L. and {Delgado-Inglada}, G.},
        title = "{Gradients of chemical abundances in the Milky Way from H II regions: distances derived from Gaia EDR3 parallaxes and temperature inhomogeneities}",
      journal = {\mnras},
         year = 2022,
        month = mar,
       volume = {510},
       number = {3},
        pages = {4436-4455},
          doi = {10.1093/mnras/stab3782},
archivePrefix = {arXiv},
       eprint = {2112.12600},
 primaryClass = {astro-ph.GA},
       adsurl = {https://ui.adsabs.harvard.edu/abs/2022MNRAS.510.4436M}
}

@ARTICLE{Evans_2022,
       author = {{Evans}, Neal J. and {Kim}, Jeong-Gyu and {Ostriker}, Eve C.},
        title = "{Slow Star Formation in the Milky Way: Theory Meets Observations}",
      journal = {\apjl},
         year = 2022,
        month = apr,
       volume = {929},
       number = {1},
          eid = {L18},
        pages = {L18},
          doi = {10.3847/2041-8213/ac6427},
archivePrefix = {arXiv},
       eprint = {2204.02314},
 primaryClass = {astro-ph.GA},
       adsurl = {https://ui.adsabs.harvard.edu/abs/2022ApJ...929L..18E}
}

@ARTICLE{Burkhart_2017,
       author = {{Burkhart}, Blakesley and {Stalpes}, Kye and {Collins}, David C.},
        title = "{The Razor{\textquoteright}s Edge of Collapse: The Transition Point from Lognormal to Power-Law Distributions in Molecular Clouds}",
      journal = {\apjl},
         year = 2017,
        month = jan,
       volume = {834},
       number = {1},
          eid = {L1},
        pages = {L1},
          doi = {10.3847/2041-8213/834/1/L1},
archivePrefix = {arXiv},
       eprint = {1609.04409},
 primaryClass = {astro-ph.GA},
       adsurl = {https://ui.adsabs.harvard.edu/abs/2017ApJ...834L...1B}
}

@ARTICLE{Burkhart_2019,
       author = {{Burkhart}, Blakesley and {Mocz}, Philip},
        title = "{The Self-gravitating Gas Fraction and the Critical Density for Star Formation}",
      journal = {\apj},
         year = 2019,
        month = jul,
       volume = {879},
       number = {2},
          eid = {129},
        pages = {129},
          doi = {10.3847/1538-4357/ab25ed},
archivePrefix = {arXiv},
       eprint = {1805.11104},
 primaryClass = {astro-ph.GA},
       adsurl = {https://ui.adsabs.harvard.edu/abs/2019ApJ...879..129B}
}

@ARTICLE{Gong_2020,
       author = {{Gong}, Munan and {Ostriker}, Eve C. and {Kim}, Chang-Goo and {Kim}, Jeong-Gyu},
        title = "{The Environmental Dependence of the X$_{CO}$ Conversion Factor}",
      journal = {\apj},
         year = 2020,
        month = nov,
       volume = {903},
       number = {2},
          eid = {142},
        pages = {142},
          doi = {10.3847/1538-4357/abbdab},
archivePrefix = {arXiv},
       eprint = {2009.14631},
 primaryClass = {astro-ph.GA},
       adsurl = {https://ui.adsabs.harvard.edu/abs/2020ApJ...903..142G}
}

@ARTICLE{Chen_2018,
       author = {{Chen}, Hope How-Huan and {Burkhart}, Blakesley and {Goodman}, Alyssa and {Collins}, David C.},
        title = "{The Anatomy of the Column Density Probability Distribution Function (N-PDF)}",
      journal = {\apj},
         year = 2018,
        month = jun,
       volume = {859},
       number = {2},
          eid = {162},
        pages = {162},
          doi = {10.3847/1538-4357/aabaf6},
archivePrefix = {arXiv},
       eprint = {1707.09356},
 primaryClass = {astro-ph.GA},
       adsurl = {https://ui.adsabs.harvard.edu/abs/2018ApJ...859..162C}
}

@ARTICLE{Alves_2017,
       author = {{Alves}, Jo{\~a}o and {Lombardi}, Marco and {Lada}, Charles J.},
        title = "{The shapes of column density PDFs. The importance of the last closed contour}",
      journal = {\aap},
         year = 2017,
        month = oct,
       volume = {606},
          eid = {L2},
        pages = {L2},
          doi = {10.1051/0004-6361/201731436},
archivePrefix = {arXiv},
       eprint = {1707.02636},
 primaryClass = {astro-ph.GA},
       adsurl = {https://ui.adsabs.harvard.edu/abs/2017A&A...606L...2A}
}

@ARTICLE{Wu_2010,
       author = {{Wu}, Jingwen and {Evans}, II, Neal J. and {Shirley}, Yancy L. and {Knez}, Claudia},
        title = "{The Properties of Massive, Dense Clumps: Mapping Surveys of HCN and CS}",
      journal = {\apjs},
         year = 2010,
        month = jun,
       volume = {188},
       number = {2},
        pages = {313-357},
          doi = {10.1088/0067-0049/188/2/313},
archivePrefix = {arXiv},
       eprint = {1004.0398},
 primaryClass = {astro-ph.GA},
       adsurl = {https://ui.adsabs.harvard.edu/abs/2010ApJS..188..313W}
}

@ARTICLE{Yuan_2022,
       author = {{Yuan}, Lixia and {Yang}, Ji and {Du}, Fujun and {Su}, Yang and {Liu}, Xunchuan and {Zhang}, Shaobo and {Sun}, Yan and {Zhou}, Xin and {Yan}, Qing-Zeng and {Ma}, Yuehui},
        title = "{Molecular Gas Structures Traced by $^{13}$CO Emission in the 18,190 $^{12}$CO Molecular Clouds from the MWISP Survey}",
      journal = {\apjs},
         year = 2022,
        month = aug,
       volume = {261},
       number = {2},
          eid = {37},
        pages = {37},
          doi = {10.3847/1538-4365/ac739f},
archivePrefix = {arXiv},
       eprint = {2205.14858},
 primaryClass = {astro-ph.GA},
       adsurl = {https://ui.adsabs.harvard.edu/abs/2022ApJS..261...37Y}
}

@ARTICLE{Gao_2004,
       author = {{Gao}, Yu and {Solomon}, Philip M.},
        title = "{The Star Formation Rate and Dense Molecular Gas in Galaxies}",
      journal = {\apj},
         year = 2004,
        month = may,
       volume = {606},
       number = {1},
        pages = {271-290},
          doi = {10.1086/382999},
archivePrefix = {arXiv},
       eprint = {astro-ph/0310339},
 primaryClass = {astro-ph},
       adsurl = {https://ui.adsabs.harvard.edu/abs/2004ApJ...606..271G}
}

@ARTICLE{Bigiel_2008,
       author = {{Bigiel}, F. and {Leroy}, A. and {Walter}, F. and {Brinks}, E. and {de Blok}, W.~J.~G. and {Madore}, B. and {Thornley}, M.~D.},
        title = "{The Star Formation Law in Nearby Galaxies on Sub-Kpc Scales}",
      journal = {\aj},
         year = 2008,
        month = dec,
       volume = {136},
       number = {6},
        pages = {2846-2871},
          doi = {10.1088/0004-6256/136/6/2846},
archivePrefix = {arXiv},
       eprint = {0810.2541},
 primaryClass = {astro-ph},
       adsurl = {https://ui.adsabs.harvard.edu/abs/2008AJ....136.2846B}
}

@ARTICLE{Kennicutt_1998,
       author = {{Kennicutt}, Jr., Robert C.},
        title = "{Star Formation in Galaxies Along the Hubble Sequence}",
      journal = {\araa},
         year = 1998,
        month = jan,
       volume = {36},
        pages = {189-232},
          doi = {10.1146/annurev.astro.36.1.189},
archivePrefix = {arXiv},
       eprint = {astro-ph/9807187},
 primaryClass = {astro-ph},
       adsurl = {https://ui.adsabs.harvard.edu/abs/1998ARA&A..36..189K}
}

@ARTICLE{Wu_2005,
       author = {{Wu}, Jingwen and {Evans}, II, Neal J. and {Gao}, Yu and {Solomon}, Philip M. and {Shirley}, Yancy L. and {Vanden Bout}, Paul A.},
        title = "{Connecting Dense Gas Tracers of Star Formation in our Galaxy to High-z Star Formation}",
      journal = {\apjl},
         year = 2005,
        month = dec,
       volume = {635},
       number = {2},
        pages = {L173-L176},
          doi = {10.1086/499623},
archivePrefix = {arXiv},
       eprint = {astro-ph/0511424},
 primaryClass = {astro-ph},
       adsurl = {https://ui.adsabs.harvard.edu/abs/2005ApJ...635L.173W}
}

@ARTICLE{Lada_2012,
       author = {{Lada}, Charles J. and {Forbrich}, Jan and {Lombardi}, Marco and {Alves}, Jo{\~a}o F.},
        title = "{Star Formation Rates in Molecular Clouds and the Nature of the Extragalactic Scaling Relations}",
      journal = {\apj},
         year = 2012,
        month = feb,
       volume = {745},
       number = {2},
          eid = {190},
        pages = {190},
          doi = {10.1088/0004-637X/745/2/190},
archivePrefix = {arXiv},
       eprint = {1112.4466},
 primaryClass = {astro-ph.GA},
       adsurl = {https://ui.adsabs.harvard.edu/abs/2012ApJ...745..190L}
}

@ARTICLE{Evans_2014,
       author = {{Evans}, II, Neal J. and {Heiderman}, Amanda and {Vutisalchavakul}, Nalin},
        title = "{Star Formation Relations in Nearby Molecular Clouds}",
      journal = {\apj},
         year = 2014,
        month = feb,
       volume = {782},
       number = {2},
          eid = {114},
        pages = {114},
          doi = {10.1088/0004-637X/782/2/114},
archivePrefix = {arXiv},
       eprint = {1401.3287},
 primaryClass = {astro-ph.GA},
       adsurl = {https://ui.adsabs.harvard.edu/abs/2014ApJ...782..114E}
}

@ARTICLE{Ma_2022,
       author = {{Ma}, Yuehui and {Wang}, Hongchi and {Zhang}, Miaomiao and {Wang}, Chen and {Zhang}, Shaobo and {Liu}, Yao and {Li}, Chong and {Zheng}, Yuqing and {Yuan}, Lixia and {Yang}, Ji},
        title = "{Gas Column Density Distribution of Molecular Clouds in the Third Quadrant of the Milky Way}",
      journal = {\apjs},
         year = 2022,
        month = sep,
       volume = {262},
       number = {1},
          eid = {16},
        pages = {16},
          doi = {10.3847/1538-4365/ac7797},
archivePrefix = {arXiv},
       eprint = {2206.03963},
 primaryClass = {astro-ph.GA},
       adsurl = {https://ui.adsabs.harvard.edu/abs/2022ApJS..262...16M}
}

@ARTICLE{Nagahama_1998,
       author = {{Nagahama}, Tomoo and {Mizuno}, Akira and {Ogawa}, Hideo and {Fukui}, Yasuo},
        title = "{A Spatially Complete \^13CO J = 1-0 Survey of the Orion A Cloud}",
      journal = {\aj},
         year = 1998,
        month = jul,
       volume = {116},
       number = {1},
        pages = {336-348},
          doi = {10.1086/300392},
       adsurl = {https://ui.adsabs.harvard.edu/abs/1998AJ....116..336N}
}

@ARTICLE{Shimajiri2011PASJ...63..105S_NobeyamaOrionA12CO,
       author = {{Shimajiri}, Yoshito and {Kawabe}, Ryohei and {Takakuwa}, Shigehisa and {Saito}, Masao and {Tsukagoshi}, Takashi and {Momose}, Munetake and {Ikeda}, Norio and {Akiyama}, Eiji and {Austermann}, Jason E. and {Ezawa}, Hajime and {Fukue}, Kei and {Hiramatsu}, Masaaki and {Hughes}, David and {Kitamura}, Yoshimi and {Kohno}, Kohtaro and {Kurono}, Yasutaka and {Scott}, Kimberly S. and {Wilson}, Grant W. and {Yoshida}, Atsumasa and {Yun}, Min S.},
        title = "{New Panoramic View of $^{12}$CO and 1.1mm Continuum Emission in the Orion A Giant Molecular Cloud. I. Survey Overview and Possible External Triggers of Star Formation}",
      journal = {\pasj},
         year = 2011,
        month = feb,
       volume = {63},
        pages = {105},
          doi = {10.1093/pasj/63.1.105},
archivePrefix = {arXiv},
       eprint = {1010.3498},
 primaryClass = {astro-ph.GA},
       adsurl = {https://ui.adsabs.harvard.edu/abs/2011PASJ...63..105S}
}

@ARTICLE{Shimajiri2014A&A...564A..68S_NobeyamaOrionA13COC18O,
       author = {{Shimajiri}, Yoshito and {Kitamura}, Yoshimi and {Saito}, Masao and {Momose}, Munetake and {Nakamura}, Fumitaka and {Dobashi}, Kazuhito and {Shimoikura}, Tomomi and {Nishitani}, Hiroyuki and {Yamabi}, Akifumi and {Hara}, Chihomi and {Katakura}, Sho and {Tsukagoshi}, Takashi and {Tanaka}, Tomohiro and {Kawabe}, Ryohei},
        title = "{High abundance ratio of $^{13}$CO to C$^{18}$O toward photon-dominated regions in the Orion-A giant molecular cloud}",
      journal = {\aap},
         year = 2014,
        month = apr,
       volume = {564},
          eid = {A68},
        pages = {A68},
          doi = {10.1051/0004-6361/201322912},
archivePrefix = {arXiv},
       eprint = {1403.2930},
 primaryClass = {astro-ph.GA},
       adsurl = {https://ui.adsabs.harvard.edu/abs/2014A&A...564A..68S}
}

@article{Yamagishi_2018_NROCygnusX,
   title={Nobeyama 45 m Cygnus-X CO Survey. I. Photodissociation of Molecules Revealed by the Unbiased Large-scale CN and C18O Maps},
   volume={235},
   ISSN={1538-4365},
   url={http://dx.doi.org/10.3847/1538-4365/aaab4b},
   DOI={10.3847/1538-4365/aaab4b},
   number={1},
   journal={The Astrophysical Journal Supplement Series},
   publisher={American Astronomical Society},
   author={Yamagishi, M. and Nishimura, A. and Fujita, S. and Takekoshi, T. and Matsuo, M. and Minamidani, T. and Taniguchi, K. and Tokuda, K. and Shimajiri, Y.},
   year={2018},
   month=mar, pages={9}
}

@INPROCEEDINGS{Minamidani2016SPIE.9914E..1ZM_FORESTreceiver,
       author = {{Minamidani}, Tetsuhiro and {Nishimura}, Atsushi and {Miyamoto}, Yusuke and {Kaneko}, Hiroyuki and {Iwashita}, Hiroyuki and {Miyazawa}, Chieko and {Nishitani}, Hiroyuki and {Wada}, Takuya and {Fujii}, Yasunori and {Takahashi}, Toshikazu and {Iizuka}, Yoshizo and {Ogawa}, Hideo and {Kimura}, Kimihiro and {Kozuki}, Yuto and {Hasegawa}, Yutaka and {Matsuo}, Mitsuhiro and {Fujita}, Shinji and {Ohashi}, Satoshi and {Morokuma-Matsui}, Kana and {Maekawa}, Jun and {Muraoka}, Kazuyuki and {Nakajima}, Taku and {Umemoto}, Tomofumi and {Sorai}, Kazuo and {Nakamura}, Fumitaka and {Kuno}, Nario and {Saito}, Masao},
        title = "{Development of the new multi-beam 100 GHz band SIS receiver FOREST for the Nobeyama 45-m Telescope}",
    booktitle = {Millimeter, Submillimeter, and Far-Infrared Detectors and Instrumentation for Astronomy VIII},
         year = 2016,
       editor = {{Holland}, Wayne S. and {Zmuidzinas}, Jonas},
       series = {Society of Photo-Optical Instrumentation Engineers (SPIE) Conference Series},
       volume = {9914},
        month = jul,
          eid = {99141Z},
        pages = {99141Z},
          doi = {10.1117/12.2232137},
       adsurl = {https://ui.adsabs.harvard.edu/abs/2016SPIE.9914E..1ZM}
}

@ARTICLE{Pety2017A&A...599A..98P_IRAMOrionB1,
       author = {{Pety}, J{\'e}r{\^o}me and {Guzm{\'a}n}, Viviana V. and {Orkisz}, Jan H. and {Liszt}, Harvey S. and {Gerin}, Maryvonne and {Bron}, Emeric and {Bardeau}, S{\'e}bastien and {Goicoechea}, Javier R. and {Gratier}, Pierre and {Le Petit}, Franck and {Levrier}, Fran{\c{c}}ois and {{\"O}berg}, Karin I. and {Roueff}, Evelyne and {Sievers}, Albrecht},
        title = "{The anatomy of the Orion B giant molecular cloud: A local template for studies of nearby galaxies}",
      journal = {\aap},
         year = 2017,
        month = jan,
       volume = {599},
          eid = {A98},
        pages = {A98},
          doi = {10.1051/0004-6361/201629862},
archivePrefix = {arXiv},
       eprint = {1611.04037},
 primaryClass = {astro-ph.GA},
       adsurl = {https://ui.adsabs.harvard.edu/abs/2017A&A...599A..98P}
}

@ARTICLE{Bik2003A&A...404..249B_OrionBHIIregion,
       author = {{Bik}, A. and {Lenorzer}, A. and {Kaper}, L. and {Comer{\'o}n}, F. and {Waters}, L.~B.~F.~M. and {de Koter}, A. and {Hanson}, M.~M.},
        title = "{Identification of the ionizing source of NGC 2024}",
      journal = {\aap},
         year = 2003,
        month = jun,
       volume = {404},
        pages = {249-254},
          doi = {10.1051/0004-6361:20030301},
archivePrefix = {arXiv},
       eprint = {astro-ph/0303029},
 primaryClass = {astro-ph},
       adsurl = {https://ui.adsabs.harvard.edu/abs/2003A&A...404..249B}
}

@ARTICLE{Marsh2017MNRAS.471.2730M_HIGAL_PPMAP,
    author = "{Marsh}, K. A. and {Whitworth}, A. P. and {Lomax}, O. and {Ragan}, S. E. and {Becciani}, U. and {Cambr{\'e}sy}, L. and {Di Giorgio}, A. and {Eden}, D. and {Elia}, D. and {Kacsuk}, P. and {Molinari}, S. and {Palmeirim}, P. and {Pezzuto}, S. and {Schneider}, N. and {Sciacca}, E. and {Vitello}, F.",
    title = "{Multitemperature mapping of dust structures throughout the Galactic Plane using the PPMAP tool with Herschel Hi-GAL data}",
    journal = "\mnras",
    year = "2017",
    month = "November",
    volume = "471",
    number = "3",
    pages = "2730-2742",
    doi = "10.1093/mnras/stx1723",
    archivePrefix = "arXiv",
    eprint = "1707.03808",
    primaryClass = "astro-ph.GA",
    adsurl = "https://ui.adsabs.harvard.edu/abs/2017MNRAS.471.2730M"
}

@ARTICLE{Molinari2010PASP..122..314M_HIGAL,
    author = {{Molinari}, S. and {Swinyard}, B. and {Bally}, J. and {Barlow}, M. and {Bernard}, J. -P. and {Martin}, P. and {Moore}, T. and {Noriega-Crespo}, A. and {Plume}, R. and {Testi}, L. and {Zavagno}, A. and {Abergel}, A. and {Ali}, B. and {Andr{\'e}}, P. and {Baluteau}, J. -P. and {Benedettini}, M. and {Bern{\'e}}, O. and {Billot}, N. P. and {Blommaert}, J. and {Bontemps}, S. and {Boulanger}, F. and {Brand}, J. and {Brunt}, C. and {Burton}, M. and {Campeggio}, L. and {Carey}, S. and {Caselli}, P. and {Cesaroni}, R. and {Cernicharo}, J. and {Chakrabarti}, S. and {Chrysostomou}, A. and {Codella}, C. and {Cohen}, M. and {Compiegne}, M. and {Davis}, C. J. and {de Bernardis}, P. and {de Gasperis}, G. and {Di Francesco}, J. and {di Giorgio}, A. M. and {Elia}, D. and {Faustini}, F. and {Fischera}, J. F. and {Fukui}, Y. and {Fuller}, G. A. and {Ganga}, K. and {Garcia-Lario}, P. and {Giard}, M. and {Giardino}, G. and {Glenn}, J. and {Goldsmith}, P. and {Griffin}, M. and {Hoare}, M. and {Huang}, M. and {Jiang}, B. and {Joblin}, C. and {Joncas}, G. and {Juvela}, M. and {Kirk}, J. and {Lagache}, G. and {Li}, J. Z. and {Lim}, T. L. and {Lord}, S. D. and {Lucas}, P. W. and {Maiolo}, B. and {Marengo}, M. and {Marshall}, D. and {Masi}, S. and {Massi}, F. and {Matsuura}, M. and {Meny}, C. and {Minier}, V. and {Miville-Desch{\^e}nes}, M. -A. and {Montier}, L. and {Motte}, F. and {M{\"u}ller}, T. G. and {Natoli}, P. and {Neves}, J. and {Olmi}, L. and {Paladini}, R. and {Paradis}, D. and {Pestalozzi}, M. and {Pezzuto}, S. and {Piacentini}, F. and {Pomar{\`e}s}, M. and {Popescu}, C. C. and {Reach}, W. T. and {Richer}, J. and {Ristorcelli}, I. and {Roy}, A. and {Royer}, P. and {Russeil}, D. and {Saraceno}, P. and {Sauvage}, M. and {Schilke}, P. and {Schneider-Bontemps}, N. and {Schuller}, F. and {Schultz}, B. and {Shepherd}, D. S. and {Sibthorpe}, B. and {Smith}, H. A. and {Smith}, M. D. and {Spinoglio}, L. and {Stamatellos}, D. and {Strafella}, F. and {Stringfellow}, G. and {Sturm}, E. and {Taylor}, R. and {Thompson}, M. A. and {Tuffs}, R. J. and {Umana}, G. and {Valenziano}, L. and {Vavrek}, R. and {Viti}, S. and {Waelkens}, C. and {Ward-Thompson}, D. and {White}, G. and {Wyrowski}, F. and {Yorke}, H. W. and {Zhang}, Q.},
    title = "{Hi-GAL: The Herschel Infrared Galactic Plane Survey}",
    journal = "\pasp",
    year = "2010",
    month = "March",
    volume = "122",
    number = "889",
    pages = "314",
    doi = "10.1086/651314",
    archivePrefix = "arXiv",
    eprint = "1001.2106",
    primaryClass = "astro-ph.GA",
    adsurl = "https://ui.adsabs.harvard.edu/abs/2010PASP..122..314M"
}

@ARTICLE{Lin2025NatAs...9..406L,
       author = {{Lin}, Lingrui and {Zhang}, Zhi-Yu and {Wang}, Junzhi and {Papadopoulos}, Padelis P. and {Shi}, Yong and {Gong}, Yan and {Sun}, Yan and {Sun}, Yichen and {Bisbas}, Thomas G. and {Romano}, Donatella and {Li}, Di and {Liu}, Hauyu Baobab and {Qiu}, Keping and {Liu}, Lijie and {Luo}, Gan and {Tsai}, Chao-Wei and {Wu}, Jingwen and {Feng}, Siyi and {Zhang}, Bo},
        title = "{Inadequate turbulent support in low-metallicity molecular clouds}",
      journal = {Nature Astronomy},
         year = 2025,
        month = mar,
       volume = {9},
        pages = {406-416},
          doi = {10.1038/s41550-024-02440-3},
archivePrefix = {arXiv},
       eprint = {2501.07636},
 primaryClass = {astro-ph.GA},
       adsurl = {https://ui.adsabs.harvard.edu/abs/2025NatAs...9..406L}
}

@ARTICLE{Lacy2017ApJ...838...66L_COabundance,
       author = {{Lacy}, John H. and {Sneden}, Christopher and {Kim}, Hwihyun and {Jaffe}, Daniel T.},
        title = "{H$_{2}$, CO, and Dust Absorption through Cold Molecular Clouds}",
      journal = {\apj},
         year = 2017,
        month = mar,
       volume = {838},
       number = {1},
          eid = {66},
        pages = {66},
          doi = {10.3847/1538-4357/aa6247},
archivePrefix = {arXiv},
       eprint = {1703.09826},
 primaryClass = {astro-ph.GA},
       adsurl = {https://ui.adsabs.harvard.edu/abs/2017ApJ...838...66L}
}

@ARTICLE{GRAVITY2021A&A...647A..59G,
       author = {{GRAVITY Collaboration} and {Abuter}, R. and {Amorim}, A. and {Baub{\"o}ck}, M. and {Berger}, J.~P. and {Bonnet}, H. and {Brandner}, W. and {Cl{\'e}net}, Y. and {Davies}, R. and {de Zeeuw}, P.~T. and {Dexter}, J. and {Dallilar}, Y. and {Drescher}, A. and {Eckart}, A. and {Eisenhauer}, F. and {F{\"o}rster Schreiber}, N.~M. and {Garcia}, P. and {Gao}, F. and {Gendron}, E. and {Genzel}, R. and {Gillessen}, S. and {Habibi}, M. and {Haubois}, X. and {Hei{\ss}el}, G. and {Henning}, T. and {Hippler}, S. and {Horrobin}, M. and {Jim{\'e}nez-Rosales}, A. and {Jochum}, L. and {Jocou}, L. and {Kaufer}, A. and {Kervella}, P. and {Lacour}, S. and {Lapeyr{\`e}re}, V. and {Le Bouquin}, J. -B. and {L{\'e}na}, P. and {Lutz}, D. and {Nowak}, M. and {Ott}, T. and {Paumard}, T. and {Perraut}, K. and {Perrin}, G. and {Pfuhl}, O. and {Rabien}, S. and {Rodr{\'\i}guez-Coira}, G. and {Shangguan}, J. and {Shimizu}, T. and {Scheithauer}, S. and {Stadler}, J. and {Straub}, O. and {Straubmeier}, C. and {Sturm}, E. and {Tacconi}, L.~J. and {Vincent}, F. and {von Fellenberg}, S. and {Waisberg}, I. and {Widmann}, F. and {Wieprecht}, E. and {Wiezorrek}, E. and {Woillez}, J. and {Yazici}, S. and {Young}, A. and {Zins}, G.},
        title = "{Improved GRAVITY astrometric accuracy from modeling optical aberrations}",
      journal = {\aap},
         year = 2021,
        month = mar,
       volume = {647},
          eid = {A59},
        pages = {A59},
          doi = {10.1051/0004-6361/202040208},
archivePrefix = {arXiv},
       eprint = {2101.12098},
 primaryClass = {astro-ph.GA},
       adsurl = {https://ui.adsabs.harvard.edu/abs/2021A&A...647A..59G}
}

@ARTICLE{Foreman-Mackey2013PASP_EMCEE,
   author = {{Foreman-Mackey}, D. and {Hogg}, D.~W. and {Lang}, D. and {Goodman}, J.
	},
    title = "{emcee: The MCMC Hammer}",
  journal = {\pasp},
archivePrefix = "arXiv",
   eprint = {1202.3665},
 primaryClass = "astro-ph.IM",
     year = 2013,
    month = mar,
   volume = 125,
    pages = {306},
      doi = {10.1086/670067},
   adsurl = {http://adsabs.harvard.edu/abs/2013PASP..125..306F}
}

@ARTICLE{Yogesh_2012arXiv1208.3524V_prebin_NPDF,
       author = {{Virkar}, Yogesh and {Clauset}, Aaron},
        title = "{Power-law distributions in binned empirical data}",
      journal = {arXiv e-prints},
         year = 2012,
        month = aug,
          eid = {arXiv:1208.3524},
        pages = {arXiv:1208.3524},
          doi = {10.48550/arXiv.1208.3524},
archivePrefix = {arXiv},
       eprint = {1208.3524},
 primaryClass = {physics.data-an},
       adsurl = {https://ui.adsabs.harvard.edu/abs/2012arXiv1208.3524V}
}

@ARTICLE{Grishin2023A&A...677A.101G_IoU,
       author = {{Grishin}, Kirill and {Mei}, Simona and {Ili{\'c}}, St{\'e}phane},
        title = "{YOLO-CL: Galaxy cluster detection in the SDSS with deep machine learning}",
      journal = {\aap},
         year = 2023,
        month = sep,
       volume = {677},
          eid = {A101},
        pages = {A101},
          doi = {10.1051/0004-6361/202345976},
archivePrefix = {arXiv},
       eprint = {2301.09657},
 primaryClass = {astro-ph.CO},
       adsurl = {https://ui.adsabs.harvard.edu/abs/2023A&A...677A.101G}
}

@ARTICLE{Burke2019MNRAS.490.3952B_IoU,
       author = {{Burke}, Colin J. and {Aleo}, Patrick D. and {Chen}, Yu-Ching and {Liu}, Xin and {Peterson}, John R. and {Sembroski}, Glenn H. and {Lin}, Joshua Yao-Yu},
        title = "{Deblending and classifying astronomical sources with Mask R-CNN deep learning}",
      journal = {\mnras},
         year = 2019,
        month = dec,
       volume = {490},
       number = {3},
        pages = {3952-3965},
          doi = {10.1093/mnras/stz2845},
archivePrefix = {arXiv},
       eprint = {1908.02748},
 primaryClass = {astro-ph.IM},
       adsurl = {https://ui.adsabs.harvard.edu/abs/2019MNRAS.490.3952B}
}

@ARTICLE{Hildebrand1983QJRAS..24..267H,
    author = "{Hildebrand}, R. H.",
    title = "{The determination of cloud masses and dust characteristics from submillimetre thermal emission.}",
    journal = "\qjras",
    year = "1983",
    month = "September",
    volume = "24",
    pages = "267-282",
    adsurl = "https://ui.adsabs.harvard.edu/abs/1983QJRAS..24..267H"
}

@BOOK{Draine2011piim.book.....D,
    author = "{Draine}, Bruce T.",
    title = "{Physics of the Interstellar and Intergalactic Medium}",
    year = "2011",
    adsurl = "https://ui.adsabs.harvard.edu/abs/2011piim.book.....D"
}

@ARTICLE{Virtanen2020NatMe..17..261V_SciPy,
       author = {{Virtanen}, Pauli and {Gommers}, Ralf and {Oliphant}, Travis E. and {Haberland}, Matt and {Reddy}, Tyler and {Cournapeau}, David and {Burovski}, Evgeni and {Peterson}, Pearu and {Weckesser}, Warren and {Bright}, Jonathan and {van der Walt}, St{\'e}fan J. and {Brett}, Matthew and {Wilson}, Joshua and {Millman}, K. Jarrod and {Mayorov}, Nikolay and {Nelson}, Andrew R.~J. and {Jones}, Eric and {Kern}, Robert and {Larson}, Eric and {Carey}, C.~J. and {Polat}, {\.I}lhan and {Feng}, Yu and {Moore}, Eric W. and {VanderPlas}, Jake and {Laxalde}, Denis and {Perktold}, Josef and {Cimrman}, Robert and {Henriksen}, Ian and {Quintero}, E.~A. and {Harris}, Charles R. and {Archibald}, Anne M. and {Ribeiro}, Ant{\^o}nio H. and {Pedregosa}, Fabian and {van Mulbregt}, Paul and {SciPy 1. 0 Contributors}},
        title = "{SciPy 1.0: fundamental algorithms for scientific computing in Python}",
      journal = {Nature Methods},
         year = 2020,
        month = feb,
       volume = {17},
        pages = {261-272},
          doi = {10.1038/s41592-019-0686-2},
archivePrefix = {arXiv},
       eprint = {1907.10121},
 primaryClass = {cs.MS},
       adsurl = {https://ui.adsabs.harvard.edu/abs/2020NatMe..17..261V}
}

@ARTICLE{Shu1991ApJ,
       author = {{Shu}, Frank H. and {Ruden}, Steven P. and {Lada}, Charles J. and {Lizano}, Susana},
        title = "{Star Formation and the Nature of Bipolar Outflows}",
      journal = {\apjl},
         year = 1991,
        month = mar,
       volume = {370},
        pages = {L31},
          doi = {10.1086/185970},
       adsurl = {https://ui.adsabs.harvard.edu/abs/1991ApJ...370L..31S}
}

@ARTICLE{McKee2007ARA&A,
       author = {{McKee}, Christopher F. and {Ostriker}, Eve C.},
        title = "{Theory of Star Formation}",
      journal = {\araa},
         year = 2007,
        month = sep,
       volume = {45},
       number = {1},
        pages = {565-687},
          doi = {10.1146/annurev.astro.45.051806.110602},
archivePrefix = {arXiv},
       eprint = {0707.3514},
 primaryClass = {astro-ph},
       adsurl = {https://ui.adsabs.harvard.edu/abs/2007ARA&A..45..565M}
}

@ARTICLE{Schneider2015MNRAS,
       author = {{Schneider}, N. and {Bontemps}, S. and {Girichidis}, P. and {Rayner}, T. and {Motte}, F. and {Andr{\'e}}, P. and {Russeil}, D. and {Abergel}, A. and {Anderson}, L. and {Arzoumanian}, D. and {Benedettini}, M. and {Csengeri}, T. and {Didelon}, P. and {di}, Francesco J. and {Griffin}, M. and {Hill}, T. and {Klessen}, R.~S. and {Ossenkopf}, V. and {Pezzuto}, S. and {Rivera-Ingraham}, A. and {Spinoglio}, L. and {Tremblin}, P. and {Zavagno}, A.},
        title = "{Detection of two power-law tails in the probability distribution functions of massive GMCs.}",
      journal = {\mnras},
         year = 2015,
        month = nov,
       volume = {453},
        pages = {L41-L45},
          doi = {10.1093/mnrasl/slv101},
archivePrefix = {arXiv},
       eprint = {1507.08869},
 primaryClass = {astro-ph.GA},
       adsurl = {https://ui.adsabs.harvard.edu/abs/2015MNRAS.453L..41S}
}

@ARTICLE{Jiao2025A&A,
       author = {{Jiao}, Sihan and {Wu}, Jingwen and {Zhang}, Zhi-Yu and {Evans}, II, Neal J. and {Tsai}, Chao-Wei and {Li}, Di and {Liu}, Hauyu Baobab and {Shi}, Yong and {Wang}, Junzhi and {Zhang}, Qizhou and {Lin}, Yuxin and {Feng}, Linjing and {Lu}, Xing and {Sun}, Yan and {Ruan}, Hao and {Deng}, Fangyuan},
        title = "{Gravitationally bound gas determines star formation in the Galaxy}",
      journal = {\aap},
         year = 2025,
        month = sep,
       volume = {701},
          eid = {A152},
        pages = {A152},
          doi = {10.1051/0004-6361/202453608},
archivePrefix = {arXiv},
       eprint = {2505.07763},
 primaryClass = {astro-ph.GA},
       adsurl = {https://ui.adsabs.harvard.edu/abs/2025A&A...701A.152J}
}

@ARTICLE{Murase_2023MNRAS.523.1373M_CygnusX_COnpdf,
       author = {{Murase}, Takeru and {Handa}, Toshihiro and {Matsusaka}, Ren and {Shimajiri}, Yoshito and {Kobayashi}, Masato I.~N. and {Kohno}, Mikito and {Nishi}, Junya and {Takeba}, Norimi and {Shibata}, Yosuke},
        title = "{Multilognormal density structure in Cygnus-X molecular clouds: a fitting for N-PDF without power law}",
      journal = {\mnras},
         year = 2023,
        month = jul,
       volume = {523},
       number = {1},
        pages = {1373-1387},
          doi = {10.1093/mnras/stad1451},
archivePrefix = {arXiv},
       eprint = {2305.07094},
 primaryClass = {astro-ph.GA},
       adsurl = {https://ui.adsabs.harvard.edu/abs/2023MNRAS.523.1373M}
}

@ARTICLE{Roy2013ApJ...763...55R_HGBS_OrionA,
       author = {{Roy}, Arabindo and {Martin}, Peter G. and {Polychroni}, Danae and {Bontemps}, Sylvain and {Abergel}, Alain and {Andr{\'e}}, Philippe and {Arzoumanian}, Doris and {Di Francesco}, James and {Hill}, Tracey and {Konyves}, Vera and {Nguyen-Luong}, Quang and {Pezzuto}, Stefano and {Schneider}, Nicola and {Testi}, Leonardo and {White}, Glenn},
        title = "{Changes of Dust Opacity with Density in the Orion A Molecular Cloud}",
      journal = {\apj},
         year = 2013,
        month = jan,
       volume = {763},
       number = {1},
          eid = {55},
        pages = {55},
          doi = {10.1088/0004-637X/763/1/55},
archivePrefix = {arXiv},
       eprint = {1211.6475},
 primaryClass = {astro-ph.GA},
       adsurl = {https://ui.adsabs.harvard.edu/abs/2013ApJ...763...55R}
}

@ARTICLE{Lada2010ApJ...724..687L,
       author = {{Lada}, Charles J. and {Lombardi}, Marco and {Alves}, Jo{\~a}o F.},
        title = "{On the Star Formation Rates in Molecular Clouds}",
      journal = {\apj},
         year = 2010,
        month = nov,
       volume = {724},
       number = {1},
        pages = {687-693},
          doi = {10.1088/0004-637X/724/1/687},
archivePrefix = {arXiv},
       eprint = {1009.2985},
 primaryClass = {astro-ph.GA},
       adsurl = {https://ui.adsabs.harvard.edu/abs/2010ApJ...724..687L}
}

@ARTICLE{Konyves2020A&A...635A..34K_HGBS_OrionB,
       author = {{K{\"o}nyves}, V. and {Andr{\'e}}, Ph. and {Arzoumanian}, D. and {Schneider}, N. and {Men'shchikov}, A. and {Bontemps}, S. and {Ladjelate}, B. and {Didelon}, P. and {Pezzuto}, S. and {Benedettini}, M. and {Bracco}, A. and {Di Francesco}, J. and {Goodwin}, S. and {Rygl}, K.~L.~J. and {Shimajiri}, Y. and {Spinoglio}, L. and {Ward-Thompson}, D. and {White}, G.~J.},
        title = "{Properties of the dense core population in Orion B as seen by the Herschel Gould Belt survey}",
      journal = {\aap},
         year = 2020,
        month = mar,
       volume = {635},
          eid = {A34},
        pages = {A34},
          doi = {10.1051/0004-6361/201834753},
archivePrefix = {arXiv},
       eprint = {1910.04053},
 primaryClass = {astro-ph.SR},
       adsurl = {https://ui.adsabs.harvard.edu/abs/2020A&A...635A..34K}
}

@ARTICLE{Konyves2015A&A...584A..91K_HGBS_Aquila,
       author = {{K{\"o}nyves}, V. and {Andr{\'e}}, Ph. and {Men'shchikov}, A. and {Palmeirim}, P. and {Arzoumanian}, D. and {Schneider}, N. and {Roy}, A. and {Didelon}, P. and {Maury}, A. and {Shimajiri}, Y. and {Di Francesco}, J. and {Bontemps}, S. and {Peretto}, N. and {Benedettini}, M. and {Bernard}, J. -Ph. and {Elia}, D. and {Griffin}, M.~J. and {Hill}, T. and {Kirk}, J. and {Ladjelate}, B. and {Marsh}, K. and {Martin}, P.~G. and {Motte}, F. and {Nguy{\^e}n Luong}, Q. and {Pezzuto}, S. and {Roussel}, H. and {Rygl}, K.~L.~J. and {Sadavoy}, S.~I. and {Schisano}, E. and {Spinoglio}, L. and {Ward-Thompson}, D. and {White}, G.~J.},
        title = "{A census of dense cores in the Aquila cloud complex: SPIRE/PACS observations from the Herschel Gould Belt survey}",
      journal = {\aap},
         year = 2015,
        month = dec,
       volume = {584},
          eid = {A91},
        pages = {A91},
          doi = {10.1051/0004-6361/201525861},
archivePrefix = {arXiv},
       eprint = {1507.05926},
 primaryClass = {astro-ph.GA},
       adsurl = {https://ui.adsabs.harvard.edu/abs/2015A&A...584A..91K}
}

@ARTICLE{Cormier2018MNRAS.475.3909C_nearbyGalaxies_1213CO,
       author = {{Cormier}, D. and {Bigiel}, F. and {Jim{\'e}nez-Donaire}, M.~J. and {Leroy}, A.~K. and {Gallagher}, M. and {Usero}, A. and {Sandstrom}, K. and {Bolatto}, A. and {Hughes}, A. and {Kramer}, C. and {Krumholz}, M.~R. and {Meier}, D.~S. and {Murphy}, E.~J. and {Pety}, J. and {Rosolowsky}, E. and {Schinnerer}, E. and {Schruba}, A. and {Sliwa}, K. and {Walter}, F.},
        title = "{Full-disc $^{13}$CO(1-0) mapping across nearby galaxies of the EMPIRE survey and the CO-to-H$_{2}$ conversion factor}",
      journal = {\mnras},
         year = 2018,
        month = apr,
       volume = {475},
       number = {3},
        pages = {3909-3933},
          doi = {10.1093/mnras/sty059},
archivePrefix = {arXiv},
       eprint = {1801.03105},
 primaryClass = {astro-ph.GA},
       adsurl = {https://ui.adsabs.harvard.edu/abs/2018MNRAS.475.3909C}
}

@ARTICLE{Torii2019PASJ...71S...2T,
       author = {{Torii}, Kazufumi and {Fujita}, Shinji and {Nishimura}, Atsushi and {Tokuda}, Kazuki and {Kohno}, Mikito and {Tachihara}, Kengo and {Inutsuka}, Shu-ichiro and {Matsuo}, Mitsuhiro and {Kuriki}, Mika and {Tsuda}, Yuya and {Minamidani}, Tetsuhiro and {Umemoto}, Tomofumi and {Kuno}, Nario and {Miyamoto}, Yusuke},
        title = "{FOREST Unbiased Galactic Plane Imaging Survey with the Nobeyama 45 m telescope (FUGIN). V. Dense gas mass fraction of molecular gas in the Galactic plane}",
      journal = {\pasj},
         year = 2019,
        month = dec,
       volume = {71},
          eid = {S2},
        pages = {S2},
          doi = {10.1093/pasj/psz033},
archivePrefix = {arXiv},
       eprint = {1809.06642},
 primaryClass = {astro-ph.GA},
       adsurl = {https://ui.adsabs.harvard.edu/abs/2019PASJ...71S...2T}
}

@ARTICLE{Beuther2025arXiv,
       author = {{Beuther}, H. and {Kuiper}, R. and {Tafalla}, M.},
        title = "{Star formation from low to high mass: A comparative view}",
      journal = {arXiv e-prints},
         year = 2025,
        month = jan,
          eid = {arXiv:2501.16866},
        pages = {arXiv:2501.16866},
          doi = {10.48550/arXiv.2501.16866},
archivePrefix = {arXiv},
       eprint = {2501.16866},
 primaryClass = {astro-ph.GA},
       adsurl = {https://ui.adsabs.harvard.edu/abs/2025arXiv250116866B}
}

@ARTICLE{Ballesteros-Paredes2011MNRAS,
       author = {{Ballesteros-Paredes}, Javier and {V{\'a}zquez-Semadeni}, Enrique and {Gazol}, Adriana and {Hartmann}, Lee W. and {Heitsch}, Fabian and {Col{\'\i}n}, Pedro},
        title = "{Gravity or turbulence? - II. Evolving column density probability distribution functions in molecular clouds}",
      journal = {\mnras},
         year = 2011,
        month = sep,
       volume = {416},
       number = {2},
        pages = {1436-1442},
          doi = {10.1111/j.1365-2966.2011.19141.x},
archivePrefix = {arXiv},
       eprint = {1105.5411},
 primaryClass = {astro-ph.GA},
       adsurl = {https://ui.adsabs.harvard.edu/abs/2011MNRAS.416.1436B}
}

@ARTICLE{Kainulainen2009A&A,
       author = {{Kainulainen}, J. and {Beuther}, H. and {Henning}, T. and {Plume}, R.},
        title = "{Probing the evolution of molecular cloud structure. From quiescence to birth}",
      journal = {\aap},
         year = 2009,
        month = dec,
       volume = {508},
       number = {3},
        pages = {L35-L38},
          doi = {10.1051/0004-6361/200913605},
archivePrefix = {arXiv},
       eprint = {0911.5648},
 primaryClass = {astro-ph.GA},
       adsurl = {https://ui.adsabs.harvard.edu/abs/2009A&A...508L..35K}
}

@ARTICLE{Lombardi2015A&A,
       author = {{Lombardi}, Marco and {Alves}, Jo{\~a}o and {Lada}, Charles J.},
        title = "{Molecular clouds have power-law probability distribution functions}",
      journal = {\aap},
         year = 2015,
        month = apr,
       volume = {576},
          eid = {L1},
        pages = {L1},
          doi = {10.1051/0004-6361/201525650},
archivePrefix = {arXiv},
       eprint = {1502.03859},
 primaryClass = {astro-ph.SR},
       adsurl = {https://ui.adsabs.harvard.edu/abs/2015A&A...576L...1L}
}

@ARTICLE{Federrath2013ApJ,
       author = {{Federrath}, Christoph and {Klessen}, Ralf S.},
        title = "{On the Star Formation Efficiency of Turbulent Magnetized Clouds}",
      journal = {\apj},
         year = 2013,
        month = jan,
       volume = {763},
       number = {1},
          eid = {51},
        pages = {51},
          doi = {10.1088/0004-637X/763/1/51},
archivePrefix = {arXiv},
       eprint = {1211.6433},
 primaryClass = {astro-ph.SR},
       adsurl = {https://ui.adsabs.harvard.edu/abs/2013ApJ...763...51F}
}

@ARTICLE{Su2019ApJS,
       author = {{Su}, Yang and {Yang}, Ji and {Zhang}, Shaobo and {Gong}, Yan and {Wang}, Hongchi and {Zhou}, Xin and {Wang}, Min and {Chen}, Zhiwei and {Sun}, Yan and {Chen}, Xuepeng and {Xu}, Ye and {Jiang}, Zhibo},
        title = "{The Milky Way Imaging Scroll Painting (MWISP): Project Details and Initial Results from the Galactic Longitudes of 25.{\textdegree}8-49.{\textdegree}7}",
      journal = {\apjs},
         year = 2019,
        month = jan,
       volume = {240},
       number = {1},
          eid = {9},
        pages = {9},
          doi = {10.3847/1538-4365/aaf1c8},
archivePrefix = {arXiv},
       eprint = {1901.00285},
 primaryClass = {astro-ph.GA},
       adsurl = {https://ui.adsabs.harvard.edu/abs/2019ApJS..240....9S}
}

@ARTICLE{Umemoto2017PASJ,
       author = {{Umemoto}, Tomofumi and {Minamidani}, Tetsuhiro and {Kuno}, Nario and {Fujita}, Shinji and {Matsuo}, Mitsuhiro and {Nishimura}, Atsushi and {Torii}, Kazufumi and {Tosaki}, Tomoka and {Kohno}, Mikito and {Kuriki}, Mika and {Tsuda}, Yuya and {Hirota}, Akihiko and {Ohashi}, Satoshi and {Yamagishi}, Mitsuyoshi and {Handa}, Toshihiro and {Nakanishi}, Hiroyuki and {Omodaka}, Toshihiro and {Koide}, Nagito and {Matsumoto}, Naoko and {Onishi}, Toshikazu and {Tokuda}, Kazuki and {Seta}, Masumichi and {Kobayashi}, Yukinori and {Tachihara}, Kengo and {Sano}, Hidetoshi and {Hattori}, Yusuke and {Onodera}, Sachiko and {Oasa}, Yumiko and {Kamegai}, Kazuhisa and {Tsuboi}, Masato and {Sofue}, Yoshiaki and {Higuchi}, Aya E. and {Chibueze}, James O. and {Mizuno}, Norikazu and {Honma}, Mareki and {Muller}, Erik and {Inoue}, Tsuyoshi and {Morokuma-Matsui}, Kana and {Shinnaga}, Hiroko and {Ozawa}, Takeaki and {Takahashi}, Ryo and {Yoshiike}, Satoshi and {Costes}, Jean and {Kuwahara}, Sho},
        title = "{FOREST unbiased Galactic plane imaging survey with the Nobeyama 45 m telescope (FUGIN). I. Project overview and initial results}",
      journal = {\pasj},
         year = 2017,
        month = oct,
       volume = {69},
       number = {5},
          eid = {78},
        pages = {78},
          doi = {10.1093/pasj/psx061},
archivePrefix = {arXiv},
       eprint = {1707.05981},
 primaryClass = {astro-ph.GA},
       adsurl = {https://ui.adsabs.harvard.edu/abs/2017PASJ...69...78U}
}

@ARTICLE{Schuller2017A&A,
       author = {{Schuller}, F. and {Csengeri}, T. and {Urquhart}, J.~S. and {Duarte-Cabral}, A. and {Barnes}, P.~J. and {Giannetti}, A. and {Hernandez}, A.~K. and {Leurini}, S. and {Mattern}, M. and {Medina}, S. -N.~X. and {Agurto}, C. and {Azagra}, F. and {Anderson}, L.~D. and {Beltr{\'a}n}, M.~T. and {Beuther}, H. and {Bontemps}, S. and {Bronfman}, L. and {Dobbs}, C.~L. and {Dumke}, M. and {Finger}, R. and {Ginsburg}, A. and {Gonzalez}, E. and {Henning}, T. and {Kauffmann}, J. and {Mac-Auliffe}, F. and {Menten}, K.~M. and {Montenegro-Montes}, F.~M. and {Moore}, T.~J.~T. and {Muller}, E. and {Parra}, R. and {Perez-Beaupuits}, J. -P. and {Pettitt}, A. and {Russeil}, D. and {S{\'a}nchez-Monge}, {\'A}. and {Schilke}, P. and {Schisano}, E. and {Suri}, S. and {Testi}, L. and {Torstensson}, K. and {Venegas}, P. and {Wang}, K. and {Wienen}, M. and {Wyrowski}, F. and {Zavagno}, A.},
        title = "{SEDIGISM: Structure, excitation, and dynamics of the inner Galactic interstellar medium}",
      journal = {\aap},
         year = 2017,
        month = may,
       volume = {601},
          eid = {A124},
        pages = {A124},
          doi = {10.1051/0004-6361/201628933},
archivePrefix = {arXiv},
       eprint = {1701.04712},
 primaryClass = {astro-ph.GA},
       adsurl = {https://ui.adsabs.harvard.edu/abs/2017A&A...601A.124S}
}

@ARTICLE{Ridge_2006AJ....131.2921R_FCRAO_CO,
       author = {{Ridge}, Naomi A. and {Di Francesco}, James and {Kirk}, Helen and {Li}, Di and {Goodman}, Alyssa A. and {Alves}, Jo{\~a}o F. and {Arce}, H{\'e}ctor G. and {Borkin}, Michelle A. and {Caselli}, Paola and {Foster}, Jonathan B. and {Heyer}, Mark H. and {Johnstone}, Doug and {Kosslyn}, David A. and {Lombardi}, Marco and {Pineda}, Jaime E. and {Schnee}, Scott L. and {Tafalla}, Mario},
        title = "{The COMPLETE Survey of Star-Forming Regions: Phase I Data}",
      journal = {\aj},
         year = 2006,
        month = jun,
       volume = {131},
       number = {6},
        pages = {2921-2933},
          doi = {10.1086/503704},
archivePrefix = {arXiv},
       eprint = {astro-ph/0602542},
 primaryClass = {astro-ph},
       adsurl = {https://ui.adsabs.harvard.edu/abs/2006AJ....131.2921R}
}

@ARTICLE{Ritchey2011ApJ...728...36R_1213_nearby,
       author = {{Ritchey}, A.~M. and {Federman}, S.~R. and {Lambert}, D.~L.},
        title = "{Interstellar CN and CH$^{+}$ in Diffuse Molecular Clouds: $^{12}$C/$^{13}$C Ratios and CN Excitation}",
      journal = {\apj},
         year = 2011,
        month = feb,
       volume = {728},
       number = {1},
          eid = {36},
        pages = {36},
          doi = {10.1088/0004-637X/728/1/36},
archivePrefix = {arXiv},
       eprint = {1012.1296},
 primaryClass = {astro-ph.GA},
       adsurl = {https://ui.adsabs.harvard.edu/abs/2011ApJ...728...36R}
}

@ARTICLE{Wang2019ApJS..243...25W,
       author = {{Wang}, Chen and {Yang}, Ji and {Su}, Yang and {Du}, Fujun and {Ma}, Yuehui and {Zhang}, Shaobo},
        title = "{Molecular Gas toward the Gemini OB1 Molecular Cloud Complex. III. Chemical Abundance}",
      journal = {\apjs},
         year = 2019,
        month = aug,
       volume = {243},
       number = {2},
          eid = {25},
        pages = {25},
          doi = {10.3847/1538-4365/ab2d2e},
       adsurl = {https://ui.adsabs.harvard.edu/abs/2019ApJS..243...25W}
}

@ARTICLE{Wang2023AJ....166..121W,
       author = {{Wang}, Chen and {Feng}, Haoran and {Yang}, Ji and {Chen}, Xuepeng and {Su}, Yang and {Yan}, Qing-Zeng and {Du}, Fujun and {Ma}, Yuehui and {Cai}, Jiajun},
        title = "{The Molecular Clouds in a Section of the Third Galactic Quadrant: Observational Properties and Chemical Abundance Ratio between CO and its Isotopologues}",
      journal = {\aj},
         year = 2023,
        month = sep,
       volume = {166},
       number = {3},
          eid = {121},
        pages = {121},
          doi = {10.3847/1538-3881/acebdd},
archivePrefix = {arXiv},
       eprint = {2308.10726},
 primaryClass = {astro-ph.GA},
       adsurl = {https://ui.adsabs.harvard.edu/abs/2023AJ....166..121W}
}

@ARTICLE{Liszt2007A&A...476..291L_CO_PDR,
       author = {{Liszt}, H.~S.},
        title = "{Formation, fractionation, and excitation of carbon monoxide in diffuse clouds}",
      journal = {\aap},
         year = 2007,
        month = dec,
       volume = {476},
       number = {1},
        pages = {291-300},
          doi = {10.1051/0004-6361:20078502},
archivePrefix = {arXiv},
       eprint = {0710.2237},
 primaryClass = {astro-ph},
       adsurl = {https://ui.adsabs.harvard.edu/abs/2007A&A...476..291L}
}

@ARTICLE{vanDishoeck1988ApJ...334..771V_COPDR,
       author = {{van Dishoeck}, Ewine F. and {Black}, John H.},
        title = "{The Photodissociation and Chemistry of Interstellar CO}",
      journal = {\apj},
         year = 1988,
        month = nov,
       volume = {334},
        pages = {771},
          doi = {10.1086/166877},
       adsurl = {https://ui.adsabs.harvard.edu/abs/1988ApJ...334..771V}
}

@ARTICLE{Goldsmith2008ApJ...680..428G,
       author = {{Goldsmith}, Paul F. and {Heyer}, Mark and {Narayanan}, Gopal and {Snell}, Ronald and {Li}, Di and {Brunt}, Chris},
        title = "{Large-Scale Structure of the Molecular Gas in Taurus Revealed by High Linear Dynamic Range Spectral Line Mapping}",
      journal = {\apj},
         year = 2008,
        month = jun,
       volume = {680},
       number = {1},
        pages = {428-445},
          doi = {10.1086/587166},
archivePrefix = {arXiv},
       eprint = {0802.2206},
 primaryClass = {astro-ph},
       adsurl = {https://ui.adsabs.harvard.edu/abs/2008ApJ...680..428G}
}

@ARTICLE{Schneider2016A&A,
       author = {{Schneider}, N. and {Bontemps}, S. and {Motte}, F. and {Ossenkopf}, V. and {Klessen}, R.~S. and {Simon}, R. and {Fechtenbaum}, S. and {Herpin}, F. and {Tremblin}, P. and {Csengeri}, T. and {Myers}, P.~C. and {Hill}, T. and {Cunningham}, M. and {Federrath}, C.},
        title = "{Understanding star formation in molecular clouds. III. Probability distribution functions of molecular lines in Cygnus X}",
      journal = {\aap},
         year = 2016,
        month = mar,
       volume = {587},
          eid = {A74},
        pages = {A74},
          doi = {10.1051/0004-6361/201527144},
archivePrefix = {arXiv},
       eprint = {1509.01082},
 primaryClass = {astro-ph.GA},
       adsurl = {https://ui.adsabs.harvard.edu/abs/2016A&A...587A..74S}
}

@ARTICLE{Goodman2009ApJ,
       author = {{Goodman}, Alyssa A. and {Pineda}, Jaime E. and {Schnee}, Scott L.},
        title = "{The ``True'' Column Density Distribution in Star-Forming Molecular Clouds}",
      journal = {\apj},
         year = 2009,
        month = feb,
       volume = {692},
       number = {1},
        pages = {91-103},
          doi = {10.1088/0004-637X/692/1/91},
archivePrefix = {arXiv},
       eprint = {0806.3441},
 primaryClass = {astro-ph},
       adsurl = {https://ui.adsabs.harvard.edu/abs/2009ApJ...692...91G}
}

@ARTICLE{Pineda2008ApJ,
       author = {{Pineda}, Jaime E. and {Caselli}, Paola and {Goodman}, Alyssa A.},
        title = "{CO Isotopologues in the Perseus Molecular Cloud Complex: the X-factor and Regional Variations}",
      journal = {\apj},
         year = 2008,
        month = may,
       volume = {679},
       number = {1},
        pages = {481-496},
          doi = {10.1086/586883},
archivePrefix = {arXiv},
       eprint = {0802.0708},
 primaryClass = {astro-ph},
       adsurl = {https://ui.adsabs.harvard.edu/abs/2008ApJ...679..481P}
}

@Article{Yanrs11030286,
    AUTHOR = {Yan, Jiangqiao and Wang, Hongqi and Yan, Menglong and Diao, Wenhui and Sun, Xian and Li, Hao},
    TITLE = {IoU-Adaptive Deformable R-CNN: Make Full Use of IoU for Multi-Class Object Detection in Remote Sensing Imagery},
    JOURNAL = {Remote Sensing},
    VOLUME = {11},
    YEAR = {2019},
    NUMBER = {3},
    ARTICLE-NUMBER = {286},
    URL = {https://www.mdpi.com/2072-4292/11/3/286},
    ISSN = {2072-4292},
    DOI = {10.3390/rs11030286}
}

@ARTICLE{Pontoppidan2024RNAAS...8...68P,
       author = {{Pontoppidan}, Klaus M. and {Evans}, Neal and {Bergner}, Jennifer and {Yang}, Yao-Lun},
        title = "{A Constrained Dust Opacity for Models of Dense Clouds and Protostellar Envelopes}",
      journal = {Research Notes of the American Astronomical Society},
         year = 2024,
        month = mar,
       volume = {8},
       number = {3},
          eid = {68},
        pages = {68},
          doi = {10.3847/2515-5172/ad303f},
       adsurl = {https://ui.adsabs.harvard.edu/abs/2024RNAAS...8...68P}
}

@ARTICLE{Feng_2026ApJ...998..224F,
       author = {{Feng}, Linjing and {Jiao}, Sihan and {Xu}, Fengwei and {Liu}, Hauyu Baobab and {Lu}, Xing and {Evans}, II, Neal J. and {Mills}, Elisabeth A.~C. and {Kov{\'a}cs}, Attila and {Zhang}, Qizhou and {Lin}, Yuxin and {Wu}, Jingwen and {Tsai}, Chao-Wei and {Li}, Di and {Zhang}, Zhi-Yu and {Yan}, Zhiqiang and {Ruan}, Hao and {Deng}, Fangyuan and {Xiong}, Yuanzhen and {Zhang}, Ruofei},
        title = "{Tails of Gravity: Persistence of Star Formation in the Central Molecular Zone}",
      journal = {\apj},
         year = 2026,
        month = feb,
       volume = {998},
       number = {2},
          eid = {224},
        pages = {224},
          doi = {10.3847/1538-4357/ae29e6},
archivePrefix = {arXiv},
       eprint = {2511.20300},
 primaryClass = {astro-ph.GA},
       adsurl = {https://ui.adsabs.harvard.edu/abs/2026ApJ...998..224F}
}

@ARTICLE{Liszt2014ApJ...780...10L,
       author = {{Liszt}, Harvey},
        title = "{N(H I)/E(B - V)}",
      journal = {\apj},
         year = 2014,
        month = jan,
       volume = {780},
       number = {1},
          eid = {10},
        pages = {10},
          doi = {10.1088/0004-637X/780/1/10},
archivePrefix = {arXiv},
       eprint = {1310.6616},
 primaryClass = {astro-ph.GA},
       adsurl = {https://ui.adsabs.harvard.edu/abs/2014ApJ...780...10L}
}

@ARTICLE{Bohlin1978ApJ...224..132B,
       author = {{Bohlin}, R.~C. and {Savage}, B.~D. and {Drake}, J.~F.},
        title = "{A survey of interstellar H I from Lalpha absorption measurements. II.}",
      journal = {\apj},
         year = 1978,
        month = aug,
       volume = {224},
        pages = {132-142},
          doi = {10.1086/156357},
       adsurl = {https://ui.adsabs.harvard.edu/abs/1978ApJ...224..132B}
}

@ARTICLE{Zucker2020A&A...633A..51Z,
       author = {{Zucker}, Catherine and {Speagle}, Joshua S. and {Schlafly}, Edward F. and {Green}, Gregory M. and {Finkbeiner}, Douglas P. and {Goodman}, Alyssa and {Alves}, Jo{\~a}o},
        title = "{A compendium of distances to molecular clouds in the Star Formation Handbook}",
      journal = {\aap},
         year = 2020,
        month = jan,
       volume = {633},
          eid = {A51},
        pages = {A51},
          doi = {10.1051/0004-6361/201936145},
archivePrefix = {arXiv},
       eprint = {2001.00591},
 primaryClass = {astro-ph.GA},
       adsurl = {https://ui.adsabs.harvard.edu/abs/2020A&A...633A..51Z}
}

@ARTICLE{Kainulainen2013A&A...549A..53K,
       author = {{Kainulainen}, J. and {Tan}, J.~C.},
        title = "{High-dynamic-range extinction mapping of infrared dark clouds. Dependence of density variance with sonic Mach number in molecular clouds}",
      journal = {\aap},
         year = 2013,
        month = jan,
       volume = {549},
          eid = {A53},
        pages = {A53},
          doi = {10.1051/0004-6361/201219526},
archivePrefix = {arXiv},
       eprint = {1210.8130},
 primaryClass = {astro-ph.GA},
       adsurl = {https://ui.adsabs.harvard.edu/abs/2013A&A...549A..53K}
}

@ARTICLE{Tremblin_2014A&A...564A.106T,
       author = {{Tremblin}, P. and {Schneider}, N. and {Minier}, V. and {Didelon}, P. and {Hill}, T. and {Anderson}, L.~D. and {Motte}, F. and {Zavagno}, A. and {Andr{\'e}}, Ph. and {Arzoumanian}, D. and {Audit}, E. and {Benedettini}, M. and {Bontemps}, S. and {Csengeri}, T. and {Di Francesco}, J. and {Giannini}, T. and {Hennemann}, M. and {Nguyen Luong}, Q. and {Marston}, A.~P. and {Peretto}, N. and {Rivera-Ingraham}, A. and {Russeil}, D. and {Rygl}, K.~L.~J. and {Spinoglio}, L. and {White}, G.~J.},
        title = "{Ionization compression impact on dense gas distribution and star formation. Probability density functions around H II regions as seen by Herschel}",
      journal = {\aap},
         year = 2014,
        month = apr,
       volume = {564},
          eid = {A106},
        pages = {A106},
          doi = {10.1051/0004-6361/201322700},
archivePrefix = {arXiv},
       eprint = {1401.7333},
 primaryClass = {astro-ph.GA},
       adsurl = {https://ui.adsabs.harvard.edu/abs/2014A&A...564A.106T}
}

@ARTICLE{Schneider_2022A&A...666A.165S,
       author = {{Schneider}, N. and {Ossenkopf-Okada}, V. and {Clarke}, S. and {Klessen}, R.~S. and {Kabanovic}, S. and {Veltchev}, T. and {Bontemps}, S. and {Dib}, S. and {Csengeri}, T. and {Federrath}, C. and {Di Francesco}, J. and {Motte}, F. and {Andr{\'e}}, Ph. and {Arzoumanian}, D. and {Beattie}, J.~R. and {Bonne}, L. and {Didelon}, P. and {Elia}, D. and {K{\"o}nyves}, V. and {Kritsuk}, A. and {Ladjelate}, B. and {Myers}, Ph. and {Pezzuto}, S. and {Robitaille}, J.~F. and {Roy}, A. and {Seifried}, D. and {Simon}, R. and {Soler}, J. and {Ward-Thompson}, D.},
        title = "{Understanding star formation in molecular clouds. IV. Column density PDFs from quiescent to massive molecular clouds}",
      journal = {\aap},
         year = 2022,
        month = oct,
       volume = {666},
          eid = {A165},
        pages = {A165},
          doi = {10.1051/0004-6361/202039610},
archivePrefix = {arXiv},
       eprint = {2207.14604},
 primaryClass = {astro-ph.GA},
       adsurl = {https://ui.adsabs.harvard.edu/abs/2022A&A...666A.165S}
}

@ARTICLE{Girichidis_2014ApJ...781...91G,
       author = {{Girichidis}, Philipp and {Konstandin}, Lukas and {Whitworth}, Anthony P. and {Klessen}, Ralf S.},
        title = "{On the Evolution of the Density Probability Density Function in Strongly Self-gravitating Systems}",
      journal = {\apj},
         year = 2014,
        month = feb,
       volume = {781},
       number = {2},
          eid = {91},
        pages = {91},
          doi = {10.1088/0004-637X/781/2/91},
archivePrefix = {arXiv},
       eprint = {1310.4346},
 primaryClass = {astro-ph.GA},
       adsurl = {https://ui.adsabs.harvard.edu/abs/2014ApJ...781...91G}
}

@ARTICLE{Lombardi_2008A&A...489..143L,
       author = {{Lombardi}, M. and {Lada}, C.~J. and {Alves}, J.},
        title = "{2MASS wide field extinction maps. II. The Ophiuchus and the Lupus cloud complexes}",
      journal = {\aap},
         year = 2008,
        month = oct,
       volume = {489},
       number = {1},
        pages = {143-156},
          doi = {10.1051/0004-6361:200810070},
archivePrefix = {arXiv},
       eprint = {0809.3740},
 primaryClass = {astro-ph},
       adsurl = {https://ui.adsabs.harvard.edu/abs/2008A&A...489..143L}
}

@ARTICLE{Froebrich_2010MNRAS.406.1350F,
       author = {{Froebrich}, Dirk and {Rowles}, Jonathan},
        title = "{The structure of molecular clouds - II. Column density and mass distributions}",
      journal = {\mnras},
         year = 2010,
        month = aug,
       volume = {406},
       number = {2},
        pages = {1350-1357},
          doi = {10.1111/j.1365-2966.2010.16769.x},
archivePrefix = {arXiv},
       eprint = {1004.0117},
 primaryClass = {astro-ph.GA},
       adsurl = {https://ui.adsabs.harvard.edu/abs/2010MNRAS.406.1350F}
}

@ARTICLE{Stutz_2015A&A...577L...6S,
       author = {{Stutz}, A.~M. and {Kainulainen}, J.},
        title = "{Evolution of column density distributions within Orion A{\ensuremath{\star}}}",
      journal = {\aap},
         year = 2015,
        month = may,
       volume = {577},
          eid = {L6},
        pages = {L6},
          doi = {10.1051/0004-6361/201526243},
archivePrefix = {arXiv},
       eprint = {1504.05188},
 primaryClass = {astro-ph.GA},
       adsurl = {https://ui.adsabs.harvard.edu/abs/2015A&A...577L...6S}
}

@ARTICLE{Schneider_2015A&A...575A..79S,
       author = {{Schneider}, N. and {Ossenkopf}, V. and {Csengeri}, T. and {Klessen}, R.~S. and {Federrath}, C. and {Tremblin}, P. and {Girichidis}, P. and {Bontemps}, S. and {Andr{\'e}}, Ph.},
        title = "{Understanding star formation in molecular clouds. I. Effects of line-of-sight contamination on the column density structure}",
      journal = {\aap},
         year = 2015,
        month = mar,
       volume = {575},
          eid = {A79},
        pages = {A79},
          doi = {10.1051/0004-6361/201423569},
archivePrefix = {arXiv},
       eprint = {1403.2996},
 primaryClass = {astro-ph.GA},
       adsurl = {https://ui.adsabs.harvard.edu/abs/2015A&A...575A..79S}
}

@ARTICLE{Lo_2009MNRAS.395.1021L,
       author = {{Lo}, N. and {Cunningham}, M.~R. and {Jones}, P.~A. and {Bains}, I. and {Burton}, M.~G. and {Wong}, T. and {Muller}, E. and {Kramer}, C. and {Ossenkopf}, V. and {Henkel}, C. and {Deragopian}, G. and {Donnelly}, S. and {Ladd}, E.~F.},
        title = "{Molecular line mapping of the giant molecular cloud associated with RCW 106 - III. Multimolecular line mapping}",
      journal = {\mnras},
         year = 2009,
        month = may,
       volume = {395},
       number = {2},
        pages = {1021-1042},
          doi = {10.1111/j.1365-2966.2009.14594.x},
archivePrefix = {arXiv},
       eprint = {0902.2452},
 primaryClass = {astro-ph.SR},
       adsurl = {https://ui.adsabs.harvard.edu/abs/2009MNRAS.395.1021L}
}

@ARTICLE{Wang_2020A&A...641A..53W,
       author = {{Wang}, Y. and {Beuther}, H. and {Schneider}, N. and {Meidt}, S.~E. and {Linz}, H. and {Ragan}, S. and {Zucker}, C. and {Battersby}, C. and {Soler}, J.~D. and {Schinnerer}, E. and {Bigiel}, F. and {Colombo}, D. and {Henning}, Th.},
        title = "{Dense gas in a giant molecular filament}",
      journal = {\aap},
         year = 2020,
        month = sep,
       volume = {641},
          eid = {A53},
        pages = {A53},
          doi = {10.1051/0004-6361/202037928},
archivePrefix = {arXiv},
       eprint = {2003.05384},
 primaryClass = {astro-ph.GA},
       adsurl = {https://ui.adsabs.harvard.edu/abs/2020A&A...641A..53W}
}

@ARTICLE{Bernard_2010A&A...518L..88B,
       author = {{Bernard}, J.-Ph. and {Paradis}, D. and {Marshall}, D.~J. and {Montier}, L. and {Lagache}, G. and {Paladini}, R. and {Veneziani}, M. and {Brunt}, C.~M. and {Mottram}, J.~C. and {Martin}, P. and {Ristorcelli}, I. and {Noriega-Crespo}, A. and {Compi{\`e}gne}, M. and {Flagey}, N. and {Anderson}, L.~D. and {Popescu}, C.~C. and {Tuffs}, R. and {Reach}, W. and {White}, G. and {Benedettini}, M. and {Calzoletti}, L. and {Digiorgio}, A.~M. and {Faustini}, F. and {Juvela}, M. and {Joblin}, C. and {Joncas}, G. and {Mivilles-Deschenes}, M.-A. and {Olmi}, L. and {Traficante}, A. and {Piacentini}, F. and {Zavagno}, A. and {Molinari}, S.},
        title = "{Dust temperature tracing the ISRF intensity in the Galaxy}",
      journal = {\aap},
         year = 2010,
        month = jul,
       volume = {518},
          eid = {L88},
        pages = {L88},
          doi = {10.1051/0004-6361/201014540},
       adsurl = {https://ui.adsabs.harvard.edu/abs/2010A&A...518L..88B}
}

@ARTICLE{Veltchev_2019MNRAS.489..788V,
       author = {{Veltchev}, Todor V. and {Girichidis}, Philipp and {Donkov}, Sava and {Schneider}, Nicola and {Stanchev}, Orlin and {Marinkova}, Lyubov and {Seifried}, Daniel and {Klessen}, Ralf S.},
        title = "{On the extraction of the power-law parts of probability density functions in star-forming clouds}",
      journal = {\mnras},
         year = 2019,
        month = oct,
       volume = {489},
       number = {1},
        pages = {788-801},
          doi = {10.1093/mnras/stz2151},
archivePrefix = {arXiv},
       eprint = {1908.00489},
 primaryClass = {astro-ph.GA},
       adsurl = {https://ui.adsabs.harvard.edu/abs/2019MNRAS.489..788V}
}

@ARTICLE{Ossenkopf_2016A&A...590A.104O,
       author = {{Ossenkopf-Okada}, V. and {Csengeri}, T. and {Schneider}, N. and {Federrath}, C. and {Klessen}, R.~S.},
        title = "{The reliability of observational measurements of column density probability distribution functions}",
      journal = {\aap},
         year = 2016,
        month = may,
       volume = {590},
          eid = {A104},
        pages = {A104},
          doi = {10.1051/0004-6361/201628095},
archivePrefix = {arXiv},
       eprint = {1603.03344},
 primaryClass = {astro-ph.IM},
       adsurl = {https://ui.adsabs.harvard.edu/abs/2016A&A...590A.104O}
}

@ARTICLE{Palmeirim_2013A&A...550A..38P,
       author = {{Palmeirim}, P. and {Andr{\'e}}, Ph. and {Kirk}, J. and {Ward-Thompson}, D. and {Arzoumanian}, D. and {K{\"o}nyves}, V. and {Didelon}, P. and {Schneider}, N. and {Benedettini}, M. and {Bontemps}, S. and {Di Francesco}, J. and {Elia}, D. and {Griffin}, M. and {Hennemann}, M. and {Hill}, T. and {Martin}, P.~G. and {Men'shchikov}, A. and {Molinari}, S. and {Motte}, F. and {Nguyen Luong}, Q. and {Nutter}, D. and {Peretto}, N. and {Pezzuto}, S. and {Roy}, A. and {Rygl}, K.~L.~J. and {Spinoglio}, L. and {White}, G.~L.},
        title = "{Herschel view of the Taurus B211/3 filament and striations: evidence of filamentary growth?}",
      journal = {\aap},
         year = 2013,
        month = feb,
       volume = {550},
          eid = {A38},
        pages = {A38},
          doi = {10.1051/0004-6361/201220500},
archivePrefix = {arXiv},
       eprint = {1211.6360},
 primaryClass = {astro-ph.SR},
       adsurl = {https://ui.adsabs.harvard.edu/abs/2013A&A...550A..38P}
}

@ARTICLE{Giannetti_2017A&A...606L..12G,
       author = {{Giannetti}, A. and {Leurini}, S. and {K{\"o}nig}, C. and {Urquhart}, J.~S. and {Pillai}, T. and {Brand}, J. and {Kauffmann}, J. and {Wyrowski}, F. and {Menten}, K.~M.},
        title = "{Galactocentric variation of the gas-to-dust ratio and its relation with metallicity}",
      journal = {\aap},
         year = 2017,
        month = oct,
       volume = {606},
          eid = {L12},
        pages = {L12},
          doi = {10.1051/0004-6361/201731728},
archivePrefix = {arXiv},
       eprint = {1710.05721},
 primaryClass = {astro-ph.GA},
       adsurl = {https://ui.adsabs.harvard.edu/abs/2017A&A...606L..12G}
}

@ARTICLE{Roueff_2021A&A...645A..26R,
       author = {{Roueff}, Antoine and {Gerin}, Maryvonne and {Gratier}, Pierre and {Levrier}, Fran{\c{c}}ois and {Pety}, J{\'e}r{\^o}me and {Gaudel}, Mathilde and {Goicoechea}, Javier R. and {Orkisz}, Jan H. and {de Souza Magalhaes}, Victor and {Vono}, Maxime and {Bardeau}, S{\'e}bastien and {Bron}, Emeric and {Chanussot}, Jocelyn and {Chainais}, Pierre and {Guzman}, Viviana V. and {Hughes}, Annie and {Kainulainen}, Jouni and {Languignon}, David and {Le Bourlot}, Jacques and {Le Petit}, Franck and {Liszt}, Harvey S. and {Marchal}, Antoine and {Miville-Desch{\^e}nes}, Marc-Antoine and {Peretto}, Nicolas and {Roueff}, Evelyne and {Sievers}, Albrecht},
        title = "{C$^{18}$O, $^{13}$CO, and $^{12}$CO abundances and excitation temperatures in the Orion B molecular cloud. Analysis of the achievable precision in modeling spectral lines within the approximation of the local thermodynamic equilibrium}",
      journal = {\aap},
         year = 2021,
        month = jan,
       volume = {645},
          eid = {A26},
        pages = {A26},
          doi = {10.1051/0004-6361/202037776},
archivePrefix = {arXiv},
       eprint = {2005.08317},
 primaryClass = {astro-ph.GA},
       adsurl = {https://ui.adsabs.harvard.edu/abs/2021A&A...645A..26R}
}

@ARTICLE{Wu_2012ApJ...756...76W,
       author = {{Wu}, Yuefang and {Liu}, Tie and {Meng}, Fanyi and {Li}, Di and {Qin}, Sheng-Li and {Ju}, Bing-Gang},
        title = "{Gas Emissions in Planck Cold Dust Clumps{\textemdash}A Survey of the J = 1-0 Transitions of $^{12}$CO, $^{13}$CO, and C$^{18}$O}",
      journal = {\apj},
         year = 2012,
        month = sep,
       volume = {756},
       number = {1},
          eid = {76},
        pages = {76},
          doi = {10.1088/0004-637X/756/1/76},
archivePrefix = {arXiv},
       eprint = {1206.7027},
 primaryClass = {astro-ph.SR},
       adsurl = {https://ui.adsabs.harvard.edu/abs/2012ApJ...756...76W}
}

@ARTICLE{Barnes_2017MNRAS.469.2263B,
    author = "{Barnes}, A. T. and {Longmore}, S. N. and {Battersby}, C. and {Bally}, J. and {Kruijssen}, J. M. D. and {Henshaw}, J. D. and {Walker}, D. L.",
    title = "{Star formation rates and efficiencies in the Galactic Centre}",
    journal = "\mnras",
    year = "2017",
    month = "August",
    volume = "469",
    number = "2",
    pages = "2263-2285",
    doi = "10.1093/mnras/stx941",
    archivePrefix = "arXiv",
    eprint = "1704.03572",
    primaryClass = "astro-ph.GA",
    adsurl = "https://ui.adsabs.harvard.edu/abs/2017MNRAS.469.2263B"
}

@ARTICLE{Compiegne_2010ApJ...724L..44C,
       author = {{Compi{\`e}gne}, M. and {Flagey}, N. and {Noriega-Crespo}, A. and {Martin}, P.~G. and {Bernard}, J. -P. and {Paladini}, R. and {Molinari}, S.},
        title = "{Dust in the Diffuse Emission of the Galactic Plane: The Herschel/Spitzer Spectral Energy Distribution Fitting}",
      journal = {\apjl},
         year = 2010,
        month = nov,
       volume = {724},
       number = {1},
        pages = {L44-L47},
          doi = {10.1088/2041-8205/724/1/L44},
archivePrefix = {arXiv},
       eprint = {1010.2774},
 primaryClass = {astro-ph.GA},
       adsurl = {https://ui.adsabs.harvard.edu/abs/2010ApJ...724L..44C}
}

@ARTICLE{Desert_1990A&A...237..215D,
       author = {{Desert}, F. -X. and {Boulanger}, F. and {Puget}, J.~L.},
        title = "{Interstellar Dust Models for Extinction and Emission}",
      journal = {\aap},
         year = 1990,
        month = oct,
       volume = {237},
        pages = {215},
       adsurl = {https://ui.adsabs.harvard.edu/abs/1990A&A...237..215D}
}

@ARTICLE{Yang_2026ApJS..282...65Y,
       author = {{Yang}, Ji and {Yan}, Qing-Zeng and {Su}, Yang and {Zhang}, Shaobo and {Zhou}, Xin and {Sun}, Yan and {Ao}, Yiping and {Chen}, Xuepeng and {Chen}, Zhiwei and {Du}, Fujun and {Fang}, Min and {Gong}, Yan and {Jiang}, Zhibo and {Jin}, Shengyu and {Ju}, Binggang and {Li}, Chong and {Li}, Yingjie and {Liu}, Yi and {Lu}, Dengrong and {Luo}, Chunsheng and {Ma}, Yuehui and {Mao}, Ruiqing and {Sun}, Jixian and {Wang}, Chen and {Wang}, Hongchi and {Wang}, Min and {Wang (Qinghai)}, Min and {Wang}, Xindong and {Xu}, Wenting and {Xu}, Ye and {Yan}, Kun and {Yan}, Ping and {Yuan}, Lixia and {Zhang}, Miaomiao and {Zhang}, Yongxing},
        title = "{The Milky Way Imaging Scroll Painting Survey: Data Release 1}",
      journal = {\apjs},
         year = 2026,
        month = feb,
       volume = {282},
       number = {2},
          eid = {65},
        pages = {65},
          doi = {10.3847/1538-4365/ae29e7},
archivePrefix = {arXiv},
       eprint = {2512.08260},
 primaryClass = {astro-ph.GA},
       adsurl = {https://ui.adsabs.harvard.edu/abs/2026ApJS..282...65Y}
}

@ARTICLE{Chapman_2009ApJ...690..496C,
       author = {{Chapman}, Nicholas L. and {Mundy}, Lee G. and {Lai}, Shih-Ping and {Evans}, II, Neal J.},
        title = "{The Mid-Infrared Extinction Law in the Ophiuchus, Perseus, and Serpens Molecular Clouds}",
      journal = {\apj},
         year = 2009,
        month = jan,
       volume = {690},
       number = {1},
        pages = {496-511},
          doi = {10.1088/0004-637X/690/1/496},
archivePrefix = {arXiv},
       eprint = {0809.1106},
 primaryClass = {astro-ph},
       adsurl = {https://ui.adsabs.harvard.edu/abs/2009ApJ...690..496C}
}

@ARTICLE{Tyagi_2025ApJ...983..110T,
       author = {{Tyagi}, Himanshu and {Manoj}, P. and {Narang}, Mayank and {Megeath}, S. Thomas and {Rocha}, Will R.~M. and {Brunken}, Nashanty and {Rubinstein}, Adam E. and {Gutermuth}, Robert and {Evans}, Neal J. and {Van Dishoeck}, Ewine F. and {Federman}, Samuel and {Watson}, Dan M. and {Neufeld}, David A. and {Anglada}, Guillem and {Beuther}, Henrik and {Caratti o Garatti}, Alessio and {Looney}, Leslie W. and {Nazari}, Pooneh and {Osorio}, Mayra and {Stanke}, Thomas and {Yang}, Yao-Lun and {Bourke}, Tyler L. and {Fischer}, William J. and {Furlan}, Elise and {Green}, Joel and {Habel}, Nolan and {Klaassen}, Pamela and {Karnath}, Nicole and {Linz}, Hendrik and {Muzerolle}, James and {Tobin}, John J. and {Atnagulov}, Prabhani and {Rahatgaonkar}, Rohan and {Sheehan}, Patrick and {Slavicinska}, Katerina and {Stutz}, Amelia M. and {Tychoniec}, Lukasz and {Wolk}, Scott and {Zakri}, Wafa},
        title = "{JWST-IPA: Chemical Inventory and Spatial Mapping of Ices in the Protostar HOPS 370{\textemdash}Evidence for an Opacity Hole and Thermal Processing of Ices}",
      journal = {\apj},
         year = 2025,
        month = apr,
       volume = {983},
       number = {2},
          eid = {110},
        pages = {110},
          doi = {10.3847/1538-4357/adb71f},
archivePrefix = {arXiv},
       eprint = {2410.06697},
 primaryClass = {astro-ph.SR},
       adsurl = {https://ui.adsabs.harvard.edu/abs/2025ApJ...983..110T}
}
\bibliographystyle{aasjournalv7}

\end{document}